\documentclass{aa501}
\usepackage{graphicx}
\usepackage{amssymb}
\usepackage{amsmath}
\usepackage{natbib}
\usepackage{rotating}

\begin{document}

\title{Integrated Spectroscopy of Bulge Globular Clusters and Fields}
\subtitle{I. The Data Base and Comparison of Individual Lick Indices\\
in Clusters and Bulge}

   \author{Thomas H. Puzia \inst{1}, Roberto P. Saglia \inst{1},
        Markus Kissler-Patig \inst{2}, Claudia Maraston \inst{3}, 
        Laura Greggio \inst{1,4},\\ Alvio Renzini \inst{2}, 
        \& Sergio Ortolani\inst{4}
        }

   \offprints{puzia@usm.uni-muenchen.de}

   \institute{Sternwarte der Ludwig-Maximilians-Universit\"at,
              Scheinerstrasse 43, D--81679 M\"unchen, Germany,\\
              email: puzia, saglia, greggio@usm.uni-muenchen.de
         \and
             European Southern Observatory, Karl-Schwarzschild-Strasse
              2, D--85748 Garching bei M\"unchen, Germany,\\
             email: mkissler, arenzini@eso.org
         \and
	     Max-Planck-Institut f\"ur Extraterrestrische Physik,
             Giessenbachstrasse, D--85748 Garching bei M\"unchen,
             Germany, email: maraston@mpe.mpg.de
	 \and
             Universit\`a di Padova, Dept. di Astronomia, 
             Vicolo dell'Osservatorio 2, 35122 Padova, Italy, \\
             email: ortolani@pd.astro.it
             }
           
   \authorrunning{Puzia et al.}  
   \titlerunning{Integrated Spectroscopy of Bulge Globular Clusters and Fields} 
           
   \date{Received June 2002; accepted ... }
   
   \abstract{We present a comprehensive spectroscopic study of the
     integrated light of metal-rich Galactic globular clusters and the
     stellar population in the Galactic bulge. We measure line indices
     which are defined by the Lick standard system and compare index
     strengths of the clusters and Galactic bulge. Both metal-rich
     globular clusters and the bulge are similar in most of the
     indices, except for the CN index. We find a significant
     enhancement in the CN$/\langle$Fe$\rangle$ index ratio in
     metal-rich globular clusters compared with the Galactic
     bulge. The mean iron index $\langle$Fe$\rangle$ of the two
     metal-rich globular clusters NGC~6528 and NGC~6553 is comparable
     with the mean iron index of the bulge. Index ratios such as
     Mgb$/\langle$Fe$\rangle$, Mg$_2/\langle$Fe$\rangle$,
     Ca4227$/\langle$Fe$\rangle$, and TiO$/\langle$Fe$\rangle$, are
     comparable in both stellar population indicating similar
     enhancements in individual elements which are traced by the
     indices. From the globular cluster data we fully empirically
     calibrate several metallicity-sensitive indices as a function of
     [Fe/H] and find tightest correlations for the Mg$_2$ index and
     the composite [MgFe] index. We find that all indices show a
     similar behavior with galactocentric radius, except for the
     Balmer series, which show a large scatter at all radii. However,
     the scatter is entirely consistent with the cluster-to-cluster
     variations in the horizontal branch morphology.}

   \maketitle
     


\section{Introduction}

Stars in globular clusters are essentially coeval and -- with very few
exceptions -- have all the same chemical composition, with only few
elements breaking the rule. As such, globular clusters are the best
approximation to {\it simple stellar populations} (SSP), and therefore
offer a virtually unique opportunity to relate the integrated spectrum
of stellar populations to age and chemical composition, and do it in a
fully empirical fashion. Indeed, the chemical composition can be
determined via high-resolution spectroscopy of cluster stars, the age
via the cluster turnoff luminosity, while integrated spectroscopy of
the cluster can also be obtained without major difficulties. In this
way, empirical relations can be established between integrated-light
line indices \citep[e.g. Lick indices as defined by][]{faber85} of the
clusters, on one hand, and their age and chemical composition on the
other hand (i.e., [Fe/H], [$\alpha$/Fe], etc.).

These empirical relations are useful in two major applications: 1) to
directly estimate the age and chemical composition of unresolved
stellar populations for which integrated spectroscopy is available
(e.g. for elliptical galaxies and spiral bulges), and 2) to provide a
basic check of population synthesis models.

Today we know of about 150 globular clusters in the Milky Way
\citep{harris96}, and more clusters might be hidden behind the
high-absorption regions of the Galactic disk. Like in the case of many
elliptical galaxies \citep[e.g.][]{harris01}, the Galactic globular
cluster system shows a bimodal metallicity distribution
\citep{freeman81,zinn85,ashman98,harris01} and consists of two major
sub-populations, the metal-rich bulge and the metal-poor halo
sub-populations.

The metal-rich ($\mathrm{[Fe/H]}>-0.8$ dex) component was initially
referred to as a ``disk'' globular cluster system \citep{zinn85}, but
it is now clear that the metal-rich globular clusters physically
reside inside the bulge and share its chemical and kinematical
properties \citep{minniti95,barbuy98,cote99}. Moreover, the best
studied metal-rich clusters (NGC 6528 and NGC 6553) appear to have
virtually the same old age as both the halo clusters and the general
bulge population \citep{ortolani95n, feltzing00, ortolani01,
zoccali01, zoccali02, feltzing02}, hence providing important clues on
the formation of the Galactic bulge and of the whole Milky Way galaxy.

Given their relatively high metallicity (up to $\sim Z_\odot$), the
bulge globular clusters are especially interesting in the context of
stellar population studies, as they allow comparisons of their
spectral indices with those of other spheroids, such as elliptical
galaxies and spiral bulges. However, while Lick indices have been
measured for a representative sample of metal-poor globular clusters
\citep{burstein84, covino95, trager98}, no such indices had been
measured for the more metal-rich clusters of the Galactic bulge. It is
the primary aim of this paper to present and discuss the results of
spectroscopic observations of a set of metal-rich globular clusters
that complement and extend the dataset so far available only for
metal-poor globulars.

Substantial progress has been made in recent years to gather the
complementary data to this empirical approach: i.e. ages and chemical
composition of the metal-rich clusters. Concerning ages, HST/WFPC2
observations of the clusters NGC~6528 and NGC~6553 have been critical
to reduce to a minimum and eventually to eliminate the contamination
of foreground disk stars (see references above), while HST/NICMOS
observations have started to extend these studies to other, more
heavily obscured clusters of the bulge \citep{ortolani01}.

High spectral-resolution studies of individual stars in these clusters
is still scanty, but one can expect rapid progress as high multiplex
spectrographs become available at 8--10m class telescopes. A few stars
in NGC~6528 and NGC~6553 have been observed at high spectral
resolution, but with somewhat discrepant results. For NGC~6528,
\cite{carretta01} and \cite{coelho01} report respectively [Fe/H]$
=+0.07$ and $-0.5$ dex (the latter value coming from low-resolution
spectra). For [M/H] the same authors derive $+0.17$ and $-0.25$ dex,
respectively. For NGC~6553 \cite{barbuy99} give [Fe/H]$ =-0.55$ dex
and [M/H]$ =-0.08$ dex, while \cite{cohen99} report [Fe/H]$ =-0.16$
dex, and \cite{origlia02} give [Fe/H]$ =-0.3$ dex, with [$\alpha$/Fe]$
=+0.3$ dex. Some $\alpha$-element enhancement has also been found
among bulge field stars, yet with apparently different
element-to-element ratios \citep{mcwilliam94}.

Hopefully these discrepancies may soon disappear, as more and better
quality high-resolution data are gathered at 8--10m class
telescopes. In summary, the overall metallicity of these two clusters
(whose color magnitude diagrams are virtually identical,
\citealt{ortolani95n}) appears to be close to solar, with an
$\alpha$-element enhancement [$\alpha$/Fe] $\simeq +0.3$ dex.

The $\alpha$-element enhancement plays an especially important role in
the present study. It is generally interpreted as the result of most
stars having formed rapidly (within less than, say $\sim 1$ Gyr), thus
having had the time to incorporate the $\alpha$-elements produced
predominantly by Type II supernovae, but failing to incorporate most
of the iron produced by the longer-living progenitors of Type Ia
supernovae. Since quite a long time, an $\alpha$-element enhancement
has been suspected for giant elliptical galaxies, inferred from the a
comparison of Mg and Fe indices with theoretical models
\citep{peletier89,worthey92, davies93, greggio97}. This interpretation
has far-reaching implications for the star formation timescale of
these galaxies, with a fast star formation being at variance with the
slow process, typical of the current hierarchical merging scenario
\citep{thomas99}. However, in principle the apparent $\alpha$-element
enhancement may also be an artifact of some flaws in the models of
synthetic stellar populations, especially at high metallicity
\citep{maraston01}. The observations presented in this paper are also
meant to provide a dataset against which to conduct a direct test of
population synthesis models, hence either excluding or straightening
the case for an $\alpha$-element enhancement in elliptical
galaxies. This aspect is extensively addressed in an accompanying
paper \citep{maraston02}.

The main goal of this work is the measurement of the Lick indices for
the metal-rich globular clusters of the bulge and of the bulge field
itself. Among others, we measure line indices of Fe, Mg, Ca, CN, and
the Balmer series which are defined in the Lick standard system
\citep{worthey97,trager98}. In \S2 we describe in detail the
observations and our data reduction which leads to the analysis and
measurement of line indices in \S3. Index ratios in globular clusters
and the bulge are presented in \S4. Index-metallicity relations are
calibrated with the new data in \S5 and \S6 discusses the index
variations as a function of galactocentric radius. \S7 closes this
work with the conclusions followed by a summary in \S8.

\begin{table*}[t!]
\centering
\caption{General properties of sample Globular Clusters. If not else
mentioned, all data were taken from the 1999 update of the McMaster
catalog of Milky Way Globular Clusters \citep{harris96}. $R_{\rm gc}$
is the globular cluster distance from the Galactic Center. $r_h$ is
the half-light radius. E$_{(B-V)}$ and $(m-M)_V$ are the reddening and
the distance modulus. $v_{\rm rad}$ the heliocentric radial
velocity. Note, that our radial-velocity errors are simple {\it
internal} errors which result from the fitting of the
cross-correlation peak. The real {\it external} errors are a factor
$\sim3-4$ larger. HBR is the horizontal-branch morphology parameter
\citep[e.g.][]{lee94}.}
\label{tab:gcprop}
\begin{tabular}{l|cccccrrr}
\hline
\noalign{\smallskip}
 GC &$R_{\rm gc}$ [kpc]
    & [Fe/H]
    &$r_h$ [arcmin]
    &E$_{(B-V)}^{\mathrm{a}}$ 
    &$(m-M)_V$
    &$v_{\rm rad}^{\mathrm{b}}$ [km s$^{-1}$] 
    &$v_{\rm rad}$ [km s$^{-1}$] 
    & HBR$^{\mathrm{c}}$ \\   
\noalign{\smallskip}
\hline
\noalign{\smallskip}
 NGC 5927      & 4.5&$-0.37$&1.15&0.45& 15.81&$ -130\pm12$&$-107.5\pm 1.0$ & $-1.00^d$ \\
 NGC 6218 (M12)& 4.5&$-1.48$&2.16&0.40& 14.02&$  -46\pm23$&$ -42.2\pm 0.5$ & $ 0.97^d$ \\
 NGC 6284      & 6.9&$-1.32$&0.78&0.28& 16.70&$    8\pm16$&$  27.6\pm 1.7$ & $ 1.00^e$ \\
 NGC 6356      & 7.6&$-0.50$&0.74&0.28& 16.77&$   35\pm12$&$  27.0\pm 4.3$ & $-1.00^d$ \\
 NGC 6388      & 4.4&$-0.60$&0.67&0.40& 16.54&$   58\pm10$&$  81.2\pm 1.2$ & $-0.70^e$ \\
 NGC 6441      & 3.5&$-0.53$&0.64&0.44& 16.62&$  -13\pm10$&$  16.4\pm 1.2$ & $-0.70^f$ \\
 NGC 6528      & 1.3&$-0.17$&0.43&0.56& 16.53&$  180\pm10$&$ 184.9\pm 3.8$ & $-1.00^d$ \\
 NGC 6553      & 2.5&$-0.34$&1.55&0.75& 16.05&$  -25\pm16$&$  -6.5\pm 2.7$ & $-1.00^d$ \\
 NGC 6624      & 1.2&$-0.42$&0.82&0.28& 15.37&$   27\pm12$&$  53.9\pm 0.6$ & $-1.00^d$ \\
 NGC 6626 (M28)& 2.6&$-1.45$&1.56&0.43& 15.12&$  -15\pm15$&$  17.0\pm 1.0$ & $ 0.90^d$ \\
 NGC 6637 (M69)& 1.6&$-0.71$&0.83&0.16& 15.16&$    6\pm12$&$  39.9\pm 2.8$ & $-1.00^d$ \\ 
 NGC 6981 (M72)&12.9&$-1.40$&0.88&0.05& 16.31&$ -360\pm18$&$-345.1\pm 3.7$ & $ 0.14^d$ \\
\noalign{\smallskip}
\hline
\end{tabular}
\begin{list}{}{}
 \item[$^{\mathrm{a}}$] taken from \cite{harris96}
 \item[$^{\mathrm{b}}$] this work
 \item[$^{\mathrm{c}}$] horizontal branch parameter, (B$-$R)/(B+V+R),
 for details see e.g. \cite{lee94}
 \item[$^{\mathrm{d}}$] taken from \cite{harris96}
 \item[$^{\mathrm{e}}$] taken from \cite{zoccali00}
 \item[$^{\mathrm{f}}$] Due to very similar HB morphologies in CMDs of
 NGC~6388 and NGC~6441 \citep[see][]{rich97}, we assume that the HBR
 parameter is similar for both globular clusters and adopt HBR$=-0.70$
 for NGC~6441.
\end{list}
\end{table*}

\section{Observations and Data Reduction}
\subsection{Observations}

\begin{figure*}
\centering 
\includegraphics[width=16.8cm]{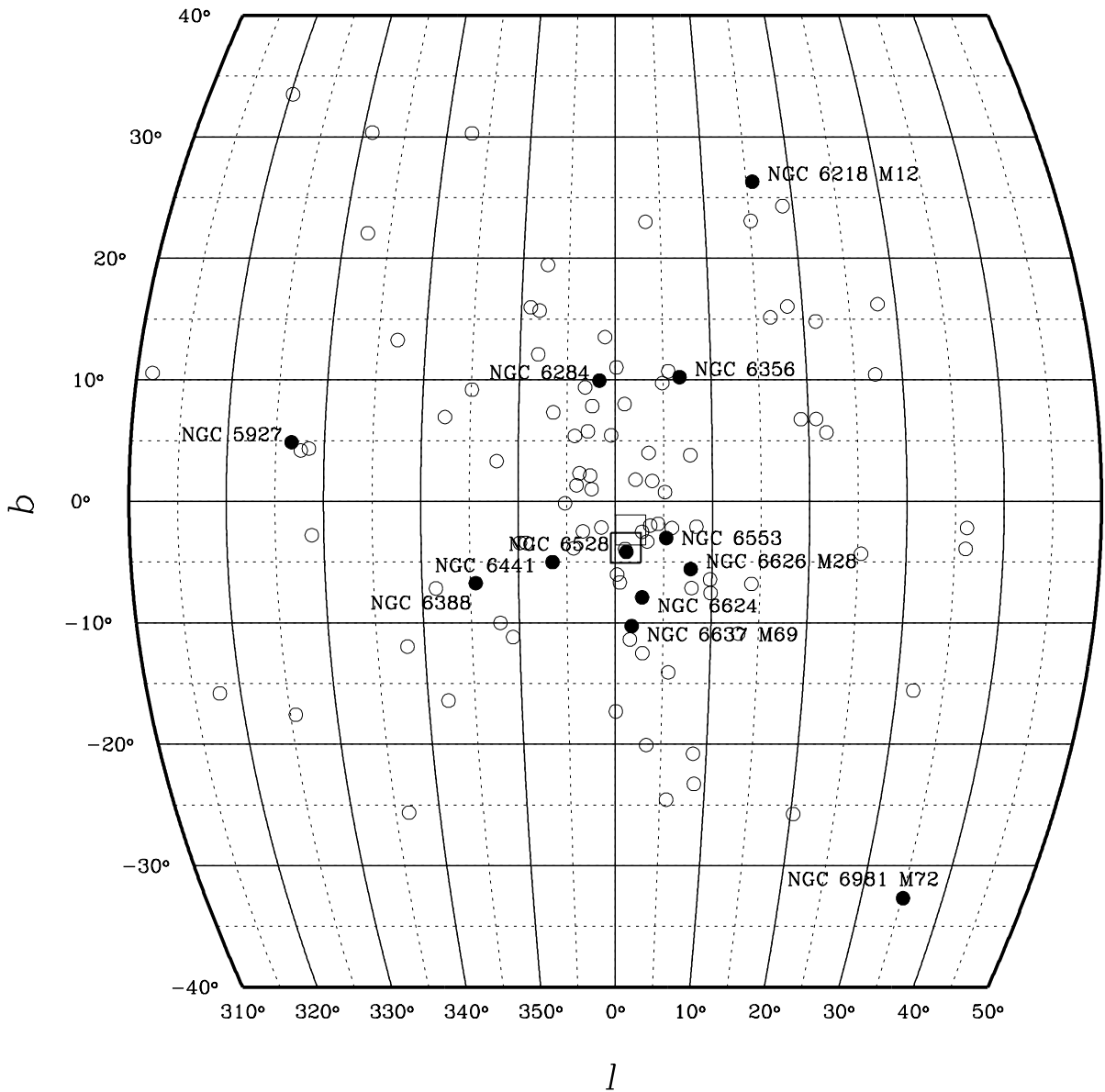}
 \caption{Distribution of galactic globular clusters as seen in
   the galactic coordinate system. The filled circles are the observed
   sample globular clusters while open circles mark the position of
   other known Milky Way globular clusters. All observed globular
   clusters are appropriately labeled. The positions were taken from
   the Globular Cluster Catalog by \cite{harris96}. Large squares show
   the positions of our three bulge fields for which spectroscopy is
   also available. Note that two of the three fields almost overlap in
   the plot.}
 \label{ps:fov}
\end{figure*}

We observed 12 Galactic globular clusters, 9 of which are located
close to the Milky-Way bulge (see Fig.~\ref{ps:fov}). Four globular
clusters belong to the halo sub-population with a mean metallicity
[Fe/H]$\leq-0.8$ dex \citep{harris96}. The other globular clusters
with higher mean metallicities are associated with the bulge. Our
sample includes the well-studied metal-rich clusters NGC 6553 and NGC
6528, which is located in Baade's Window. Several relevant cluster
properties are summarized in Table~\ref{tab:gcprop}. Our cluster
sample was selected to maximize the number of high-metallicity
clusters and to ensure a high enough signal--to--noise ratio (S/N) of
the resulting spectra.
    
Long-slit spectra were taken on three nights in July 5th to 7th 1999
with the Boller \&\ Chivens Spectrograph of ESO's 1.52~m on La
Silla. We used grating \#23 with 600 grooves per mm yielding a
dispersion of 1.89~\AA /pix with a spectral range from $\sim3400$~\AA\
to $\sim7300$~\AA. We used the detector CCD \#39, a Loral
2048$\times$2048 pix$^{2}$ chip, with a pixel size of 15~$\mu m$ and a
scale of 0.82\arcsec/pix. Its readout noise is 5.4 $e^-$ and the gain
was measured with 1.2 $e^-/$ADU. In order to check the dark current we
also obtained dark images which resulted in a negligible average dark
current of 0.0024 $e^{-}{\rm s}^{-1}{\rm pix}^{-1}$. The total slit
length of the spectrograph covers 4.5\arcmin\ on the sky. For the
benefit of light sampling the slit width was fixed at 3\arcsec , which
guarantees an instrumental resolution ($\sim6.7$~\AA ) which is
smaller than the average resolution ($\ga8$~\AA) of the Lick standard
system \citep{worthey94etal, trager98}. The mean seeing during the
observing campaign varied between 0.8\arcsec\ and 1.6\arcsec ,
resulting in seeing-limited spectra. Consequently, the stellar disks
are smeared over 1--2 pixel along the spatial axis.

To ensure a representative sampling of the underlying stellar
population we obtained several spectra with slightly offset pointings.
In general three long-slit spectra were taken for each of our target
clusters (see Table \ref{tab:obslog} for details). The observing
pattern was optimized in time (i.e.~in airmass) to obtain one spectrum
of the nuclear region and spectra of adjacent fields by shifting the
telescope a few arc seconds (i.e. $\sim2$ slit widths) to the North
and South. Exposure times were adjusted according to the surface
brightness of each globular cluster to reach an statistically secure
luminosity sampling of the underlying stellar population. Before and
after each block of science exposures, lamp spectra were taken for
accurate wavelength calibration.

In addition to the globular cluster data, we obtained long-slit
spectra of three stellar fields near the Galactic center (see
Fig.~\ref{ps:fov}). Two of them are located in Baade's Window. The
exposure time for a single bulge spectrum is 1800 seconds. Five
slightly offset pointings have been observed in each field resulting
in 15 exposures of 30 min. each.

During each night Lick and flux standard stars were observed for later
index and flux calibrations. Table \ref{tab:obslog} shows the
observing log of all three nights. Figure \ref{ps:fov} gives the
positions of all observed globular clusters (filled dots) and bulge
fields (open squares) in the galactic coordinate system.

\begin{table*}
\begin{center}
\caption{Journal of all performed observations.}
\label{tab:obslog}
\begin{tabular}[angle=0,width=\textwidth]{l|llccrr}
\hline
\noalign{\smallskip}
 Night & Targets &  Exptime & RA(J2000)& DEC (J2000) &$l$[$^{\rm o}$]&$b $[$^{\rm o}$]\\ 
\noalign{\smallskip}
\hline
\noalign{\smallskip}
 5.7.1999 
&NGC 5927 & 3$\times$600s & 15h 28m 00.5s & $-50^{\rm o}$ 40' 22''&326.60&$  4.86$ \\
&NGC 6388 & 3$\times$600s & 17h 36m 17.0s & $-44^{\rm o}$ 44' 06''&345.56&$ -6.74$ \\
&NGC 6528 & 3$\times$600s & 18h 04m 49.6s & $-30^{\rm o}$ 03' 21''&  1.14&$ -4.17$ \\
&NGC 6624 & 3$\times$600s & 18h 23m 40.5s & $-30^{\rm o}$ 21' 40''&  2.79&$ -7.91$ \\
&NGC 6981 & 1$\times$1320s& 20h 53m 27.9s & $-12^{\rm o}$ 32' 13''& 35.16&$-32.68$ \\
&Bulge1   & 5$\times$1800s& 18h 03m 12.1s & $-29^{\rm o}$ 52' 06''&  1.13&$  3.78$ \\
\noalign{\smallskip}
\hline
\noalign{\smallskip}
 6.7.1999 
&NGC 6218 & 3$\times$1200s& 16h 47m 14.5s & $-01^{\rm o}$ 56' 52''& 15.72&$ 26.31$ \\
&NGC 6441 & 3$\times$600s & 17h 50m 12.9s & $-37^{\rm o}$ 03' 04''&353.53&$ -5.01$ \\
&NGC 6553 & 3$\times$720s & 18h 09m 15.6s & $-25^{\rm o}$ 54' 28''&  5.25&$ -3.02$ \\
&NGC 6626 & 3$\times$600s & 18h 24m 32.9s & $-24^{\rm o}$ 52' 12''&  7.80&$ -5.58$ \\
&NGC 6981 & 1$\times$1800s& 20h 53m 27.9s & $-12^{\rm o}$ 32' 13''& 35.16&$-32.68$ \\
&Bulge2   & 5$\times$1800s& 18h 05m 21.3s & $-29^{\rm o}$ 58' 38''&  1.26&$  4.23$ \\
\noalign{\smallskip}
\hline
\noalign{\smallskip}
 7.7.1999 
&NGC 6284 & 3$\times$600s & 17h 04m 28.8s & $-24^{\rm o}$ 45' 53''&358.35&$  9.94$ \\
&NGC 5927 & 2$\times$600s & 15h 28m 00.5s & $-50^{\rm o}$ 40' 22''&326.60&$  4.86$ \\
&NGC 6356 & 3$\times$900s & 17h 23m 35.0s & $-17^{\rm o}$ 48' 47''&  6.72&$ 10.22$ \\
&NGC 6637 & 3$\times$900s & 18h 31m 23.2s & $-32^{\rm o}$ 20' 53''&  1.72&$-10.27$ \\
&NGC 6981 & 1$\times$1800s& 20h 53m 27.9s & $-12^{\rm o}$ 32' 13''& 35.16&$-32.68$ \\
&Bulge3   & 5$\times$1800s& 17h 58m 38.3s & $-28^{\rm o}$ 43' 33''&  1.63&$  2.35$ \\
\noalign{\smallskip}
\hline
\end{tabular}
\end{center}
\end{table*}

\subsection{Data Reduction}
\label{ln:datared}
We homogeneously applied standard reduction techniques to the whole
data set using the IRAF\footnote{IRAF is distributed by the National
  Optical Astronomy Observatories, which are operated by the
  Association of Universities for Research in Astronomy, Inc., under
  cooperative agreement with the National Science Foundation.}
platform \citep{tody93}. The basic data reduction was performed for
each night individually. In brief, a masterbias was subtracted from
the science images followed by a division by a normalized masterflat
spectrum which has been created from five quarz-lamp exposures. The
quality, i.e. the flatness, of the spectra along the spatial axis was
checked on the sky spectra after flatfielding. Any gradients along the
spatial axis were found to be smaller than $\lesssim5$\%.

He-Ne-Ar-Fe lines were used to calibrate all spectra to better than
0.13~\AA\ (r.m.s.). Unfortunately, the beam of the calibration lamp
covers only the central 3.3\arcmin\ along the slit's spatial axis
(perpendicular to the dispersion direction), which allows no precise
wavelength calibration for the outer parts close to the edge of the
CCD chip. We tried, however, to extrapolate a 2-dim.
$\lambda$-calibration to the edges of the long-slit and found a
significant increase in the r.m.s. up to an unacceptable
0.7~\AA. Hence, to avoid calibration biases we use data only from
regions which are covered by the arc lamp beam. Our effective slit
length is therefore 3.3\arcmin\ with a slit width of 3\arcsec . For
each single pixel row along the dispersion axis an individual
wavelength solution was found and subsequently applied to each object,
bulge, and sky spectrum.  After wavelength calibration the signal
along the spatial axis was averaged in $\lambda$-space, i.e. the flux
of 3.3\arcmin\ was averaged to obtain the final spectrum of a single
pointing.

Finally, spectrophotometric standard stars, Feige~56, Feige~110, and
Kopff~27 \citep{stone83,baldwin84} were used to convert counts into
flux units.

\subsection{Radial velocities}
\label{ln:rv}
All radial velocity measurements were carried out after the
subtraction of a background spectrum (see Sect.~\ref{ln:bkgestim})
using cross-correlation with high-S/N template spectra of two globular
clusters in M31 (i.e. 158--213 and 225--280, see \citealt{huchra82}
for nomenclature). Both globular clusters have metallicities which
match the average metallicity of our globular cluster sample. We
strictly followed the recipe of the Fourier cross-correlation which is
implemented in the {\sc fxcor} task of {\tt IRAF} (see {\tt IRAF}
manual for details). Table \ref{tab:gcprop} summarizes the results
including the {\it internal} uncertainties of our measurements
resulting from the fitting of the cross-correlation peak.

Following the rule of thumb, by which $1/10$ of the instrumental
resolution ($\sim6.7$~\AA) transforms into the radial velocity
resolution, we estimate for our spectra a resolution of
$\sim40$~km~s$^{-1}$. In order to estimate the {\it real} uncertainty
we compare the radial velocity measurements of one globular cluster
(NGC~6981) which was observed in all three nights. We find a
dispersion in radial velocity $\sigma_v\approx17$~km~s$^{-1}$ and a
maximal deviation of 32.4~km~s$^{-1}$. A comparison of measured radial
velocities of all our Lick standard stars with values taken from the
literature gives a dispersion of $\sigma_v\approx40$ km s$^{-1}$ which
matches the earlier rough estimate. In the case of NGC~6981, the {\it
internal} error estimate ($\Delta_{\rm cc} v_{\rm rad}=18.4$ km
s$^{-1}$) underestimates the {\it real} radial velocity uncertainty
assumed to be of the order of $\sim40$ ~km~s$^{-1}$ by a factor of
$\sim2$. Note however, that data of lower S/N will produce larger
radial velocity uncertainties. Moreover, taking into account the slit
width of 3\arcsec\ the maximum possible radial velocity error for a
star positioned at the edge of the slit is $\sim200$ km s$^{-1}$. For
high surface-brightness fluctuations inside the slit, this would
inevitably result in larger radial velocity errors than originally
expected from the calibration quality. Since we sum up all the flux
along the slit, we most effectively eliminate this surface-brightness
fluctuation effect. In fact, after a check of all our single spectra,
we find no exceptionally bright star inside the slit aperture, which
could produce a systematic deviation from the mean radial velocity.

After all, we estimate that our {\it real} radial velocity
uncertainties are larger by a factor $\sim2-4$ than the values given
in Table \ref{tab:gcprop}.

\subsection{Transformation to the Lick System}
\label{ln:licktrafo}
The Lick standard system was initially introduced by \cite{burstein84}
in order to study element abundances from low-resolution integrated
spectra of extragalactic stellar systems. It has recently been updated
and refined by several authors \citep{gonzalez93, worthey94etal,
worthey97, trager98}. The Lick system defines line indices for
specific atomic and molecular absorption features, such as Fe, Mg, Ca
and CN, CH, TiO, in the optical range from $\sim4100$~\AA\ to
$\sim6100$~\AA. The definitions of a line index are given in Appendix
\ref{ln:lickcode}. We implemented the measuring procedure in a
software and tested it extensively on original Lick spectra (see
App.~\ref{ln:lickcode} for details). This code is used for all further
measurements.

The Lick system provides two sets of index passband definitions. One
set of 21 passband definitions was published in \cite{worthey94etal}
to which we will refer as the {\it old} set. A {\it new} and refined
set of passband definitions is given in \cite{trager98} which is
supplemented by the Balmer index definitions of \cite{worthey97}. This
new set of 25 indices is used throughout the subsequent
analysis. However, we also provide Lick indices based on the old
passband definitions (see Appendix~\ref{ln:indexmeasurements_old})
which enables a consistent comparison with predictions from SSP models
which make use of fitting functions based on the old set of passband
definitions. Note that indices and model predictions which use two
different passband definition sets are prone to systematic
offsets. This point will be discussed in the second paper of the
series \citep{maraston02}.

Before measuring indices, one has carefully to degrade spectra with
higher resolution to adapt to the resolution of the Lick system. We
strictly followed the approach of \citet{worthey97} and degraded our
spectra to the wavelength-dependent Lick resolution ($\sim11.5$~\AA\
at 4000~\AA, 8.4~\AA\ at 4900~\AA, and 9.8~\AA\ at 6000~\AA). The
effective resolution (FWHM) of our spectra has been determined from
calibration-lamp lines and isolated absorption features in the object
spectra. The smoothing of our data is done with a wavelength-dependent
Gaussian kernel with the width
\begin{equation}
\sigma_{\rm smooth}(\lambda)=\left(\frac{\mathrm{FWHM}(\lambda)_{\rm
    Lick}^2- \mathrm{FWHM}(\lambda)_{\rm data}^2}{8\, {\rm
    ln}2}\right)^{\frac{1}{2}}.
\end{equation}
We tested the shape of absorption lines in our spectra and found that
they are very well represented by a Gaussian. \citeauthor{worthey97}
tested the shape of the absorption lines in the Lick spectra and found
also no deviation from a Gaussian. Both results justify the use of a
Gaussian smoothing kernel.

The smoothing kernel for the bulge stellar fields is generally
narrower since one has to account for the non-negligible velocity
dispersion of bulge field stars. A typical line-of-sight velocity
dispersion $\sigma_{\rm LOS}\approx 100$ km s$^{-1}$ was assumed for
the bulge data \citep[e.g.][]{spaenhauer92}. We do not correct for the
mean velocity dispersion of the globular clusters \citep[$\sigma_{\rm
LOS}\approx 10$ km s$^{-1}$][]{pryor93}.

Another point of concern for low-S/N spectra (S/N$\lesssim$10 per
resolution element) is the slope of the underlying continuum
\citep[see][for detailed discussion of this effect]{beasley00} which
influences the pseudo-continuum estimate for broad features and biases
the index measurement. However, since all our spectra are of high S/N
($\gtrsim 50$ per resolution element), we are not affected by a noisy
continuum.

\begin{figure*}[ht!]
 \centering
  \includegraphics[width=16.0cm]{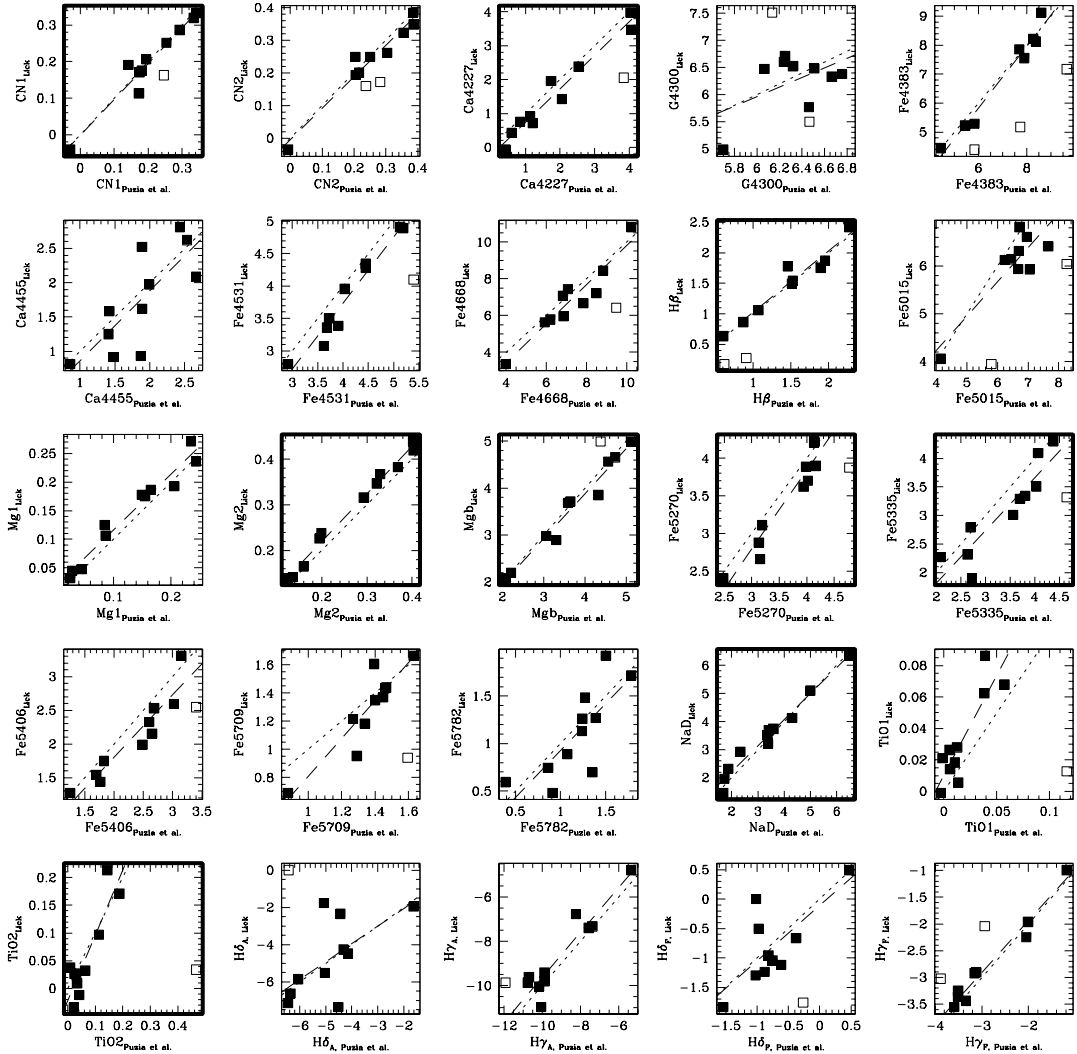}
  \caption{Comparison of passband measurements from our spectra
    and original Lick data for 12 Lick standard stars. The dotted line
    shows the one-to-one relation, whereas the dashed line is a
    least-square fit to the filled squares. Data, which have been
    discarded from the fit because of too large errors or deviations,
    are shown as open squares. Bold frames indicate some of the
    widely used Lick indices which are also analysed in this work.}
 \label{ps:idxcomp}
\end{figure*}

After taking care of the resolution corrections, one has to correct
for systematic, higher-order effects. These variations are mainly due
to imperfect smoothing and calibration of the spectra. To correct the
small deviations 12 index standard stars from the list of
\cite{worthey94etal} have been observed throughout the observing run.
Figure \ref{ps:idxcomp} shows the comparison between the Lick data and
our index measurements for all passbands. Least-square fits using a
$\kappa$-$\sigma$-clipping (dashed lines) are used to parameterize the
deviations from the Lick system as a function of wavelength. The
functional form of the fit is
$${\rm EW}_{\rm cal} = \alpha + (1+\beta)\cdot{\rm EW}_{\rm raw},$$ where
${\rm EW}_{\rm cal}$ and ${\rm EW}_{\rm raw}$ are the calibrated and
raw indices, respectively. Table~\ref{tab:indexlicktrafo} summarizes
the individual coefficients $\alpha$ and $\beta$. This correction
functions are applied to all further measurements. The corresponding
coefficients for index measurements using the {\it old} passband
definitions are documented in Table~\ref{tab:indexlicktrafo_old}.

\begin{table}
\begin{center}
\caption{Summary of the coefficients $\alpha$ and
  $\beta$ for all 1st and 2nd-order index corrections.}
\label{tab:indexlicktrafo}
\begin{tabular}[angle=0,width=\textwidth]{l|rrrc}
\hline
\noalign{\smallskip}
 index & $\alpha$ & $\beta$ & r.m.s. & units\\ 
\noalign{\smallskip}
\hline
\noalign{\smallskip}
CN$_1$   & $-0.0017$ & $-0.0167$ & $0.0251$ &mag \\
CN$_2$   & $-0.0040$ & $-0.0389$ & $0.0248$ &mag \\
Ca4227   & $-0.2505$ & $-0.0105$ & $0.2582$ &\AA \\
G4300    & $ 0.6695$ & $-0.1184$ & $0.4380$ &\AA \\
Fe4384   & $-0.5773$ & $ 0.0680$ & $0.2933$ &\AA \\
Ca4455   & $-0.1648$ & $ 0.0249$ & $0.4323$ &\AA \\
Fe4531   & $-0.3499$ & $ 0.0223$ & $0.1566$ &\AA \\
Fe4668   & $-0.8643$ & $ 0.0665$ & $0.5917$ &\AA \\
H$\beta$ & $ 0.0259$ & $ 0.0018$ & $0.1276$ &\AA \\
Fe5015   & $ 1.3494$ & $-0.2799$ & $0.3608$ &\AA \\
Mg$_1$   & $ 0.0176$ & $-0.0165$ & $0.0160$ &mag \\
Mg$_2$   & $ 0.0106$ & $ 0.0444$ & $0.0112$ &mag \\
Mgb      & $ 0.0398$ & $-0.0392$ & $0.1789$ &\AA \\
Fe5270   & $-0.3608$ & $ 0.0514$ & $0.1735$ &\AA \\
Fe5335   & $-0.0446$ & $-0.0725$ & $0.3067$ &\AA \\
Fe5406   & $-0.0539$ & $-0.0730$ & $0.2054$ &\AA \\
Fe5709   & $-0.5416$ & $ 0.3493$ & $0.1204$ &\AA \\
Fe5782   & $-0.0610$ & $-0.0116$ & $0.2853$ &\AA \\
NaD      & $ 0.3620$ & $-0.0733$ & $0.2304$ &\AA \\
TiO$_1$  & $ 0.0102$ & $ 0.2723$ & $0.0133$ &mag \\
TiO$_2$  & $-0.0219$ & $ 0.1747$ & $0.0342$ &mag \\
H$\delta_A$&$-0.1525$& $-0.0465$ & $1.5633$ &\AA \\
H$\gamma_A$&$ 0.4961$& $ 0.0117$ & $0.6288$ &\AA \\
H$\delta_F$&$-0.1127$& $-0.0639$ & $0.4402$ &\AA \\
H$\gamma_F$&$-0.0062$& $-0.0343$ & $0.1480$ &\AA \\
\noalign{\smallskip}
\hline
\end{tabular}
\end{center}
\end{table}

Note, that most passbands require only a small linear offset, but no
offset as a function of index strength. While the former is simply due
to a small variation in the wavelength calibration, the latter is
produced by over/under-smoothing of the spectra. Absorption lines for
which the smoothing pushes the wings outside narrowly defined feature
passbands are mostly affected by this non-linear effect. However, for
passbands of major interest (such as CN, H$\beta$, Fe5270, Fe5335,
Mgb, and Mg$_2$) the Lick indices are satisfactorily reproduced by a
simple offset (no tilt) in the index value (see
Fig.~\ref{ps:idxcomp}).

\section{Analysis of the Spectra}

\subsection{Estimating the background light}
\label{ln:bkgestim}
Long-slit spectroscopy of extended objects notoriously suffers from
difficulties in estimating the contribution of the sky and background
light. Since we observe globular clusters near the Galactic Bulge,
their spectra will be contaminated by an unknown fraction of the bulge
light, depending on the location on the sky (see Fig.~\ref{ps:fov}).
In order to estimate the contribution of the background, two different
approaches have been applied. The first approach was to estimate the
sky and bulge contribution from separately taken sky and bulge spectra
(hereafter ``background modeling''). The other technique was to
extract the total background spectrum from low-intensity regions at
the edges of the spatial axis in the object spectrum itself
(henceforth ``background extraction''). While the first technique
suffers from the unknown change of the background spectrum between the
position of the globular clusters and the background fields, the
second one suffers from lower S/N. However, tests have shown that the
``background extraction'' allows a more reliable estimate of the
background spectrum.

We compare both background subtraction techniques in Table
\ref{tab:gcrop}. We find that the ``background modeling''
systematically overestimates the background light contribution as one
goes to larger galactocentric radii. The index differences increase
between spectra which have been cleaned using ``background modeling''
and ``background extraction''. This is basically due to an
overestimation of the background light from single background spectra
which were taken at intermediate galactocentric radii. We, therefore,
drop the ``background modeling'' and proceed for all subsequent
analyses with the ``background extraction'' technique. In summary, the
crucial drawback of the ``background modeling'' is that it requires a
prediction of the bulge light fraction from separate spectra which is
strongly model-dependent. The bulge light contains changing scale
heights for different stellar populations \citep[see][and references
therein]{frogel88, wyse97}. The background light at the cluster
position includes an unknown mix of bulge and disk stellar populations
\citep{frogel88, frogel90, feltzing00}, an unknown contribution from
the central bar \citep{unavane98a, unavane98b}, and is subject to
differential reddening on typical scales of $\sim90$\arcsec\
\citep{frogel99b} which complicates the modeling. Clearly, with
presently available models \citep[e.g.][]{kent91,freudenreich98} it is
impossible to reliably predict a spectrum of the galactic bulge as a
function of galactic coordinates. The ''background extraction''
technique naturally omits model predictions and allows to obtain the
total background spectrum, including sky {\it and} bulge light, from
the object spectrum itself.

We selected low-luminosity outer sections in the slit's intensity
profile (see Fig.~\ref{ps:profiles1}) to derive the background
spectrum for each globular cluster. Only those regions which show flat
and locally lowest intensities and are located outside the half-light
radius $r_h$ \citep{trager95} are selected. We sum the spectra of the
background light of all available pointings to create one high-S/N
background spectrum for each globular cluster. All globular clusters
were corrected using this background spectrum. The
background-to-cluster light ratio depends on galactic coordinates, and
is $\la0.1$ for NGC~6388 and $\sim1$ for NGC~6528. In order to lower
this ratio, only regions inside $r_h$ are used to create the final
globular-cluster spectrum. This restriction decreases the
background-to-cluster ratio by a factor of $\ga2$. In the case of
NGC~6218, NGC~6553, and NGC~6626 the half-light diameter $2r_h$ is
larger or comparable to the spatial dimensions of the slit, so that no
distinct background regions can be defined. For these clusters we
estimate the background from flat, low-luminosity parts along the
spatial axis inside $r_h$ but avoid the central regions (see
Fig.~\ref{ps:profiles1}).

\subsection{Contamination by Bright Objects}
\label{ln:contam}
To check if bright foreground stars inside the slit contaminate the
globular cluster light, we plot the intensity profile along the slit's
spatial axis. The profiles of each pointing are documented in
Figure~\ref{ps:profiles1}. Since we use the light only inside one
half-light radius (indicated by the shaded region) and therefore
maximize the cluster-to-background ratio, the probability for a
significant contamination by bright non-member objects is very
low. Even very bright foreground stars will contribute only a small
fraction to the total light.
\begin{figure*}[!ht]
 \centering
 \includegraphics[width=7.45cm]{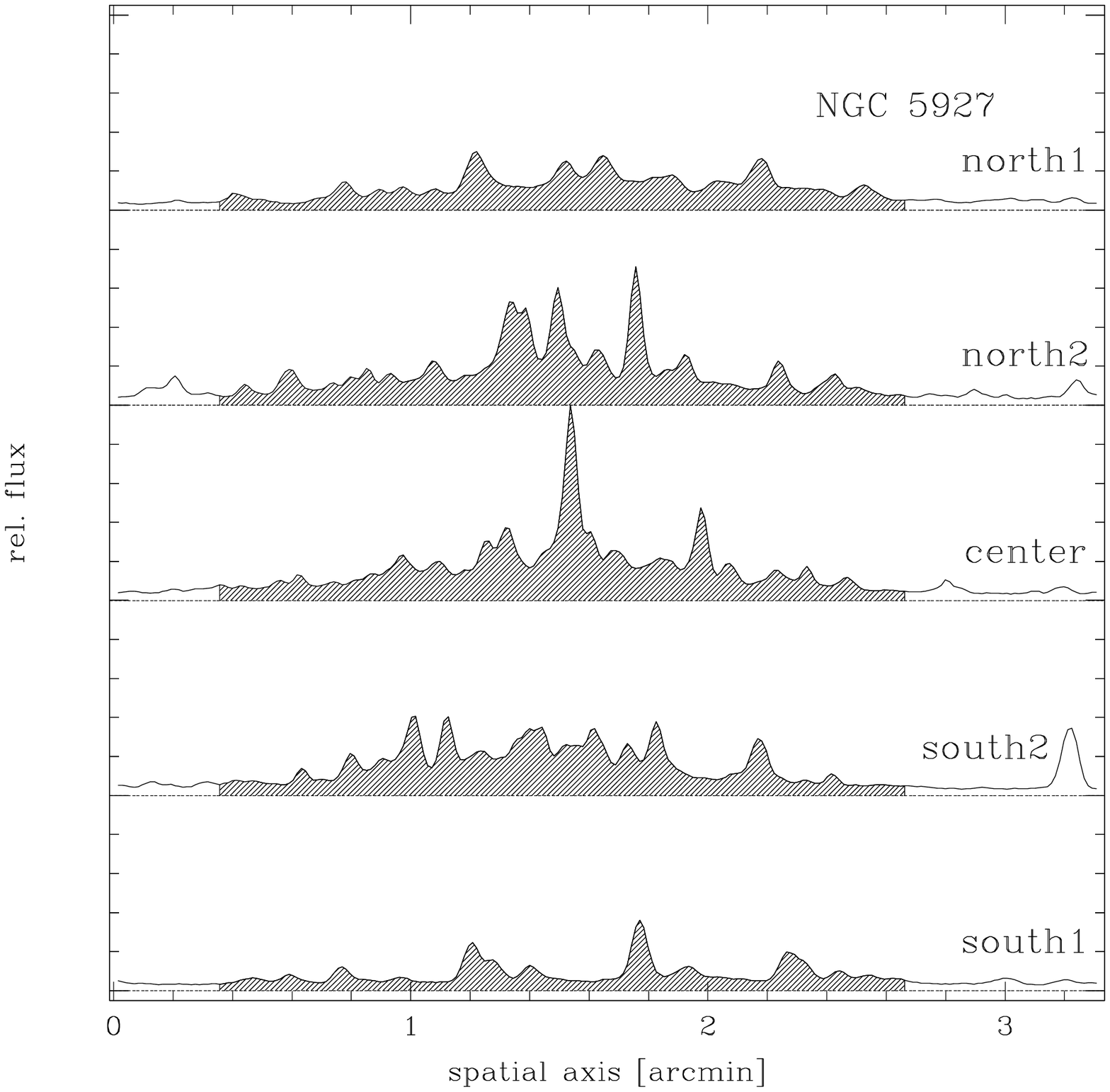}
 \includegraphics[width=7.45cm]{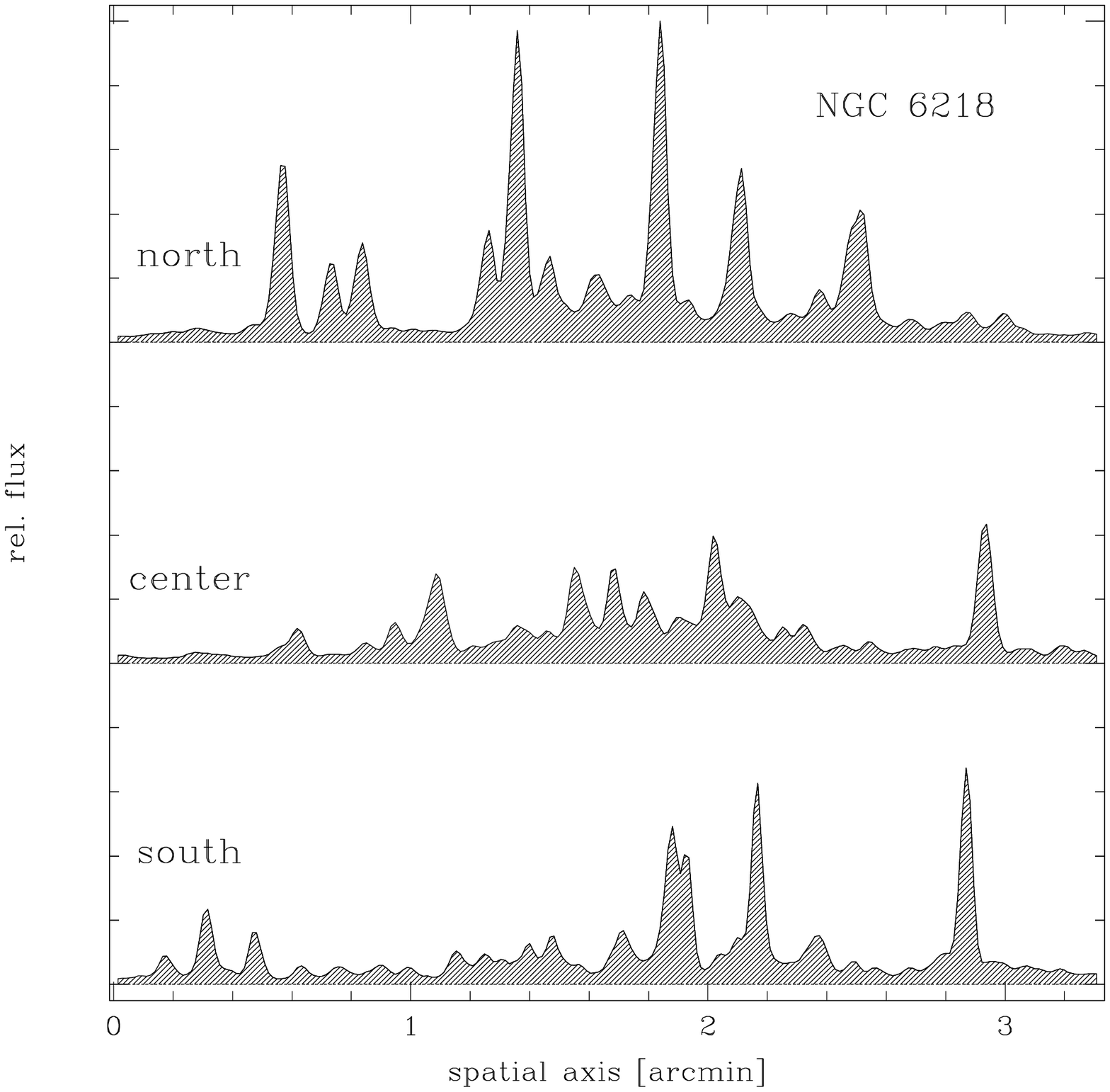}
 \includegraphics[width=7.45cm]{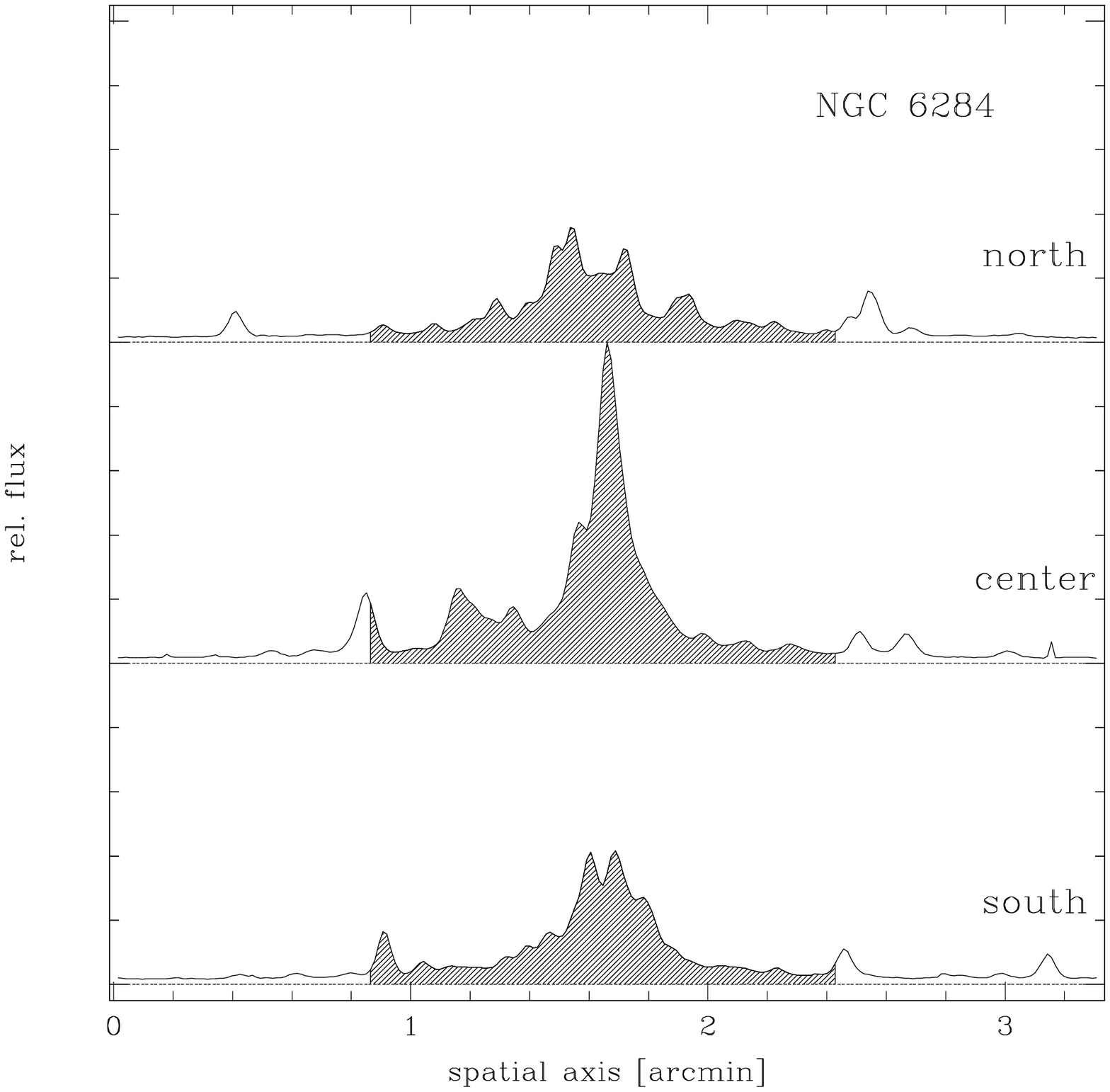}
 \includegraphics[width=7.45cm]{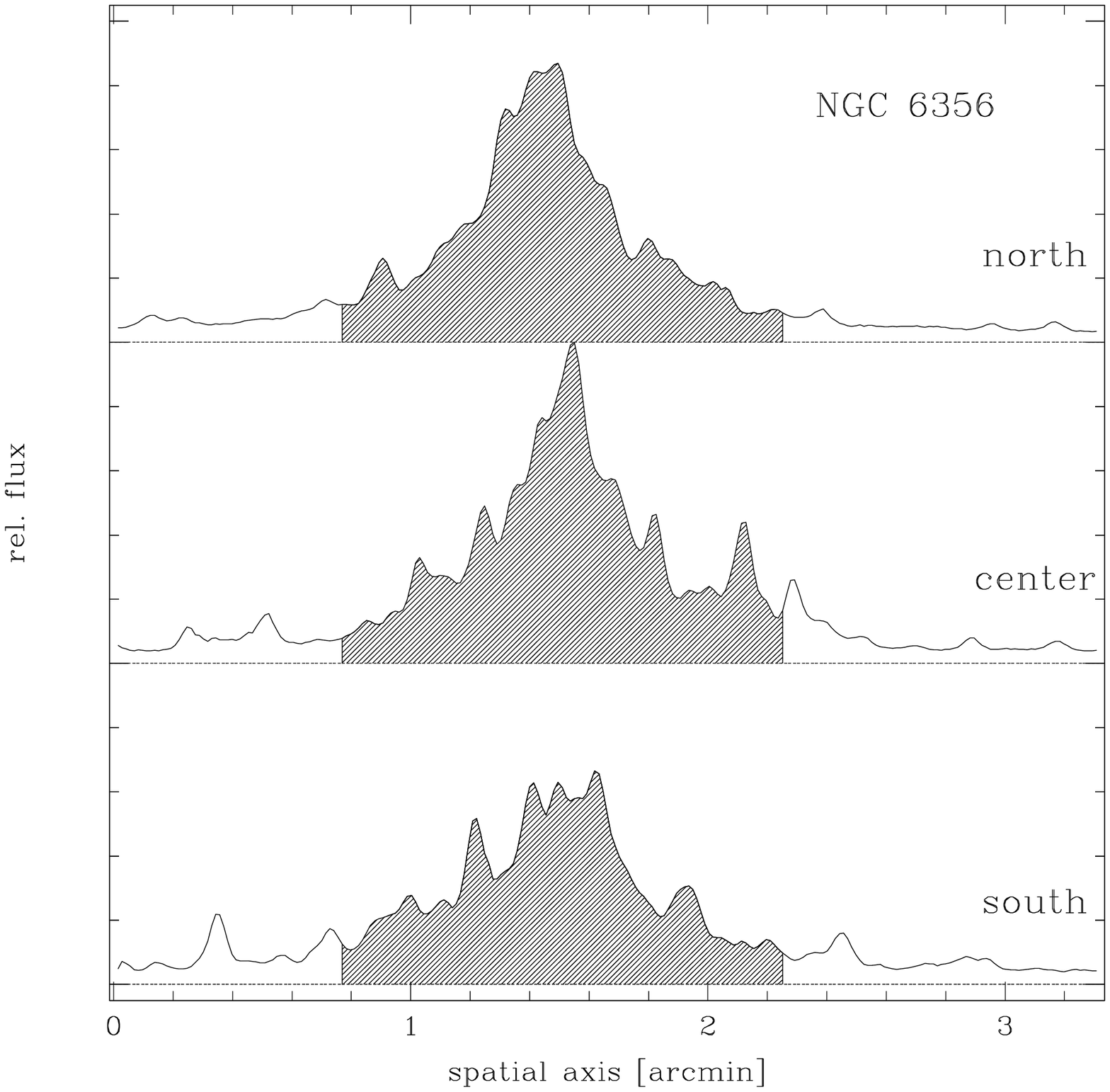}
 \includegraphics[width=7.45cm]{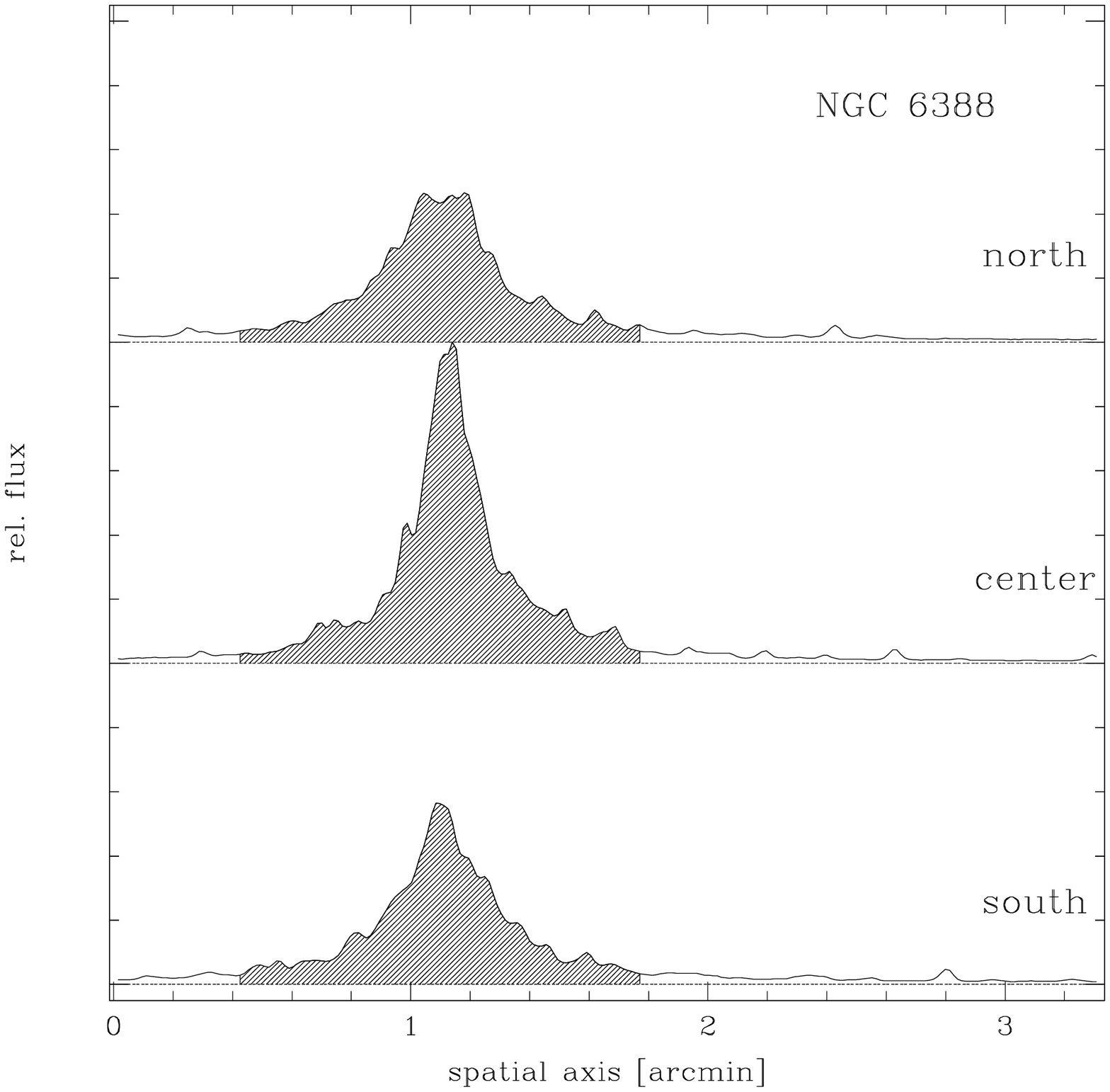}
 \includegraphics[width=7.45cm]{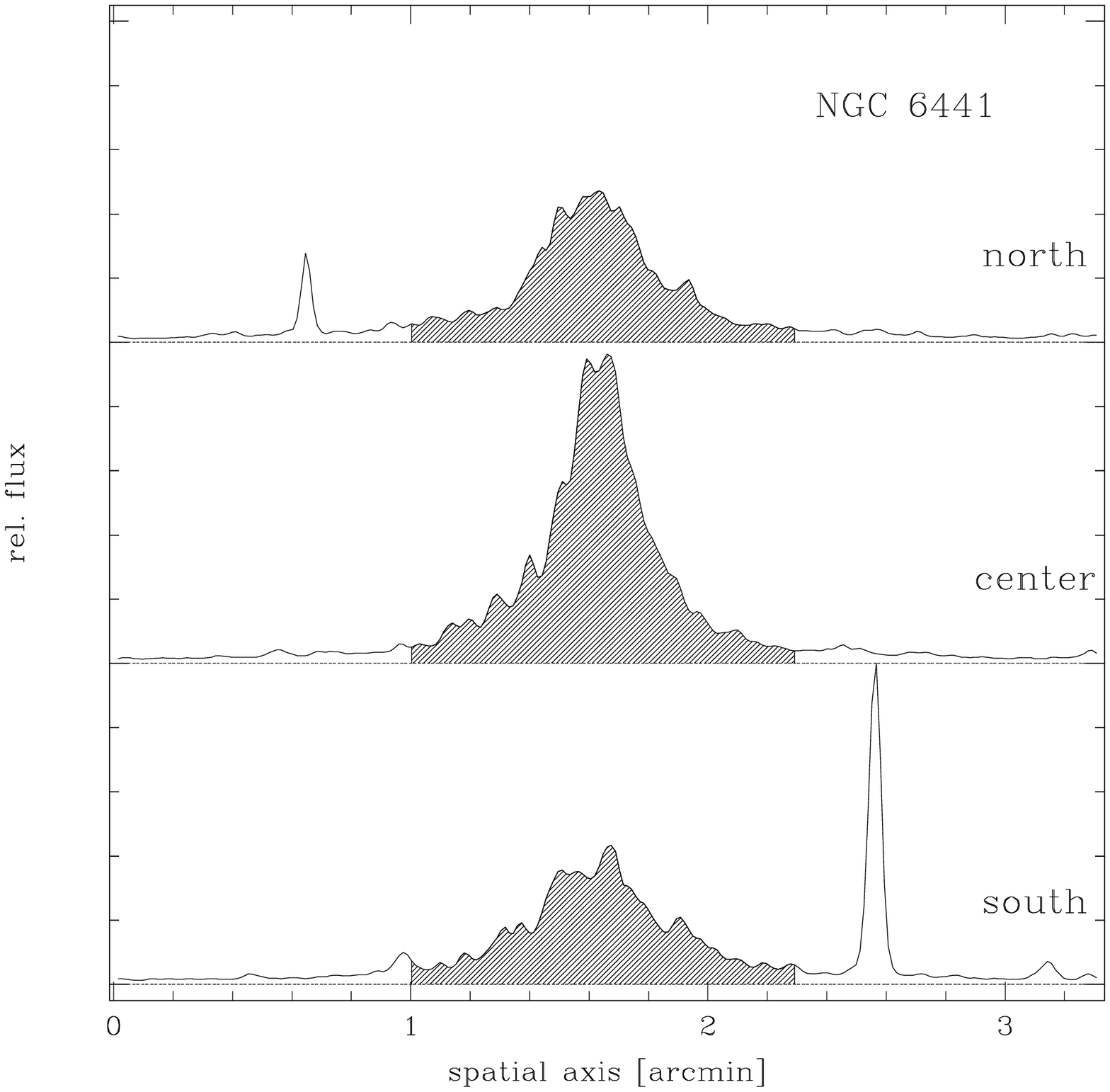}
 \caption{Intensity profiles of each pointing for all sample globular
   clusters. The fraction of the profile which was used to create the
   final globular cluster spectrum is shaded. Each cluster has at
   least three pointings which are shifted by a few slit widths to the
   north and south. Note that clusters with a sampled luminosity less
   than $10^4 L_\odot$ and relatively large half-light radii (i.e.
   see Sect.~\ref{ln:lumsample} and Tab.~\ref{tab:gcprop}) have
   strongly fluctuating profiles.}
\label{ps:profiles1}
\end{figure*}
\addtocounter{figure}{-1}
\begin{figure*}[!ht]
 \centering
 \includegraphics[width=7.45cm]{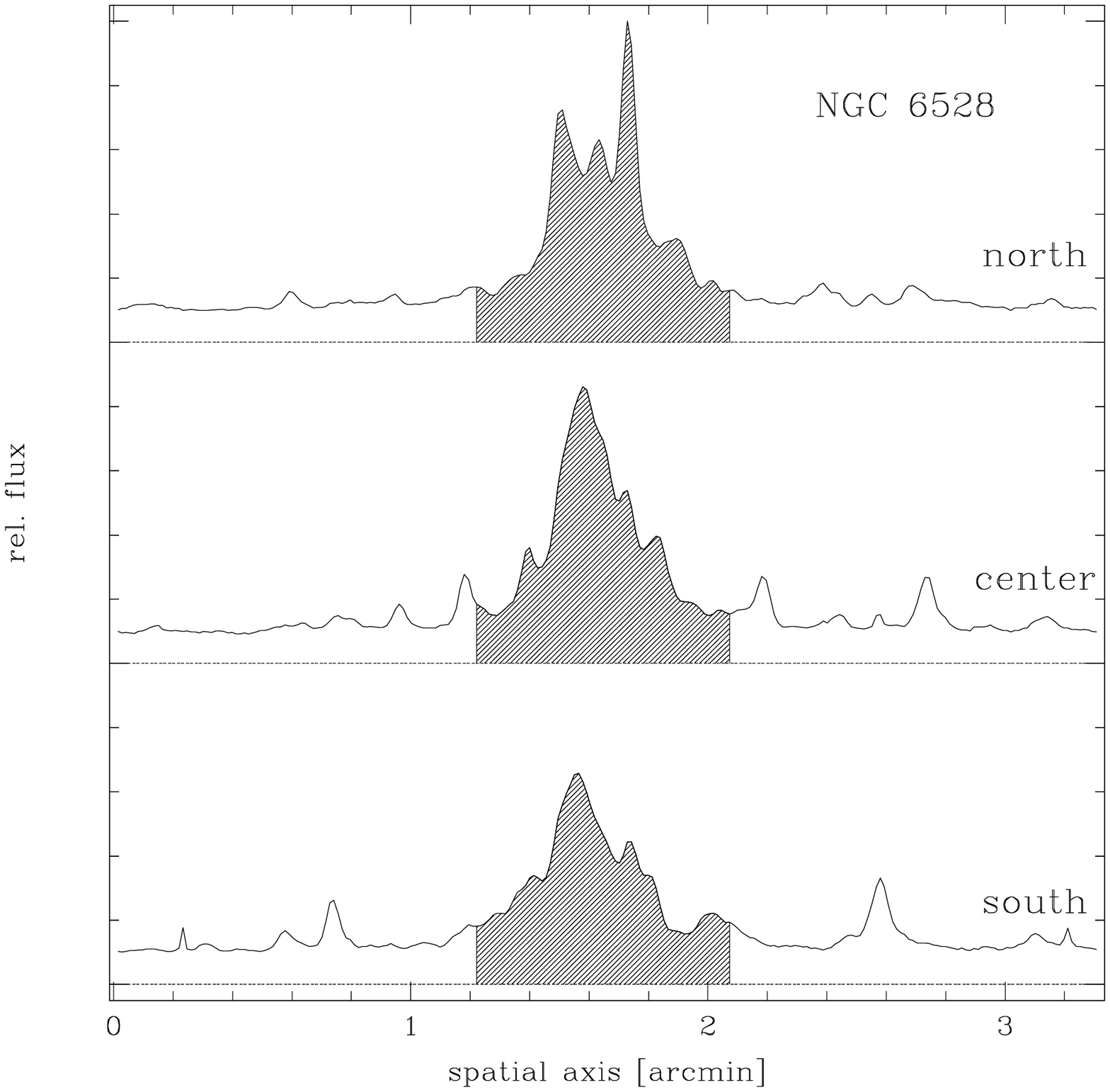}
 \includegraphics[width=7.45cm]{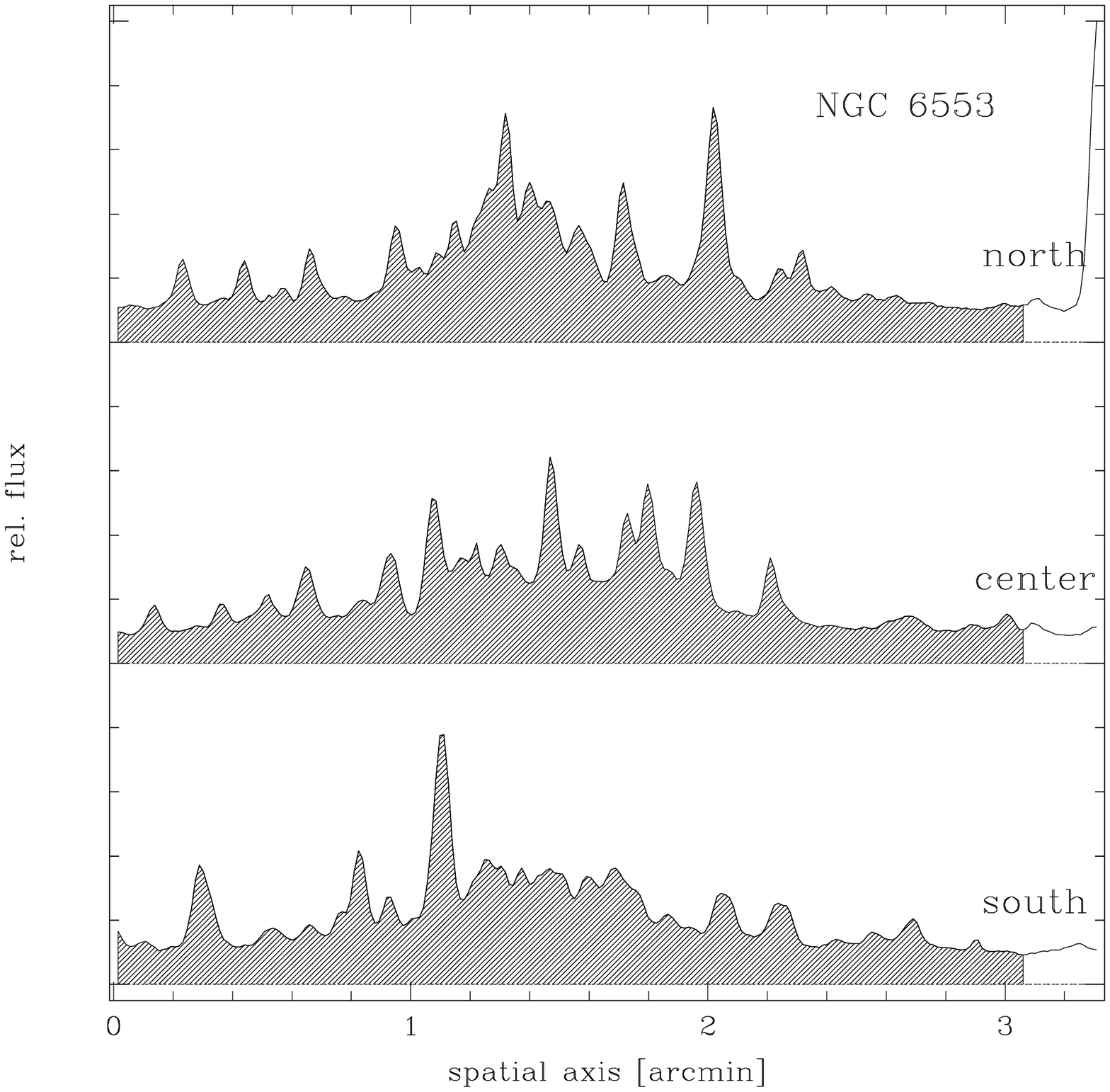}
 \includegraphics[width=7.45cm]{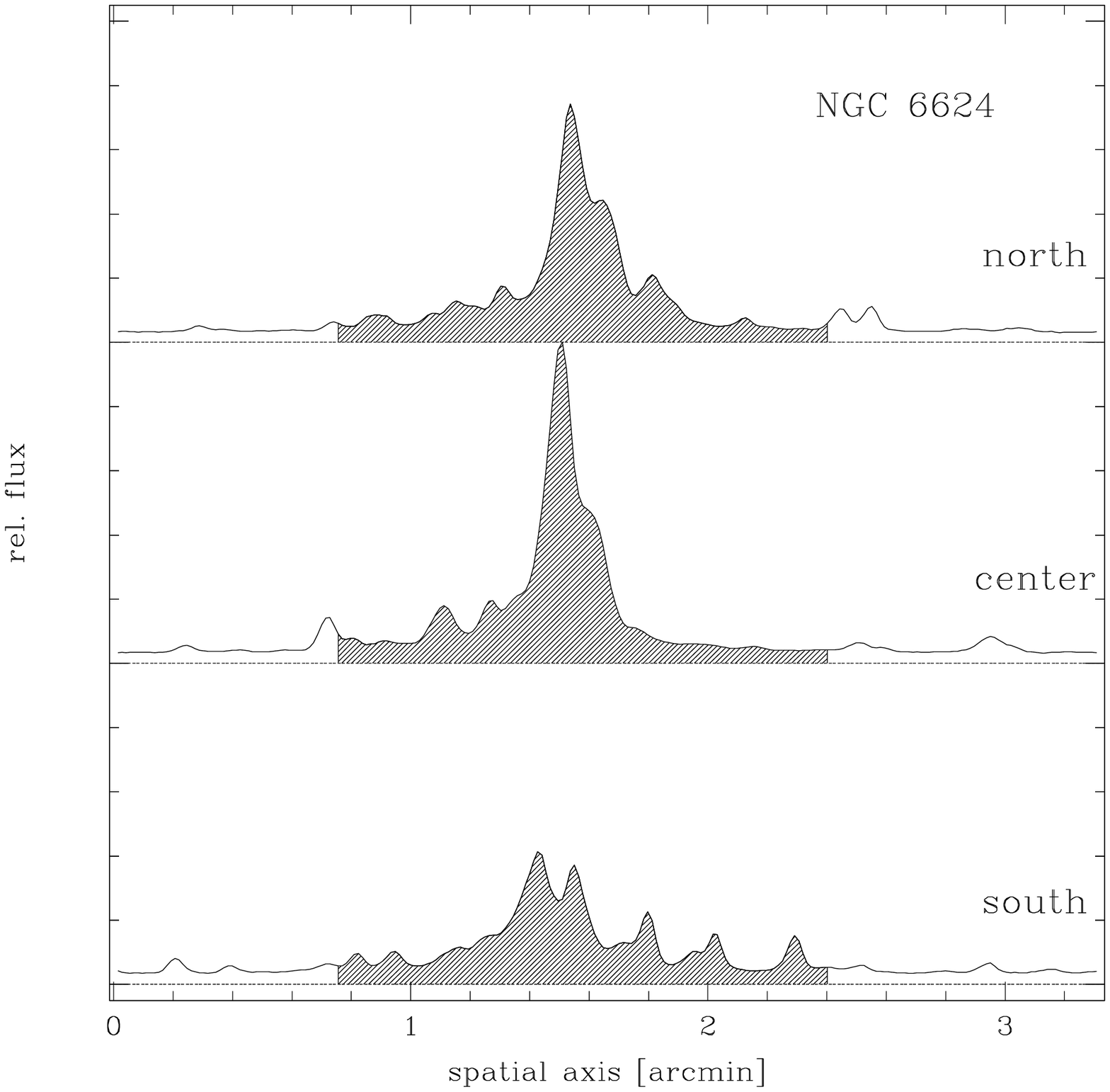}
 \includegraphics[width=7.45cm]{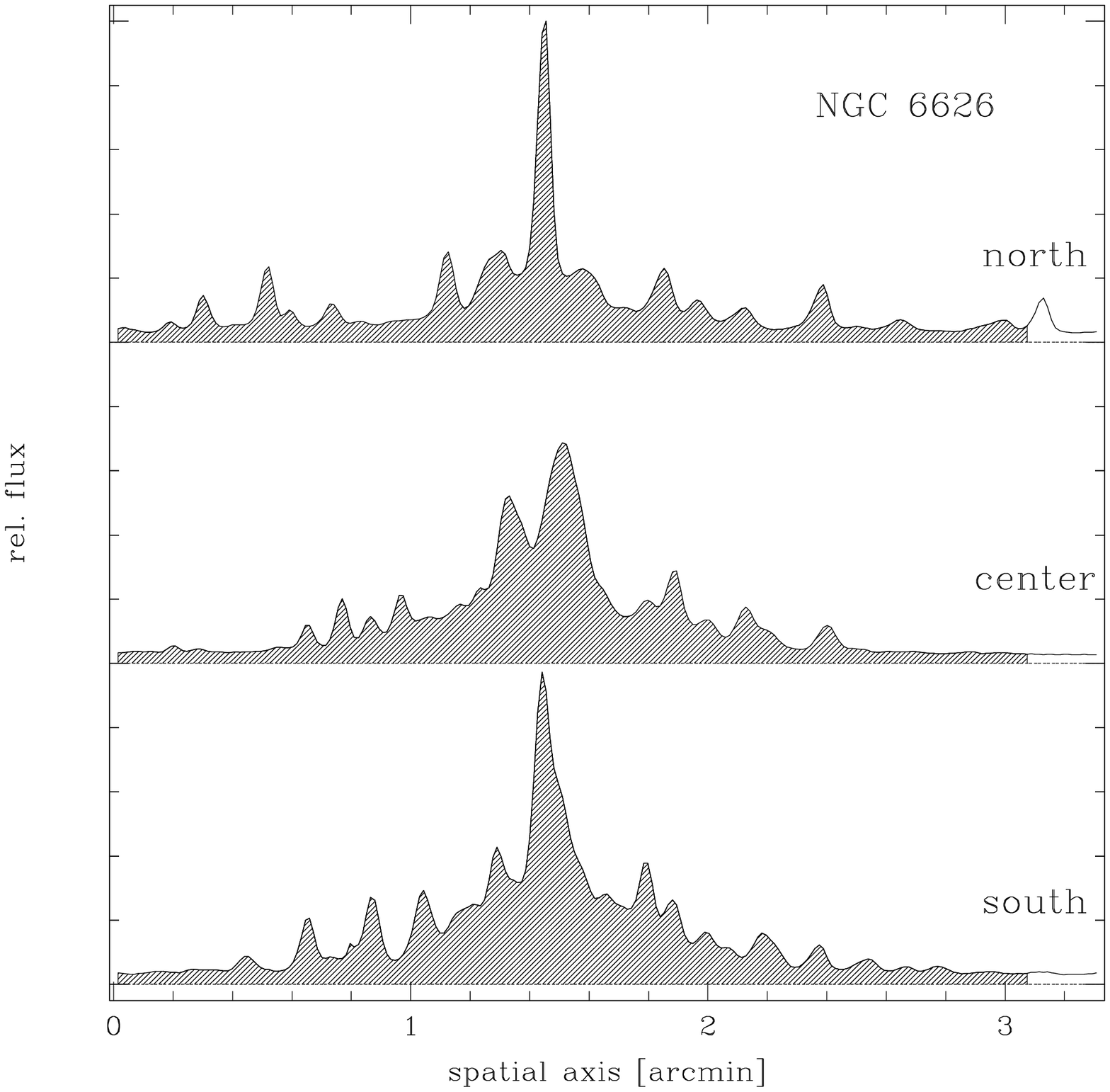}
 \includegraphics[width=7.45cm]{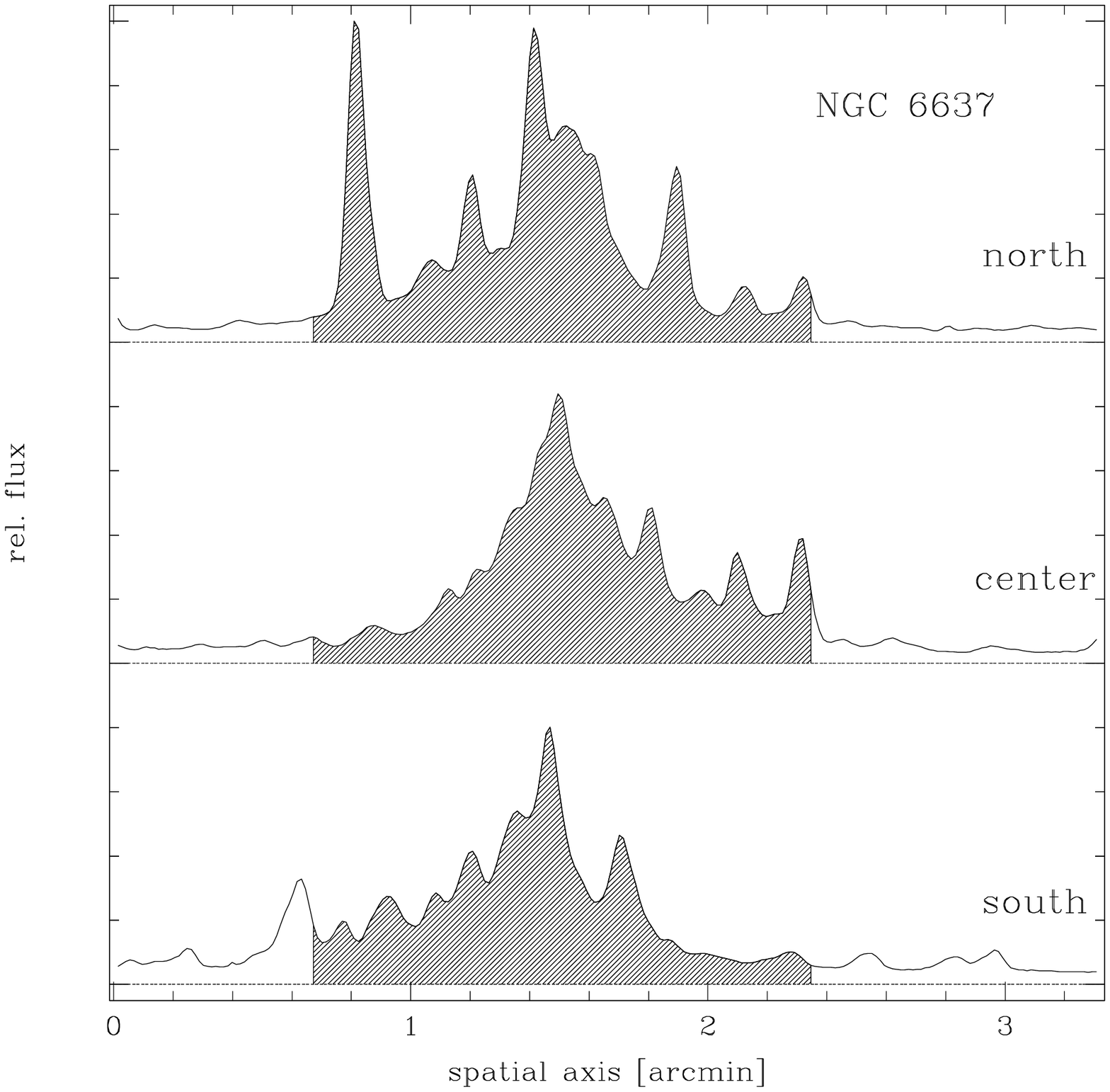}
 \includegraphics[width=7.45cm]{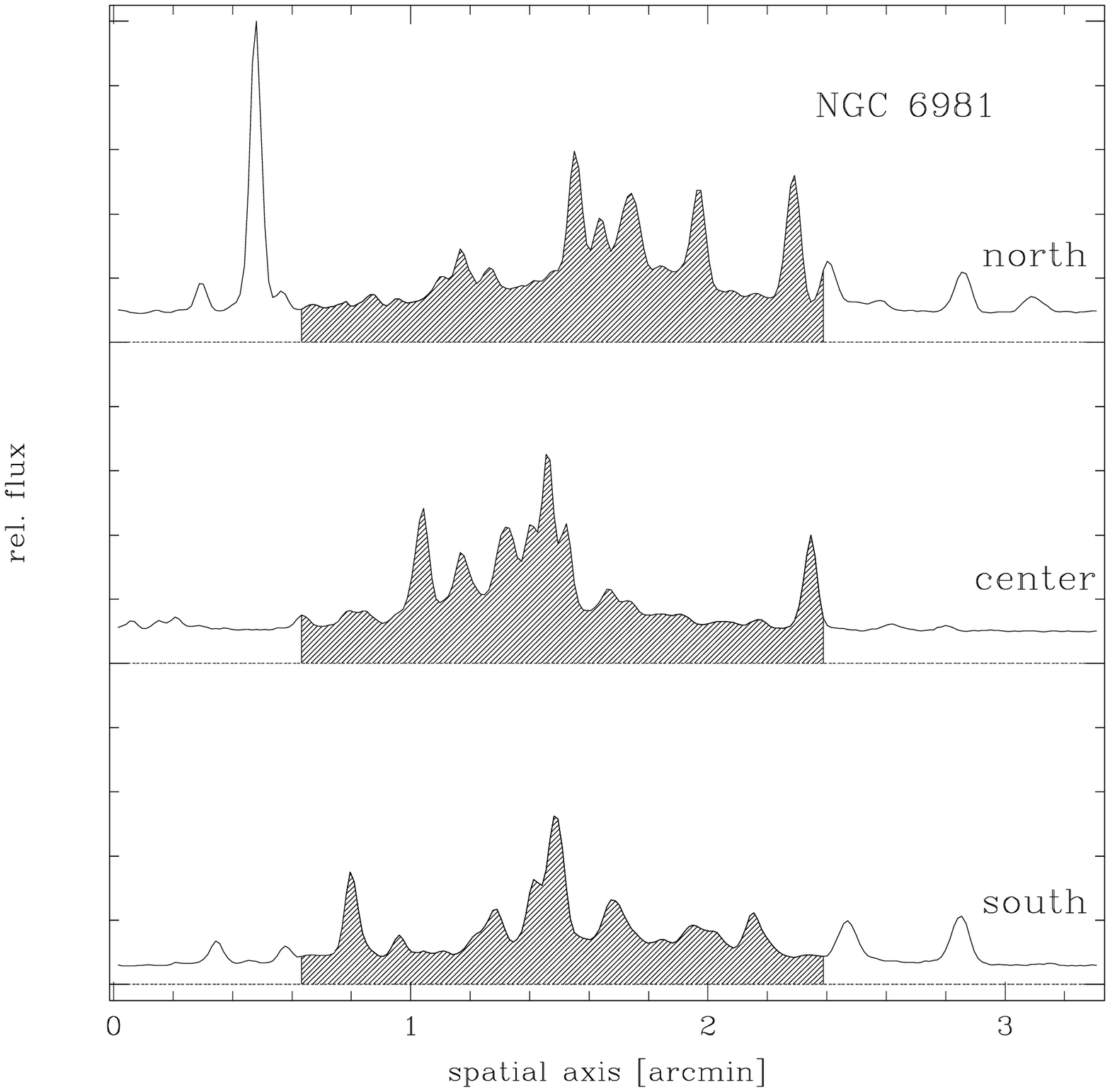}
 \caption{ -- continued.}
\end{figure*}

However, three of our sample globular clusters (NGC~6218, NGC~6553,
and NGC~6626) are extended and their half-light diameter are just or
not entirely covered by the slit. The low radial velocity resolution
of our spectra does not allow to distinguish between globular cluster
stars and field stars inside the slit. Galactic stellar-population
models \citep[e.g.][]{robin96} predict a maximum cumulative amount of
4 stars with magnitudes down to $V=19.5$ (all stars with $V=18.5-19.5$
mag) towards the Galactic center inside the equivalent of three
slits. This maximum estimate applies only to the Baade's Window
globular clusters NGC~6528 and NGC~6553. All other fields have
effectively zero probability to be contaminated by foreground stars.
Nonetheless, even in the worst-case scenario, if 4 stars of 19th
magnitude would fall inside one slit, their fractional contribution to
the total light would be $\la1.2\cdot10^{-4}$. For globular clusters
at larger galactocentric radii this fraction is even lower. Hence, we
do not expect a large contamination by foreground disk stars.

One critical case is the northern pointing of NGC~6637 in which a
bright star falls inside the half-light radius (see upper panel of the
NGC~6637 profile in Fig.~\ref{ps:profiles1}). This star contributes
$\la10$\% to the total light of the sampled globular cluster and its
radial velocity is indistinguishable from the one of NGC~6637. An
inspection of DSS images shows that the NGC~6637 field contains more
such bright stars which are concentrated around the globular cluster
center and are therefore likely to be cluster members. We therefore
assume that the star is a member of NGC~6637 and leave it in the
spectrum.

\subsection{Comparison with Previous Measurements}
 
\begin{table}[t!]
\begin{center}
\caption{Offsets and dispersion of the residuals between our data and the
  literature. Dispersions are 1 $\sigma$ scatter of the residuals.}
\label{tab:literatureoffsets}
\begin{tabular}[angle=0,width=\textwidth]{l|rrr}
\hline
\noalign{\smallskip}
 index & offset & dispersion & units\\ 
\noalign{\smallskip}
\hline
\noalign{\smallskip}
G4300    & $ 0.45$ & 0.70 & \AA \\
H$\beta$ & $ 0.27$ & 0.57 & \AA \\
Mg$_2$   & $0.009$ & 0.014& mag \\
Mgb      & $-0.01$ & 0.27 & \AA \\
Fe5270   & $-0.33$ & 0.44 & \AA \\
Fe5335   & $ 0.12$ & 0.27 & \AA \\
\noalign{\smallskip}
\hline
\end{tabular}
\end{center}
\end{table}

\begin{figure*}[!ht]
 \centering \includegraphics[bb=18 338 592 700,clip,width=14.5cm]{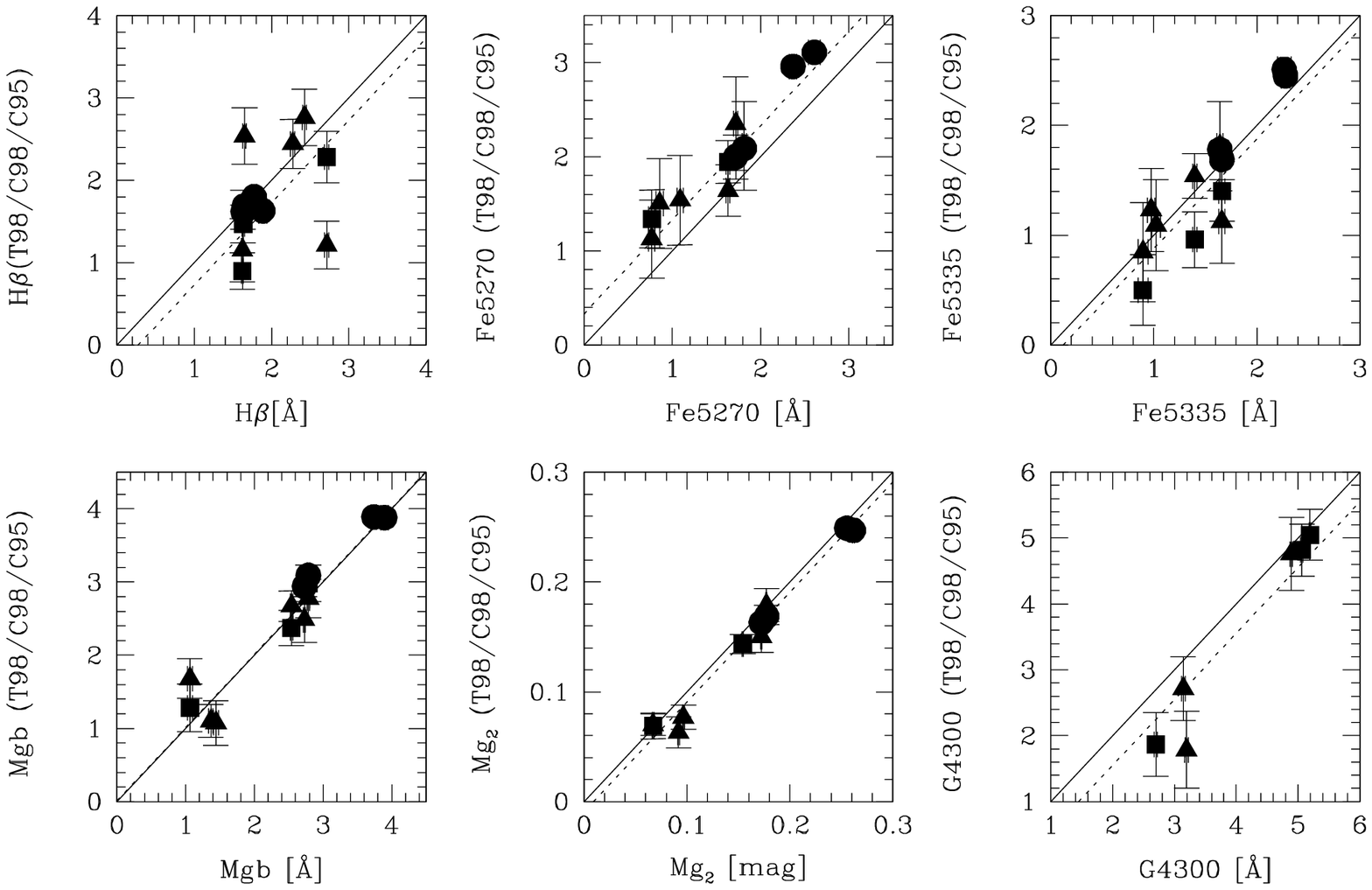}
   \caption{Comparison of index measurements of \cite{trager98},
	marked by squares, \cite{cohen98}, marked by circles (without
	errors for the \citeauthor{cohen98} data), and
	\cite{covino95}, indicated by triangles, with our data. Solid
	lines mark the one-to-one relation and dashed lines the mean
	offsets.}
\label{ps:dataidxcomp_cls}
\end{figure*}

Lick indices\footnote{We point out the work of \cite{bica86} who
performed a spectroscopic study of 63 LMC, SMC, Galactic globular and
compact open clusters. However, the final resolution of their spectra
is too low (11 \AA ) to allow an analysis of standard Lick indices.}
are available in the literature for a few globular clusters in our
sample, as we intentionally included these clusters for
comparison. The samples of \cite{trager98} and \cite{covino95} and
\cite{cohen98} have, respectively, three, six, and four clusters in
common with our data. Note that the indices of \cite{covino95} and
\cite{cohen98} were measured with the older passband definitions of
\cite{burstein84} and are subject to potential systematic offsets.
Where necessary we also converted the values of \citeauthor{covino95}
to the commonly used \AA-scale for atomic indices and kept the
magnitude scale for molecular bands. Table \ref{tab:gcrop} summarizes
all measurements, including our data. Figure \ref{ps:dataidxcomp_cls}
shows the comparison of some indices between the previously mentioned
data sets and ours. The mean offset in the sense EW$_{\rm data} -
$EW$_{\rm lit.}$ and the dispersion are given in Table
\ref{tab:literatureoffsets}. Most indices agree well with the
literature values and have offsets smaller than the dispersion.

Only the Fe5270 index is $0.75\, \sigma$ higher for our data compared
with the literature. This is likely to be due to imperfect smoothing
of the spectra in the region of $\sim5300$ \AA . Our smoothing kernel
is adjusted according to the Lick resolution given by the linear
relations in \cite{worthey97}. This relations are fit to individual
line resolution data which show a significant increase in scatter in
the spectral range around 5300 \AA\ \citep[see Figure 7
in][]{worthey97}. Hence even if our smoothing is correctly applied,
the initial fitting of the Lick resolution data by
\citeauthor{worthey97} might introduce biases which cannot be
accounted for a posteriori. However, the offset between the literature
and our data is reduced by the use of the synthetic
$\langle$Fe$\rangle$ index which is a combination of the Fe5270 and
Fe5335 index. The $\langle$Fe$\rangle$ index partly cancels out the
individual offsets of the former two indices.

\subsection{Estimating the Sampled Luminosity}
\label{ln:lumsample}
The spectrograph slit samples only a fraction of the total light of a
globular cluster's stellar population. The less light is sampled the
higher the chance that a spectrum is dominated by a few bright
stars. In general, globular cluster spectra of less than $10^4
L_\odot$ are prone to be dominated by statistical fluctuations in the
number of high-luminosity stars (such as RGB and AGB stars, etc.). For
a representative spectrum it is essential to adequately map all
evolutionary states in a stellar population, such that no large
statistical fluctuations for the short-living phases are expected. We
therefore estimate the total sampled luminosity of the underlying
stellar population 1) from spectrophotometry of the flux-calibrated
spectra and 2) from the integration of globular cluster surface
brightness profiles.

As a basic condition of the first method we confirm that all three
nights have had photometric conditions using the ESO database for
atmospheric conditions at La
Silla\footnote{http://www.ls.eso.org/lasilla/dimm/}. We use the flux
at 5500 \AA\ in the co-added and background-subtracted spectra and
convert it to an apparent magnitude with the relation
\begin{equation}
m_V = -2.5\cdot\log(F_\lambda) - (19.79\pm0.24)
\end{equation}
where $F_\lambda$ is the flux in erg cm$^{-2}$s$^{-1}$\AA$^{-1}$. The
zero point was determined from five flux-standard spectra, which have
been observed in every night. Its uncertainty is the 1$\sigma$
standard deviation of all measurements. After correcting for the
distance, the absolute magnitudes were de-reddened using the values
given in \cite{harris96}.\footnote{These reddening values were derived
from CMD studies of individual globular cluster and are a reliable
estimate of the effective reddening, in contrast to coarse survey
reddening maps such as the COBE/DIRBE reddening maps by
\cite{schlegel98}. These maps tend to overestimate the reddening in
high-extinction regions.} The reddening values are given in Table
\ref{tab:gcprop} along with the distance modulus
\citep{harris96}. Using the absolute magnitude of the combined
globular cluster spectrum, we calculate the total sampled luminosity
\begin{equation}
L_T=BC_V\cdot10^{-0.4\cdot(m_V-(m-M)_V-M_\odot-3.1\cdot\mathrm{E}_{(B-V)})}
\end{equation}
where $M_\odot=4.82$ mag is the absolute solar magnitude in the V band
\citep{hayes85, neckel86a, neckel86b}. With the bolometric correction
${\rm BC}_V$ \citep{renzini98, maraston98} we obtain the total
bolometric luminosity $L_T$. The total globular cluster luminosity is
compared to the sampled flux and tabulated in Table~\ref{tab:fluxprop}
as $L_{\rm slit}$.

\begin{table*}[ht!]
 \caption{Sampled and total luminosities of observed globular clusters
 and bulge. All values have been determined from the co-added spectra
 of all available pointings. For the co-added bulge spectrum we
 adopted a mean metallicity of [Fe/H]$\,\approx\, -0.33$ dex
 \citep{zoccali02}.}  \label{tab:fluxprop}
\begin{center}
\begin{tabular}{l|ccccccccrr}
\hline
\noalign{\smallskip}
 cluster 
 & $F_\lambda$(@5500\AA)$^{\mathrm{a}}$ 
 & $M_V^{\mathrm{b}}$
 & $M_V^{\mathrm{c}}$ 
 & $BC_{\rm V}^{\mathrm{d}}$
 & $L_{\mathrm{prof}}^{\mathrm{e}}$
 & $L_{\mathrm{slit}}^{\mathrm{f}}$ 
 & $L_{\mathrm{GC}}^{\mathrm{g}}$ 
 & $\frac{L_{\mathrm{slit}}}{L_{\mathrm{GC}}}$ 
 & $N_{\rm RGB}^{\mathrm{h}}$
 & $N_{\rm uRGB}^{\mathrm{i}}$\\    
\noalign{\smallskip}
\hline
\noalign{\smallskip}
NGC~5927&$(3.6\pm0.2)\cdot10^{-13}$&$ -5.88$&$ -7.80$& 1.57&$1.7\cdot10^4$ &$(3.0\pm0.8)\cdot10^4$ & $ 1.8\cdot10^5$& 0.171&  359&   9\\ 
NGC~6218&$(2.0\pm0.1)\cdot10^{-13}$&$ -2.65$&$ -7.32$& 1.29&$4.0\cdot10^3$ &$(1.3\pm0.3)\cdot10^3$ & $ 9.3\cdot10^4$& 0.014&   15&   0\\ 
NGC~6284&$(3.7\pm0.1)\cdot10^{-13}$&$ -6.27$&$ -7.87$& 1.32&$1.9\cdot10^4$ &$(3.6\pm0.9)\cdot10^4$ & $ 1.6\cdot10^5$& 0.230&  435&  11\\ 
NGC~6356&$(6.4\pm0.1)\cdot10^{-13}$&$ -6.94$&$ -8.52$& 1.51&$4.8\cdot10^4$ &$(7.6\pm1.8)\cdot10^4$ & $ 3.3\cdot10^5$& 0.233&  913&  23\\ 
NGC~6388&$(2.8\pm0.1)\cdot10^{-12}$&$ -8.68$&$ -9.82$& 1.47&$1.6\cdot10^5$ &$(3.7\pm1.0)\cdot10^5$ & $ 1.1\cdot10^6$& 0.351& 4430& 111\\ 
NGC~6441&$(2.0\pm0.1)\cdot10^{-12}$&$ -8.52$&$ -9.47$& 1.49&$1.3\cdot10^5$ &$(3.2\pm0.9)\cdot10^5$ & $ 7.8\cdot10^5$& 0.417& 3894& 97\\ 
NGC~6528&$(4.9\pm0.2)\cdot10^{-13}$&$ -7.28$&$ -6.93$& 1.66&$2.3\cdot10^4$ &$(1.1\pm0.3)\cdot10^5$ & $ 8.3\cdot10^4$& 1.376$^{\mathrm{j}}$& 1376&  34\\ 
NGC~6553&$(2.0\pm0.1)\cdot10^{-13}$&$ -6.41$&$ -7.99$& 1.59&$1.5\cdot10^4$ &$(4.9\pm1.4)\cdot10^4$ & $ 2.1\cdot10^5$& 0.234&  593&  15\\ 
NGC~6624&$(8.0\pm0.7)\cdot10^{-13}$&$ -5.78$&$ -7.50$& 1.54&$1.8\cdot10^4$ &$(2.7\pm0.8)\cdot10^4$ & $ 1.3\cdot10^5$& 0.205&  322&   8\\ 
NGC~6626&$(5.6\pm0.1)\cdot10^{-13}$&$ -5.61$&$ -8.33$& 1.30&$1.4\cdot10^4$ &$(1.9\pm0.5)\cdot10^4$ & $ 2.4\cdot10^5$& 0.082&  231&   6\\ 
NGC~6637&$(8.0\pm1.4)\cdot10^{-14}$&$ -2.70$&$ -7.52$& 1.43&$1.5\cdot10^4$ &$(1.5\pm0.6)\cdot10^3$ & $ 1.2\cdot10^5$& 0.012&   17&   0\\ 
NGC~6981&$(1.2\pm0.1)\cdot10^{-13}$&$ -3.95$&$ -7.04$& 1.31&$7.7\cdot10^3$ &$(4.2\pm1.3)\cdot10^3$ & $ 7.3\cdot10^4$& 0.058&   50&   1\\ 
Bulge   &$(4.0\pm0.3)\cdot10^{-13}$&$ -5.14$&  \dots & 1.59&  \dots        &$(1.5\pm0.7)\cdot10^4$ &    \dots       & \dots&  180&   5\\
 \noalign{\smallskip}
 \hline
\end{tabular}
\end{center}
\begin{list}{}{}
\item[$^{\mathrm{a}}$] sampled flux at 5500 \AA\ in erg cm$^{-2}$s$^{-1}$\AA$^{-1}$
\item[$^{\mathrm{b}}$] absolute magnitude of the sampled light
\item[$^{\mathrm{c}}$] absolute globular cluster magnitude \citep{harris96}
\item[$^{\mathrm{d}}$] V-band bolometric correction for a 12 Gyr old
stellar population calculated for the according cluster metallicity
(see Table~\ref{tab:gcprop}). The values were taken from
\cite{maraston98, maraston02s}.
\item[$^{\mathrm{e}}$] sampled bolometric luminosity $L_T$ in $L_\odot$ from
  the integration of King surface brightness profiles of
  \cite{trager95}
\item[$^{\mathrm{f}}$] sampled bolometric luminosity $L_T$ in $L_\odot$
  calculated from the total light sampled by all slit pointings
\item[$^{\mathrm{g}}$] globular cluster's total bolometric luminosity
$L_T$ in $L_\odot$
\item[$^{\mathrm{h}}$] expected number of RGB stars contributing to
the sampled luminosity
\item[$^{\mathrm{i}}$] expected number of upper RGB stars ($\Delta
M_{\rm Bol}\leq 2.5$ mag down from the tip of the RGB) contributing to
the sampled luminosity
\item[$^{\mathrm{j}}$] see Section~\ref{ln:lumsample}
\end{list}
\end{table*}

For the integration of the surface brightness profiles we use the data
from \cite{trager95} who provide the parameters of single-mass,
non-rotating, isotropic King profiles \citep{king66} for all sample
globular clusters. The integrated total V-band luminosities have been
transformed to $L_T$ and are included in Table~\ref{tab:fluxprop} as
$L_{\rm prof}$. Note that for most globular clusters the results from
both techniques agree well. However, for some globular clusters the
integration of the surface brightness profile gives systematically
larger values. This is due to the fact that the profiles were
calculated from the flux of all stars in a given radial interval
whereas the slits sample a small fraction of the flux at a given
radius. Hence, the likelihood to sample bright stars which dominate
the surface brightness profile falls rapidly with radius. Since bright
stars are point sources the slit will most likely sample a smaller
total flux than predicted by the surface brightness profile. This
effect is most prominent for globular clusters with relatively large
half-light radii and waggly intensity profiles (cf.
Fig.~\ref{ps:profiles1}).

Among the values reported in Table~\ref{tab:fluxprop}, the case of
NGC~6528 is somewhat awkward, as the estimated luminosity sampled by
the slit is apparently higher than the total luminosity of the
cluster, which obviously cannot be. This cluster projects on a very
dense bulge field, and therefore the inconsistency probably arises
from either an underestimate of the field contribution that we have
subtracted from the cluster+field co-added spectrum, or to an
underestimate of the total luminosity of the cluster as reported in
\cite{harris96}, or from a combination of these two effects.

From the sampled flux $L_{\mathrm{slit}}$ we estimate the number of
red giant stars contributing to the total light. \cite{renzini98}
gives the expected number of stars for each stellar evolutionary phase
of a $\sim15$ Gyr old, solar-metallicity simple stellar population.
In general, in this stellar population the brightest stars which
contribute a major fraction of the flux to the integrated light are
found on the red giant branch (RGB) which contributes $\sim40$\%
\citep{renzini88} to the total light. The last two columns of Table
\ref{tab:fluxprop} give the expected number of RGB and upper RGB stars
in the sampled light. Upper RGB stars are defined here as those within
2.5 bolometric magnitudes from the RGB tip. The RGB and upper RGB
lifetimes are $\sim6\cdot 10^8$ and $\sim1.5\cdot 10^7$ years,
respectively.

Due to the small expected number of RGB and upper RGB stars
contributing to the spectra of NGC~6218 and NGC~6637, both spectra are
prone to be dominated by a few bright stars. In fact, for both
clusters the intensity profiles (see Figure~\ref{ps:profiles1}) show
single bright stars. However, the contribution of the brightest single
object is $\la10$\% (see Sect.~\ref{ln:contam}) for all spectra. All
other spectra contain enough RGB stars to be unaffected by statistical
fluctuations in the number of bright stars.

The sampled luminosity of the bulge fields is more difficult to
estimate. Uncertain sky subtraction (see problems with ``background
modeling'' in Sect.~\ref{ln:bkgestim}), and patchy extinction in
combination with the bulge's spatial extension along the line of sight
make the estimate of the sampled luminosity quite uncertain. Here we
simply give upper and lower limits including all available
uncertainties. The average extinction in Baade's Window is $\langle
A_V\rangle\approx1.7$ mag and varies between 1.3 and 2.8 mag
\citep{stanek96}. The more recent reddening maps of \cite{schlegel98}
confirm the previous measurements and give for our three Bulge fields
the extinction in the range $1.6\la A_{V}\la 2.1$ mag. We adopt a
distance of $8-9$ kpc to the Galactic center and use the faintest and
brightest sky spectrum to estimate the flux at 5500 \AA. The total
sampled luminosity $L_T$ of the final co-added Bulge spectrum is
$(1.3-2.6)\cdot 10^4 L_\odot$. Our value is in good agreement with the
sampled luminosity derived from surface brightness estimates in
Baade's Window and several fields at higher galactic latitudes by
\cite{terndrup88}. According to his V-band surface brightness
estimates for Baade's Window and a field at the galactic coordinates
$l=0.1^{\rm o}$ and $b=-6^{\rm o}$, the sampled luminosity in an area
equivalent to all our bulge-field pointings in one of the two fields
is $(2.6\pm0.5)\cdot 10^4 L_\odot$ and $(1.2\pm0.3)\cdot 10^4
L_\odot$, respectively.

\section{Index Ratios in Globular Clusters and Bulge Fields}

\begin{figure*}
\centering
 \includegraphics[width=14.8cm]{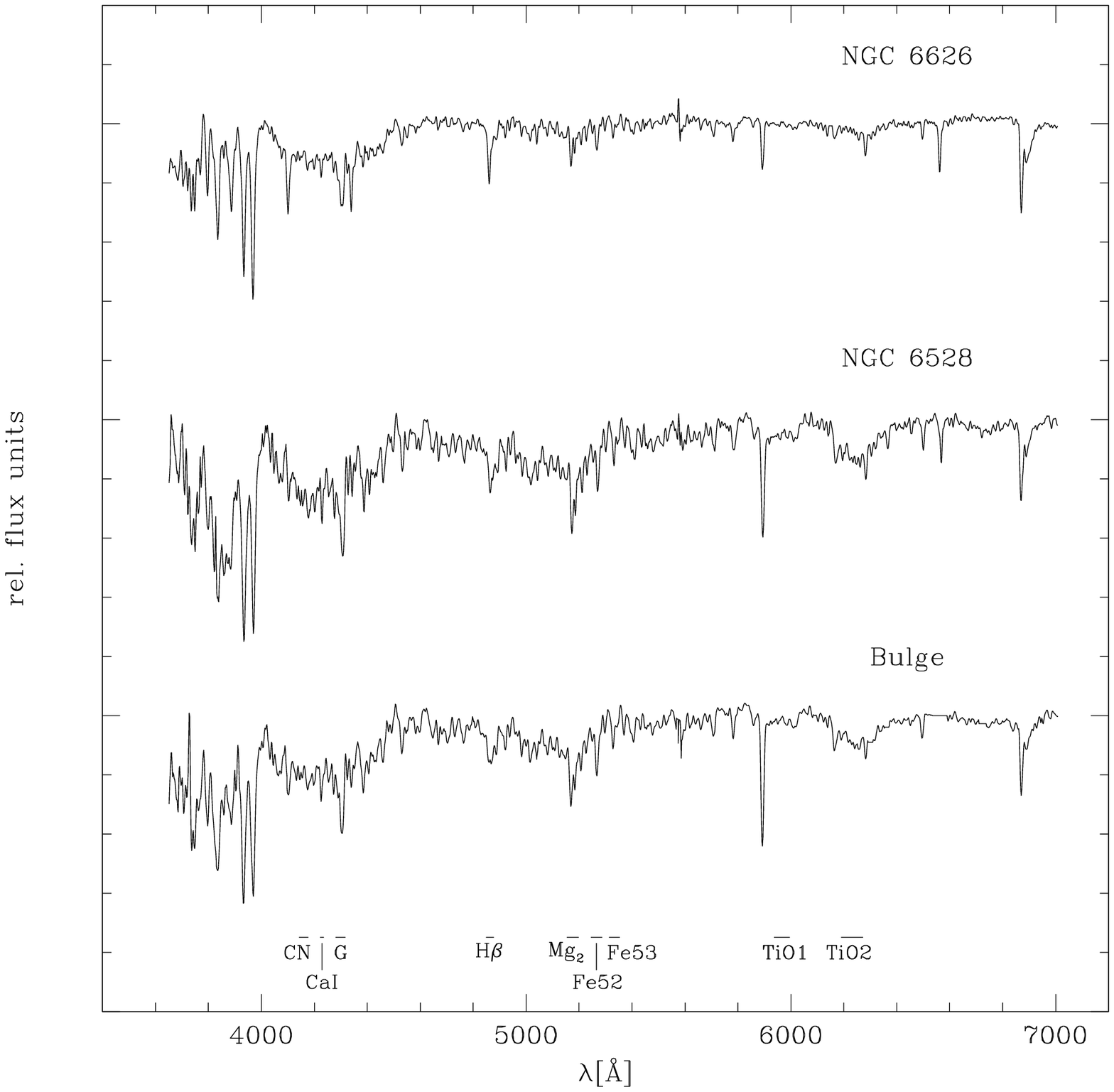}
 \caption{Representative spectra of two globular clusters,
   i.e. NGC~6626 and NGC~6528, and the Galactic bulge. The two
   clusters represent the limits of the metallicity range which is
   covered by our sample. NGC~6626 has a mean metallicity
   [Fe/H]$=-1.45$ dex. NGC~6528, on the other hand, has a mean
   metallicity [Fe/H]$=-0.17$ dex \citep{harris96}. Note the
   similarity between the bulge and the NGC~6528 spectrum. Important
   Lick-index passbands are indicated at the bottom of the panel.}
 \label{ps:plotspec} 
\end{figure*}

Figure \ref{ps:plotspec} shows two representative spectra of a
metal-poor (NGC~6626) and a metal-rich (NGC~6528) globular cluster,
together with the co-added spectrum from the 15 bulge pointings.

In the following we focus on the comparison of index ratios between
globular clusters and the field stellar population in the Galactic
bulge. We include the data of \cite{trager98} who measured Lick
indices for metal-poor globular clusters and use our index
measurements (due to higher S/N) whenever a globular cluster is a
member of both data sets.

All Lick indices are measured on the cleaned and co-added
globular-cluster and bulge spectra. Statistical uncertainties are
determined in bootstrap tests (see App.~\ref{ln:lickcoderr} for
details). We additionally determine the statistical slit-to-slit
variations between the different pointings for each globular cluster
and estimate the maximum systematic error due to the uncertainty in
radial velocity. All line indices and their statistical and systematic
uncertainties are documented in Table \ref{tab:gcindices1}.

It is worth to mention that the slit-to-slit fluctuations of index
values, which are calculated from different pointings (3 and 5 for
globular clusters and 15 for the bulge), are generally larger than the
Poisson noise of the co-added spectra. Such variations are expected
from Poisson fluctuations in the number of bright stars inside the
slit and the sampled luminosities of the single spectra correlate well
with the slit-to-slit index variations for each globular cluster. More
pointings are required to solidify this correlation and to search for
other effects such as radial index changes.

\subsection{The $\alpha$-element Sensitive Indices vs. $\langle$Fe$\rangle$}
\begin{figure*}[!ht]
  \centering 
  \includegraphics[width=8.9cm]{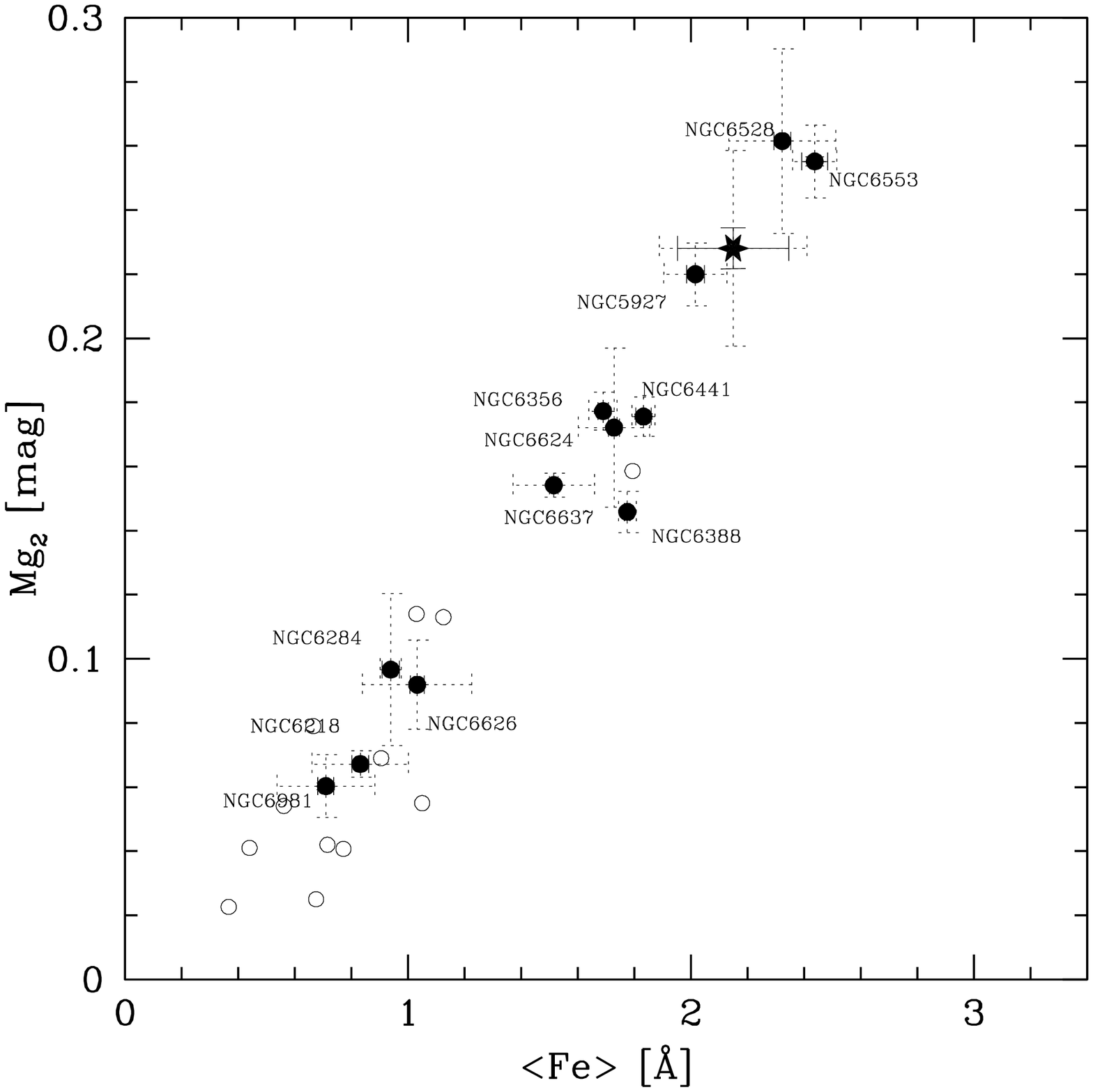}
  \includegraphics[width=8.9cm]{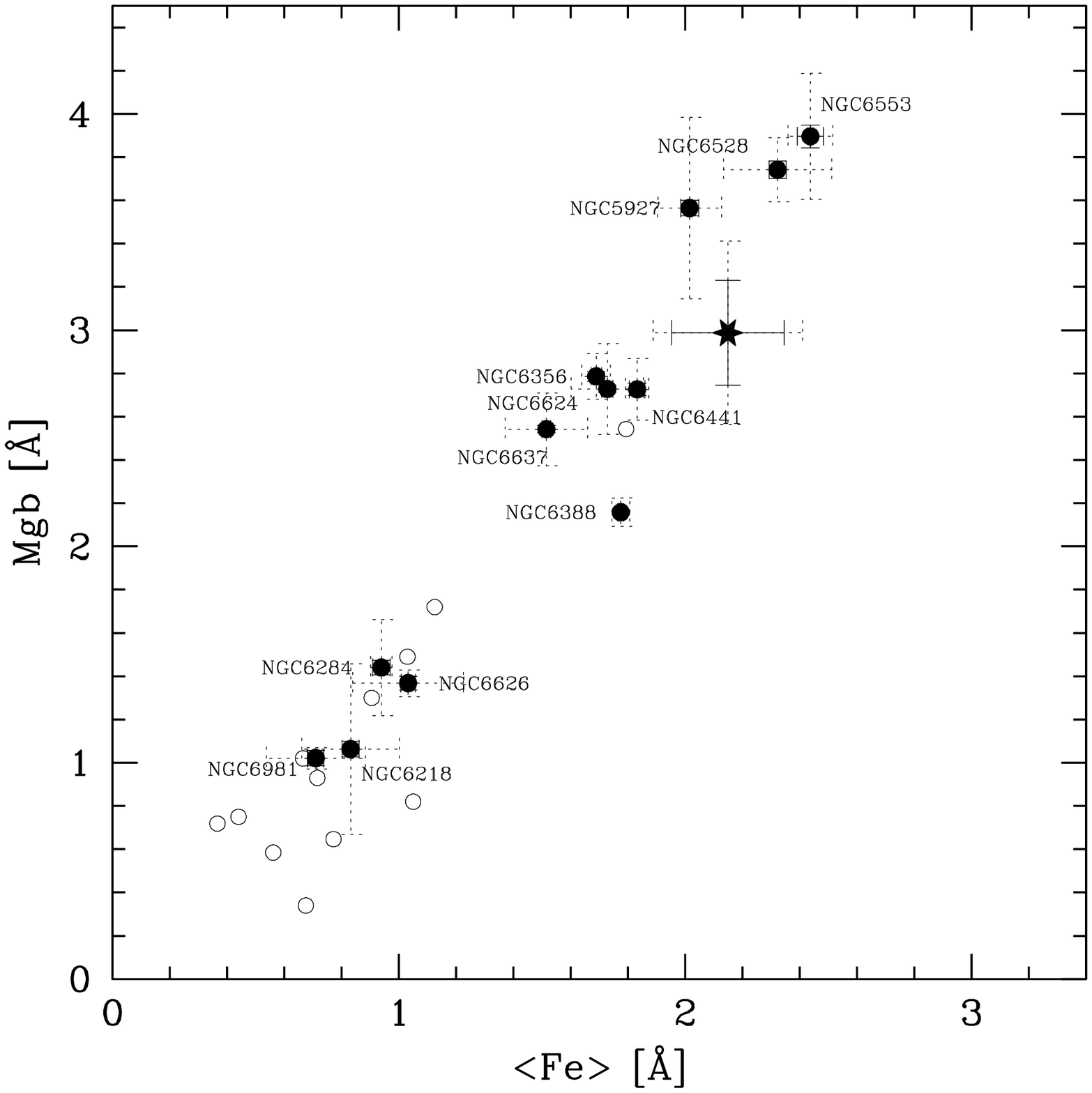}
  \includegraphics[width=8.9cm]{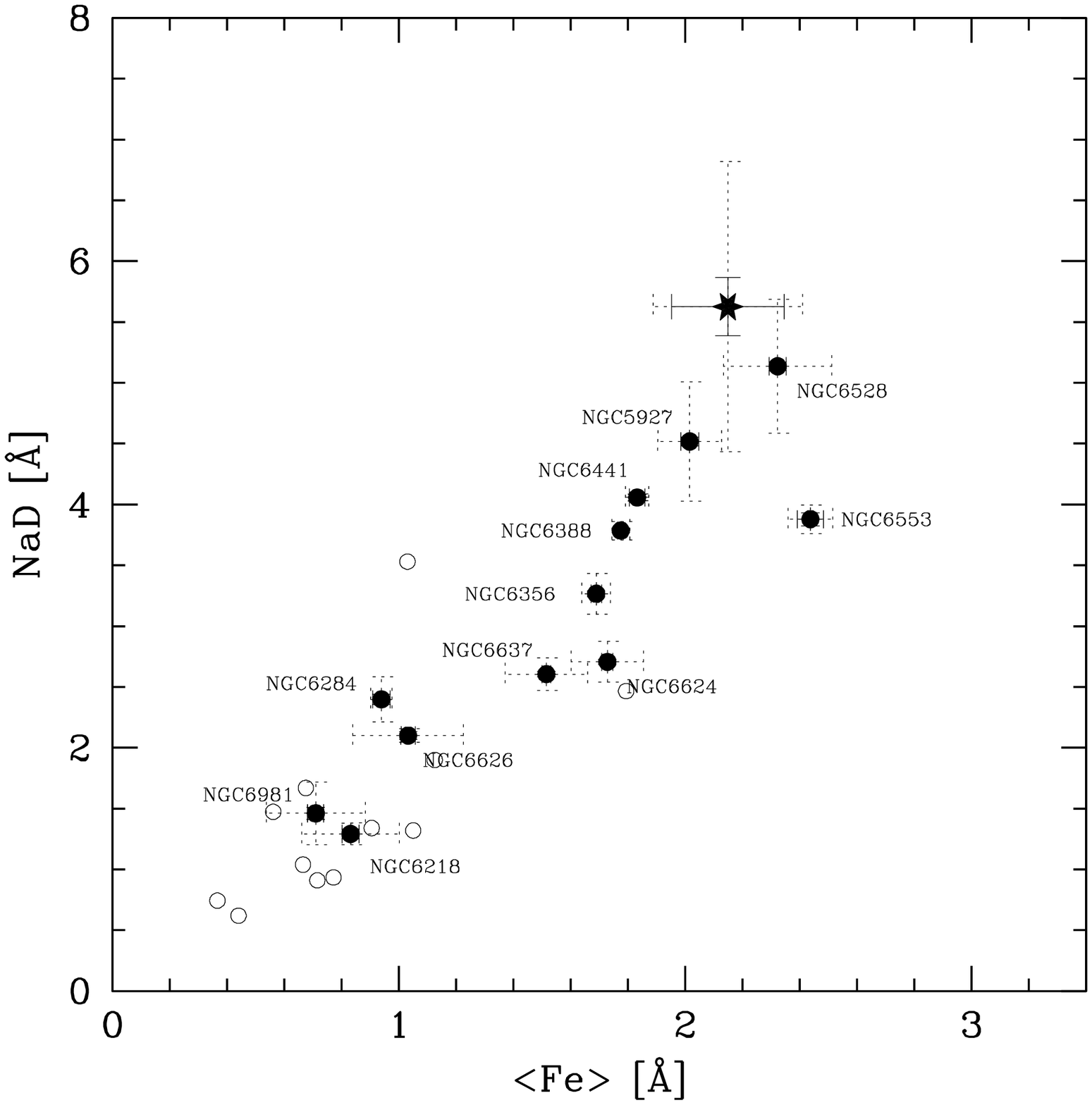}
  \includegraphics[width=8.9cm]{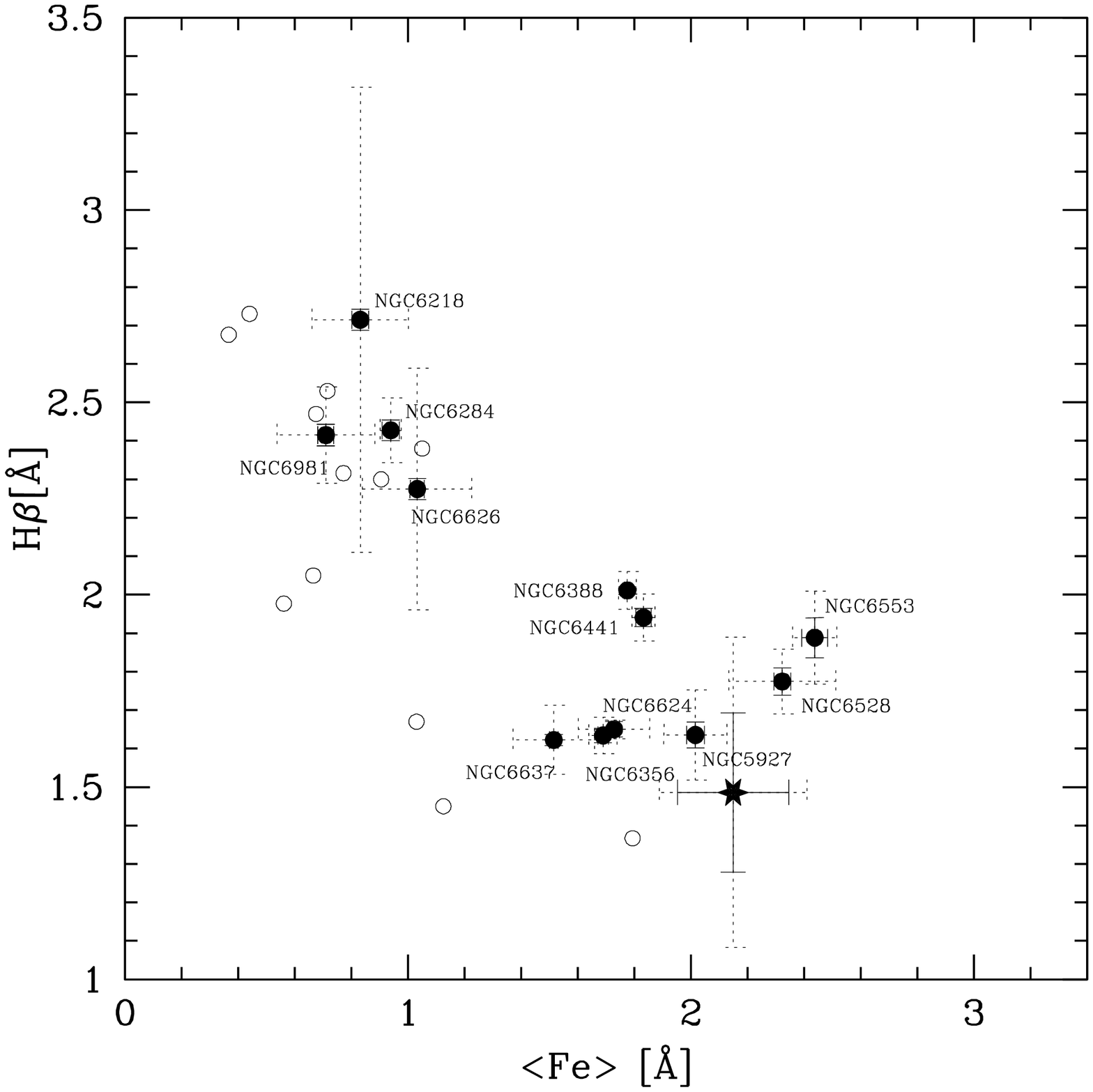}
\caption{Lick-index ratios for Mg$_2$, Mgb, NaD, H$\beta$ versus the
  mean iron index $\langle$Fe$\rangle=(\mathrm{Fe5270} +
  \mathrm{Fe5335})/2$. Filled dots show the index measurements of our
  sample globular clusters, whilst open circles show the data of
  \cite{trager98}. A solid star indicates the index values derived from the
  co-added spectrum of the Galactic bulge. Solid error bars show bootstrap
  errors which represent the total uncertainty due to the intrinsic
  noise of the co-added spectra. Statistical slit-to-slit fluctuations
  between different pointings are shown as dotted error bars. Systematic
  radial velocity errors are not plotted, but given in Table
  \ref{tab:gcindices1}. For clarity reasons no error bars are plotted
  for the \citeauthor{trager98} sample which are generally an order of
  magnitude larger than the intrinsic noise of our spectra. The mean
  errors of the \citeauthor{trager98} data are 0.3 \AA\ for the 
  $\langle$Fe$\rangle$ index, 0.01 mag for Mg$_2$, and 0.3 \AA\ 
  for Mgb, NaD, and H$\beta$.}
 \label{ps:plotsalpha}
\end{figure*}
\addtocounter{figure}{-1}
\begin{figure*}[!ht]
  \centering
  \includegraphics[width=8.9cm]{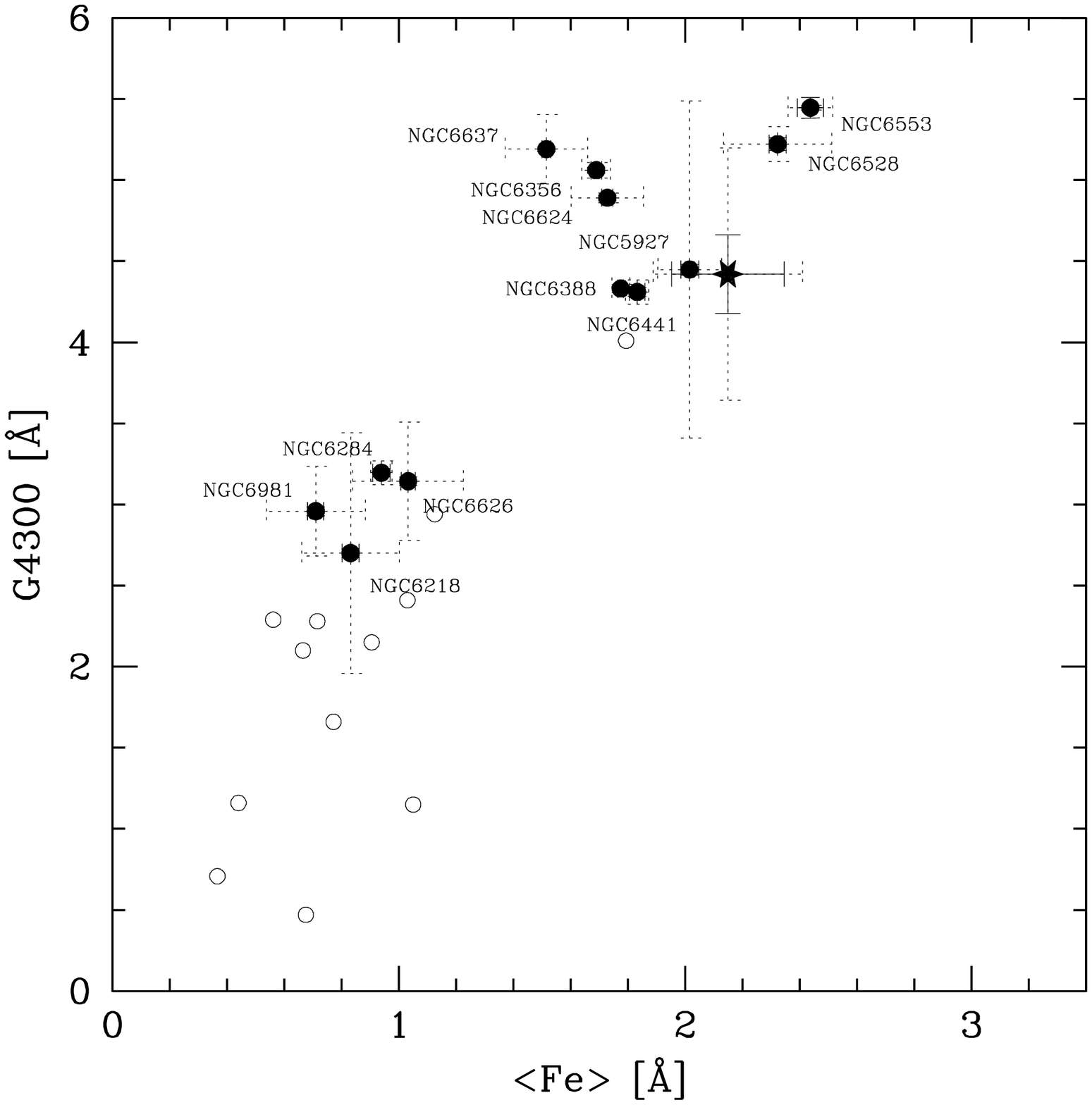}
  \includegraphics[width=8.9cm]{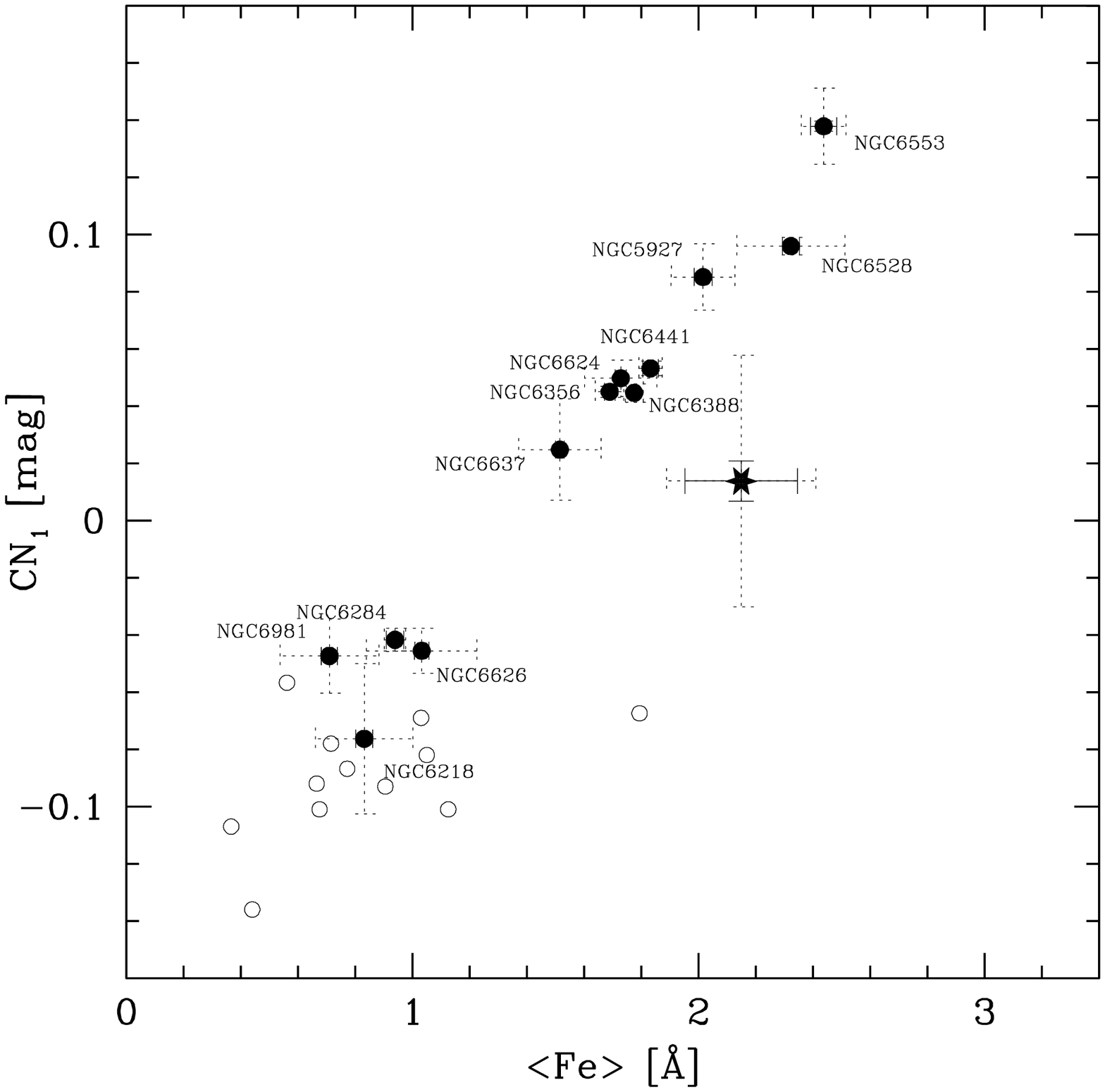}
  \includegraphics[width=8.9cm]{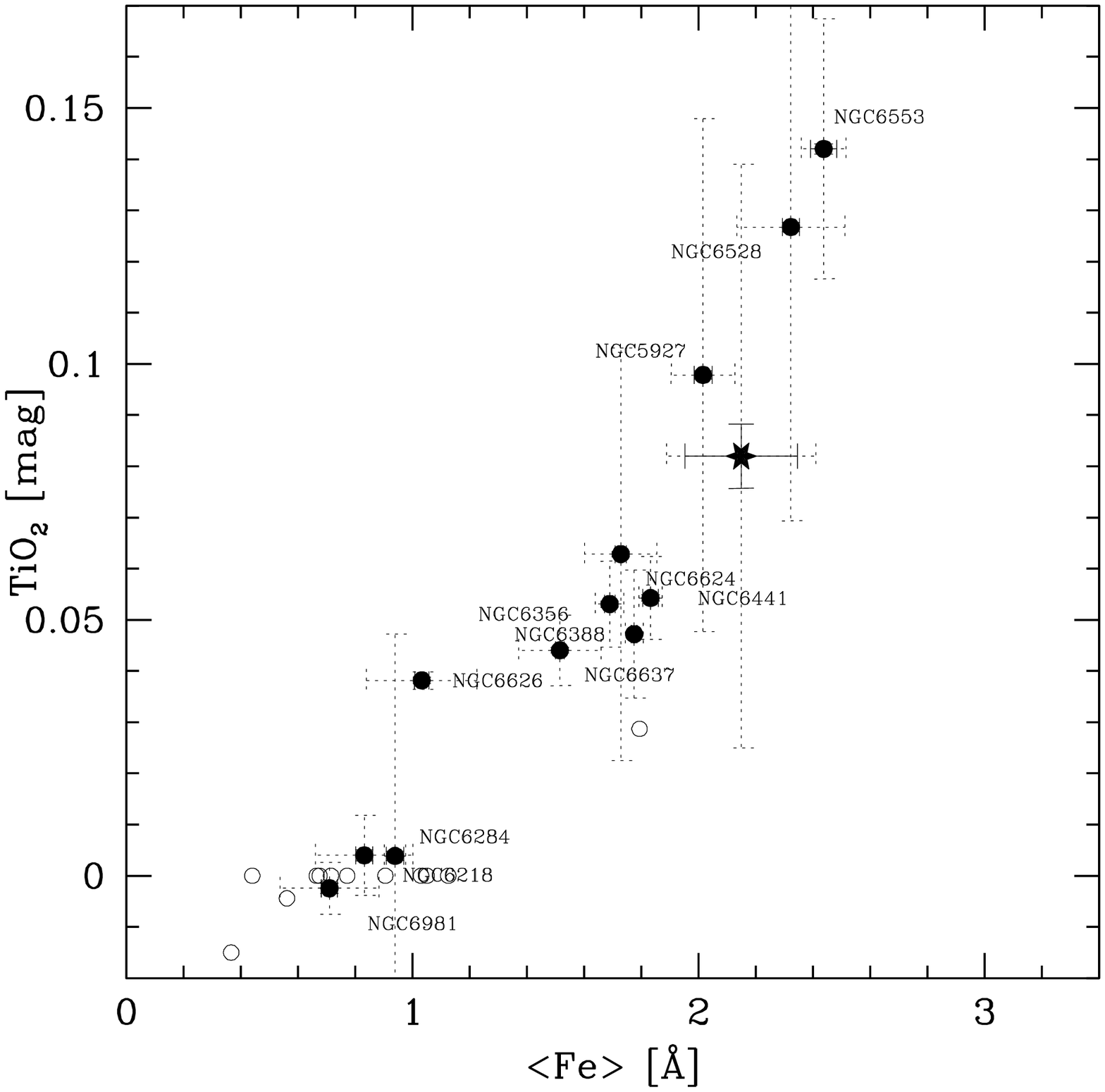}
  \includegraphics[width=8.9cm]{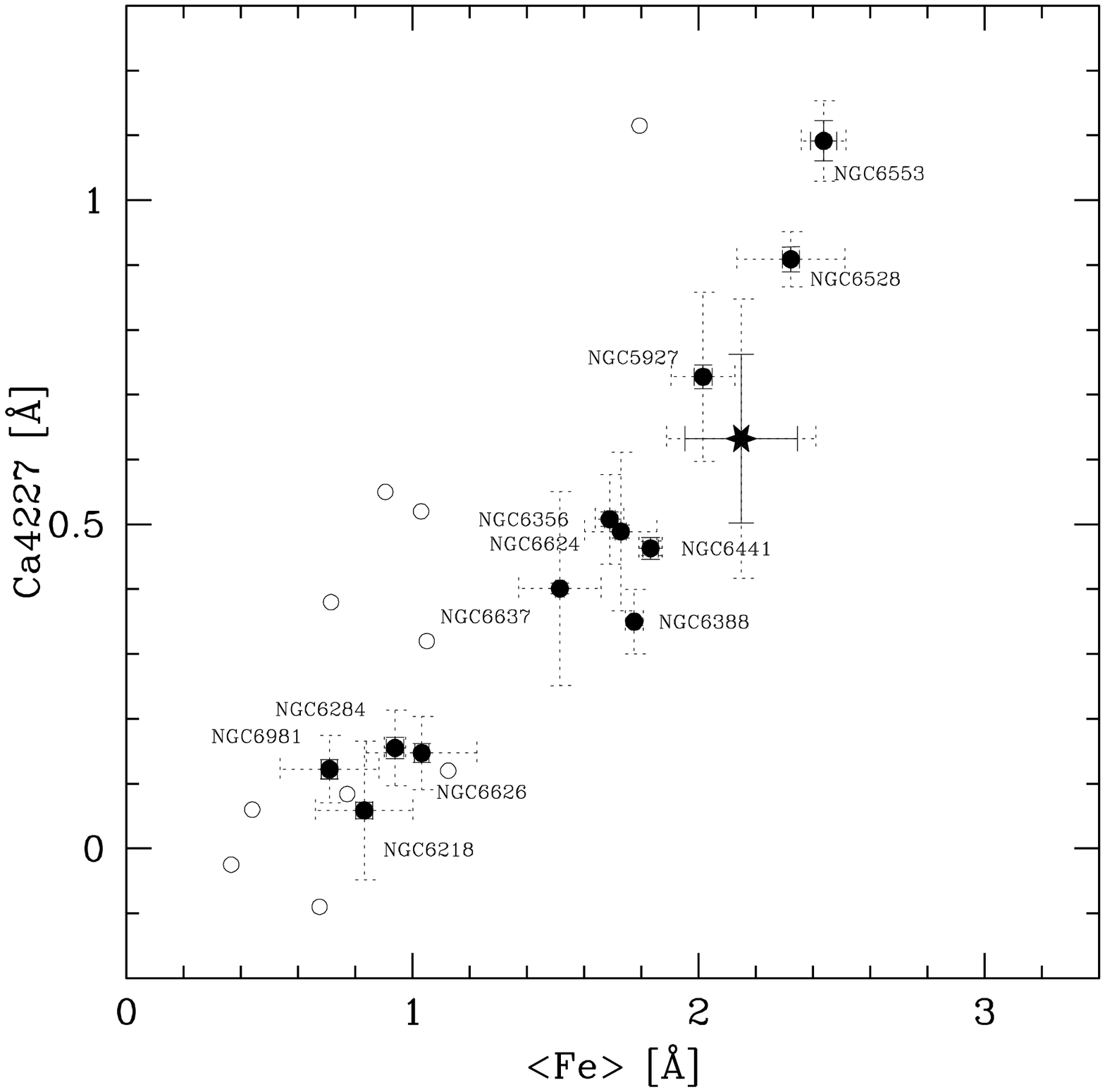}
 \caption{-- continued: G4300, CN$_1$, TiO$_2$, and Ca4227
   versus $\langle$Fe$\rangle$. The mean errors of the
  \citeauthor{trager98} data are 0.3 \AA\ for the $\langle$Fe$\rangle$
  index, and 0.4 \AA, 0.03 mag, 0.01 mag, and 0.4 \AA\ for the G4300,
  CN$_1$, TiO$_2$, and Ca4227 index, respectively.}  
\end{figure*}

$\alpha$-particle capture elements with even atomic numbers (C, O, Mg,
Si, Ca, etc.) are predominantly produced in type II supernovae
\citep{tsujimoto95, woosley95, thomas98}. The progenitors of SNe~II
are massive stars, which explode and pollute the interstellar medium
after their short lifetime of some $10^7$ years. The ejecta of SNe~II
have a mean [$\alpha$/Fe]$\sim 0.4$ dex. On the other hand, type Ia
supernovae eject mainly iron-peak elements ([$\alpha$/Fe]$\sim -0.3$
dex) $\sim1$ Gyr after the formation of their progenitor stars.
Stellar populations which have been created on short timescales are
likely to show [$\alpha$/Fe] enhancement. The [$\alpha$/Fe] ratio is
therefore potentially a strong discriminator of star-formation
histories. Alternative explanations, however, include a changing IMF
slope and/or a changing binary fraction.

Such enhancements have already been suspected and observed in the
stellar populations in giant elliptical galaxies \citep{worthey92},
the Galactic bulge \citep{mcwilliam94}, and for disk and halo stars in
the Milky Way \citep{edvardsson93, fuhrmann98}. A detailed discussion
of the [$\alpha$/Fe] ratio in our sample globular clusters and their
assistance to parameterize simple stellar population models for
varying [$\alpha$/Fe] ratios will be presented in the second paper of
this series \citep{maraston02}.

To search for any trends in the index($\alpha$)/index(Fe) ratio in the
globular cluster population and the bulge we plot supposedly
$\alpha$-element sensitive indices against the mean iron index
$\langle$Fe$\rangle$. Figure \ref{ps:plotsalpha} shows some
representative index measurements for globular clusters and bulge
fields. Generally, all the correlations between $\alpha$-sensitive
indices and the mean iron index are relatively tight. For our sample
globular clusters a Spearman rank test yields values between 0.87 and
0.97 (1 indicates perfect correlation, $-1$ anti-correlation) for the
indices CN$_1$, TiO$_2$, Ca4227, Mgb, Mg$_2$. The CN$_1$ and TiO$_2$
indices show the tightest correlation with $\langle$Fe$\rangle$,
followed by Mg$_2$ and Ca4227. All correlations are linear (no
higher-order terms are necessary) and hold to very high metallicities
of the order of the mean bulge metallicity (filled star in Figure
\ref{ps:plotsalpha}). The three most metal-rich globular clusters in
our sample, i.e. NGC~5927, NGC~6528, and NGC~6553, have roughly the
same mean iron index as the stellar populations in the Galactic bulge
indicating similar [Fe/H]. This was also found in recent photometric
CMD studies of the two latter globular clusters and the bulge
\citep{ortolani95,zoccali02}. Ranking by the $\langle$Fe$\rangle$ and
Mg indices, which are among the best metallicity indicators in the
Lick sample of indices (see Sect.~\ref{ln:indmet}), the most
metal-rich globular cluster in our sample is NGC~6553, followed by
NGC~6528 and NGC~5927.

The comparison of some $\alpha$-sensitive indices of globular clusters
and the bulge requires some further words. The Ca4227, Mgb, and Mg$_2$
index of the bulge light is in good agreement with the sequence formed
by globular clusters. All deviations from this sequence are of the
order of $\la1\sigma$ according to the slit-to-slit variations. One
exception is the CN index which is significantly higher in metal-rich
globular clusters than in the bulge. We discuss this important point
in Section \ref{ln:cn}. In general, our data show that the ratio of
$\alpha$-sensitive to iron-sensitive indices is comparable in
metal-rich globular clusters and in the stellar population of the
Galactic bulge.

Likely super-solar [$\alpha$/Fe] ratios in globular clusters and the
bulge were shown in numerous high-resolution spectroscopy
studies. From a study of 11 giants in Baade's window
\cite{mcwilliam94} report an average [Mg/Fe]$\approx 0.3$ dex, while
\cite{barbuy99} and \cite{carretta01} find similar [Mg/Fe] ratios in
two red giants in NGC 6553 and in four red horizontal branch stars in
NGC~6528. Similarly, \citeauthor{mcwilliam94} find [Ca/Fe]$\approx0.2$
dex, which is reflected by the former observations in globular
clusters. Although the studied number of stars is still very low, the
first high-resolution spectroscopy results point to a similar
super-solar $\alpha$-element abundance in both Milky Way globular
clusters and the bulge which is supported by our data.

\subsection{CN vs. $\langle$Fe$\rangle$}
\label{ln:cn}
The CN index measures the strength of the CN absorption band at 4150
\AA. The Lick system defines two CN indices, CN$_1$ and CN$_2$ which
differ slightly in their continuum passband definitions. The
measurements for both indices give very similar results, but we prefer
the CN$_1$ index due to its smaller calibration biases (see
Fig.~\ref{ps:idxcomp}) and refer in the following to CN$_1$ as the CN
index.

Like for most other indices, the CN index of globular clusters
correlates very tightly with the $\langle$Fe$\rangle$ index, following
a linear relation (see Figure \ref{ps:plotsalpha}). A Spearman rank
test yields 0.97 as a correlation coefficient. The apparent gap at CN
$\sim0$ mag is a result of the bimodal distribution of metallicity in
our cluster sample, and similar gaps are recognizable in all other
index vs. $\langle$Fe$\rangle$ diagrams.

Quite striking is the comparison of the bulge value of the CN index
with that of globular clusters at the same value of the
$\langle$Fe$\rangle$ index: the CN index of the bulge is significantly
offset to a lower value by $\sim 0.05$ mag, corresponding to at least
a 2$\sigma$ effect. This is also evident from Figure
\ref{ps:plotspec}, showing that the CN feature is indeed much stronger
in the cluster NGC~6528 than in the bulge spectrum. We also note that
the CN index of NGC~6528 and NGC~6553 is as strong as in the most
metal-rich clusters in M31 studied by \cite{burstein84}.

It is well known that globular cluster stars often exhibit so-called
{\it CN anomalies}, with stars in a cluster belonging either to a
CN-strong or a CN-weak group (see \citealt{kraft94} for an extended
review). Among the various possibilities to account for these
anomalies, accretion of AGB ejecta during the early phases of the
cluster evolution appears now the most likely explanation
\citep{kraft94, ventura01}, as originally proposed by \cite{dantona83}
and \cite{renzini83}. In this scenario, some $\sim 30\times 10^6$
years after cluster formation (corresponding to the lifetime of $\sim
8\, M_\odot$ stars) the last Type II supernovae explode and AGB stars
begin to appear in the cluster. Then the low-velocity AGB wind and
super-wind materials may accumulate inside the potential well of the
cluster, and are highly enriched in carbon and/or nitrogen from the
combined effect of the third dredge-up and envelope-burning processes
\citep{renzini81}. Conditions are then established for the low-mass
stars (now still surviving in globular clusters) having a chance to
accrete carbon and/or nitrogen-enriched material, thus preparing the
conditions for the CN anomalies we observe in today clusters. One of
the arguments in favor of the accretion scenario is that field stars
do not share the CN anomalies of their cluster counterparts
\citep{kraft82}. Indeed, contrary to the case of clusters, in the
field no localized, high-density accumulation of AGB ejecta could take
place, and low-mass stars would have not much chance to accrete AGB
processed materials. In the case of the bulge, its much higher
velocity dispersion ($\sim 100$ km s$^{-1}$) compared to that of
clusters (few km s$^{-1}$) would make accretion even less likely. In
conclusion, we regard the lower CN index of the bulge relative to
metal-rich globular clusters as consistent with -- and actually
supporting -- the accretion scenario already widely entertained for
the origin of CN anomalies in globular-cluster stars.

\subsection{H$\beta$ vs. $\langle$Fe$\rangle$} Figure
\label{ln:hbeta}
\ref{ps:plotsalpha} shows a plot of H$\beta$ vs.
$\langle$Fe$\rangle$. The Spearman rank coefficient for the globular
cluster sequence is $-0.52$ indicating a mild anti-correlation. At
high $\langle$Fe$\rangle$, the H$\beta$ index of globular clusters is
slightly stronger than that of the bulge field. However, the values
are consistent with each other within $\sim1\sigma$, with the large
slit-to-slit variations exhibited by the bulge spectra being a result
of the lower luminosity sampling due to the lower surface brightness
in Baade's Window compared to globular clusters.

The two clusters NGC~6441 and NGC~6388 show somewhat stronger H$\beta$
compared to clusters with similar $\langle$Fe$\rangle$ index. This
offset is probably caused by the conspicuous blue extension of the HB
of these two clusters, a so far unique manifestation of the ``second
parameter'' effect among the metal-rich population of bulge globular
clusters \citep{rich97}. Contrary to NGC~6441 and NGC~6388, the other
globular clusters with comparable $\langle$Fe$\rangle$ indices
(i.e. NGC~5927, NGC~6356, NGC~6624, and NGC~6637) have without
exception purely red horizontal branches (HBR$=-1.0$).

Also the two most metal-rich clusters in our sample, NGC~6553 and
NGC~6528, appear to have a somewhat stronger H$\beta$ compared to a
linear extrapolation of the trend from lower values of the
$\langle$Fe$\rangle$ index. In this case, however, the relatively
strong H$\beta$ cannot be ascribed to the HB morphology, since the HB
of these two clusters is purely red \citep{ortolani95n, zoccali01}. In
principle, a younger age would produce a higher H$\beta$ index, but
optical and near-infrared HST color-magnitude diagrams of these two
clusters indicate they are virtually coeval with halo clusters
\citep{ortolani95n, ortolani01, zoccali01, feltzing02}. So, we are
left without an obvious interpretation of the relatively strong
H$\beta$ feature in the spectra of these clusters.  Perhaps the effect
is just due to insecure sampling, i.e., to statistical fluctuations in
the stars sampled by the slit in either the cluster or in the adjacent
bulge field used in the background subtraction. Another reason for the
offset might be the increasing dominance of metallic lines inside the
H$\beta$ feature passband which could artificially increase the index
value.

\subsection{Other Indices vs. $\langle$Fe$\rangle$}
\label{ln:otherind}

NaD -- The correlation coefficient for this index pair is
0.94. Globular clusters and bulge compare well within the errors. Both
stellar populations follow, within their uncertainties, the same
trend. A clear exception from this correlation is NGC~6553, which
shows a significantly lower NaD index for its relatively high
$\langle$Fe$\rangle$ than the sequence of all other globular
clusters. The reason for this offset is unclear.

G4300 -- The G4300 index predominantly traces the carbon abundance in
the G band. For giants, its sensitivity to oxygen is about $1/3$ of
that to carbon \citep{tripicco95}. The metal-rich globular clusters
fall in the same region as the bulge data. In combination with the CN
index which mainly traces the CN molecule abundance, this implies that
the offset between bulge and globular clusters in the CN
vs. $\langle$Fe$\rangle$ plot is most likely due to an offset in the
nitrogen abundance between bulge and clusters.

TiO -- The TiO abundance is measured by the TiO$_1$ and TiO$_2$
indices. Both indices do not differ in their correlation with the mean
iron index (Spearman rank coefficient 0.96), but we use TiO$_2$
because of its better calibration. In Figure \ref{ps:plotsalpha} we
plot TiO$_2$ vs. $\langle$Fe$\rangle$ which shows the strongest
indices for NGC~6553 and NGC~6528, followed by NGC~5927 and the bulge.

The absorption in the TiO band sensitively depends on $T_{\rm eff}$
which is very low for very metal-rich RGB stars. While the strongest
TiO bands are observed in metal-rich M-type giants almost no
absorption is seen in metal-rich K-type RGB stars. As $T_{\rm eff}$
decreases towards the RGB tip, a large increase in the TiO-band
absorption occurs which drives the observed bending of the upper RGB
in color-magnitude diagrams, in particular those which use V-band
magnitudes \citep{carretta98, saviane00}. In fact, the most metal-rich
globular clusters in the Milky Way, e.g. NGC~6553 and NGC~6528, show
the strongest bending of the RGBs \citep[e.g.][]{ortolani91, cohen95}.
Figure \ref{ps:plotsalpha} shows that the slit-to-slit scatter is
extremely large for the metal-rich data. This is likely reflecting the
sparsely populated upper RGB. In other words, for metal-rich stellar
populations the TiO index is prone to be dominated by single bright
stars which increase the slit-to-slit scatter due to statistically
less significant sampling (see also the high slit-to-slit scatter of
NGC~6218 due to its small luminosity sampling). Another Ti-sensitive
index in the Lick system is Fe4531 \citep{gorgas93}. It shows similar
behaviour as a function of $\langle$Fe$\rangle$.

\begin{table*}[t!]
 \caption{Coefficients of the index vs. [Fe/H] relations. The
   r.m.s. ($\sqrt{\chi^2 /n}$) is given in the units of the
   parameterization (in dex in equation \ref{eqn:indmet1} and in \AA\
   or mag in equation \ref{eqn:indmet2}).}
 \label{tab:indmetcoef}
\begin{tabular}{l|cccc|cccc}
\hline
\noalign{\smallskip}
 index & $a$ & $b$ & $c$ & r.m.s.
       & $d$ & $e$ & $f$ & r.m.s. \\    
\noalign{\smallskip}
\hline
\noalign{\smallskip}
Mg$_2$& $-2.46\pm0.10$ &$16.24\pm1.81$ & $-29.88\pm6.52$ &0.151
&       $ 0.29\pm0.01$ & $0.22\pm0.02$ & $  0.05\pm0.01$ &0.016 \\ 
Mgb   & $-2.53\pm0.14$ & $1.11\pm0.16$ & $ -0.14\pm0.04$ &0.182
&       $ 4.46\pm0.19$ & $3.51\pm0.35$ & $  0.79\pm0.14$ &0.254 \\
$\langle$Fe$\rangle$
      & $-2.83\pm0.21$ & $1.91\pm0.36$ & $ -0.35\pm0.13$ &0.199
&       $ 2.68\pm0.12$ & $1.85\pm0.23$ & $  0.39\pm0.09$ &0.167 \\   
{[MgFe]}&$-2.76\pm0.14$& $1.59\pm0.20$ & $ -0.26\pm0.06$ &0.150
&       $ 3.45\pm0.13$ & $2.55\pm0.24$ & $  0.55\pm0.10$ &0.173 \\
H$\beta$&$-1.99\pm2.26$& $2.09\pm2.24$ & $ -0.78\pm0.54$ &0.384
&       $ 1.55\pm0.20$ &$-0.33\pm0.37$ & $  0.08\pm0.15$ &0.271 \\
CN$_1$& $-0.83\pm0.11$ & $6.84\pm0.86$ & $-17.12\pm13.47$&0.314
&       $ 0.16\pm0.02$ & $0.26\pm0.04$ & $  0.06\pm0.02$ &0.032 \\  
 \noalign{\smallskip}
 \hline
\end{tabular}
\end{table*}

\section{Index-Metallicity Relations}
\label{ln:indmet}

\begin{figure*}[!ht]
 \centering 
 \includegraphics[width=18.5cm]{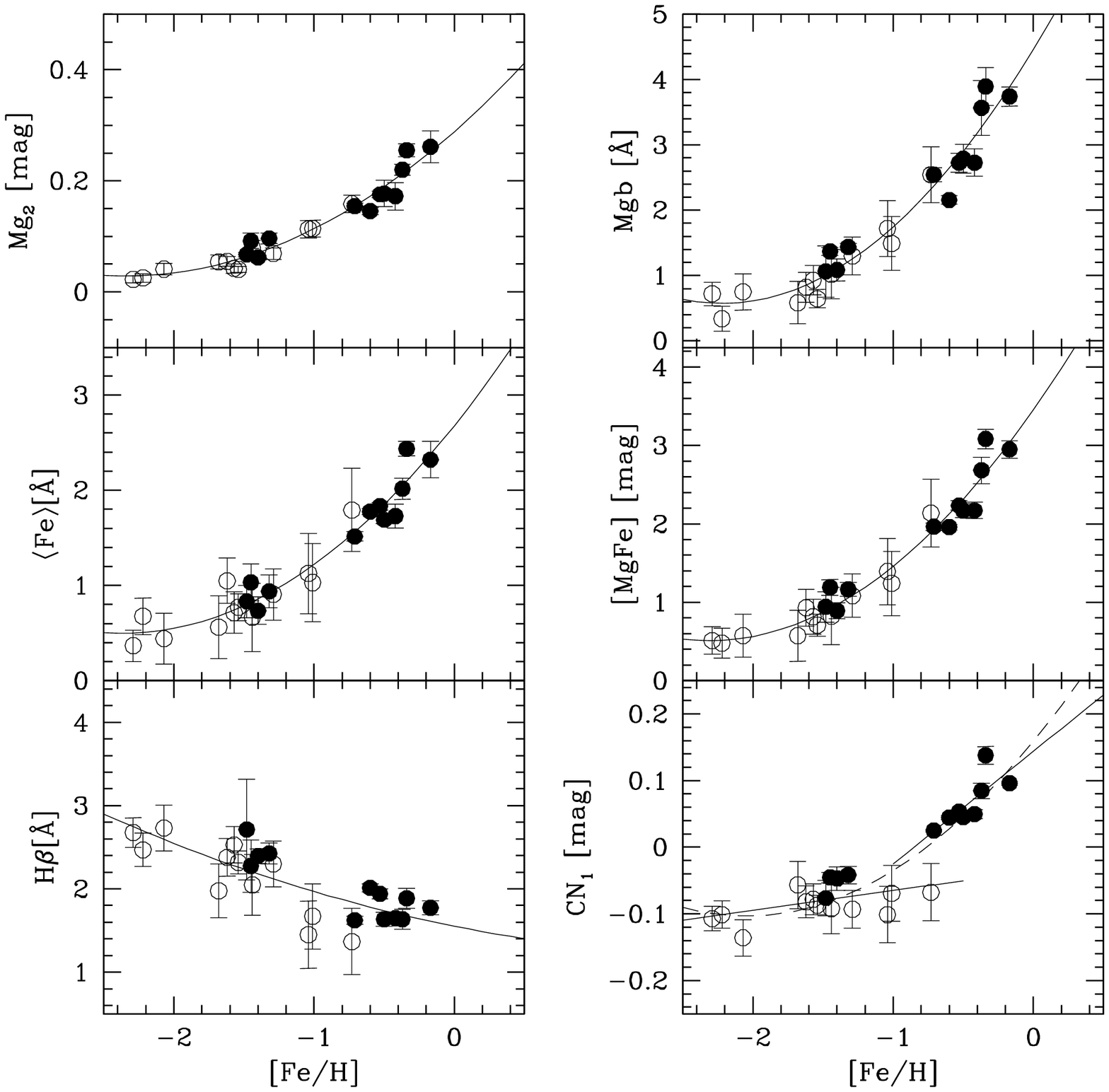}
 \caption{Line indices as a function of mean globular cluster
   metallicity. Our sample globular clusters are shown as filled
   circles while the open circles denote the globular cluster data of
   \cite{trager98}.}
 \label{ps:indmet}
\end{figure*}

We use the mean [Fe/H] values from the 1999 update of the McMaster
catalog \citep{harris96} to create parabolic relations between line
indices and the globular cluster metallicity as expressed by [Fe/H],
based on the Zinn-West scale\footnote{Note that to derive [Z/H] from
[Fe/H], the [$\alpha$/Fe] of the globular clusters needs to be
accounted for.} \citep{zinn84}. Together with the globular cluster
data of \cite{trager98} the sample comprises 21 Galactic globular
cluster with metallicities $-2.29\leq$[Fe/H]$\leq-0.17$. Figure
\ref{ps:indmet} shows six indices as a function of [Fe/H] most of
which show tight correlations. Least-square fitting of second-order
polynomials
\begin{eqnarray}
\label{eqn:indmet1}
{\rm[Fe/H]}& = & a + b\cdot ({\rm EW}) + c\cdot({\rm EW})^2 \\
\label{eqn:indmet2}
{\rm EW}   & = & d + e\cdot {\rm[Fe/H]} + f\cdot{\rm[Fe/H]}^2
\end{eqnarray}
where EW is the index equivalent width in Lick units, allows a simple
parameterization of these sequences as index vs. [Fe/H] and vice
versa. The obtained coefficients are summarized in Table
\ref{tab:indmetcoef}. Higher-order terms improve the fits only
marginally and are therefore unnecessary. 

These empirical relations represent metallicity calibrations of Lick
indices with the widest range in [Fe/H] ever obtained. Note that the
best metallicity indicators in Table \ref{tab:indmetcoef} are the
[MgFe] and Mg$_2$ indices both with a r.m.s. of 0.15 dex. Leaving out
globular clusters with poor luminosity sampling and relatively
uncertain background subtraction (i.e. NGC~6218, NGC~6553, NGC~6626,
and NGC~6637) changes the coefficients only little within their error
limits. In particular, the high-metallicity part of all relations is
not driven by the metal-rich globular cluster NGC~6553.

We point out that all relations could be equally well fit by
first-order polynomials if the metal-rich clusters are
excluded. Consequently, such linear relations would overestimate the
metallicity for a given index value at high metallicities (except for
H$\beta$ which would underestimate [Fe/H]; however, H$\beta$ is anyway
not a good metallicity indicator). This clearly emphasizes the caution
one has to exercise when deriving mean metallicities from SSP models
which have been extrapolated to higher metallicities. The current
sample enables a natural extension of the metallicity range for which
Lick indices can now be calibrated. In the second paper of the series
\citep{maraston02} we compare the data with the predictions of SSP
models.

We also point out that the fitting of the CN index improves when
CN$>0$ and CN$<0$ data are fit separately by first-order polynomials.
The lines are indicated in Figure \ref{ps:indmet}. Their functional
forms are
\begin{eqnarray*}
{\rm CN} =&  (0.14\pm0.03)+(0.17\pm0.06)\cdot {\rm [Fe/H]}&{\rm : CN}>0 \\
{\rm CN} =& (-0.04\pm0.02)+(0.03\pm0.01)\cdot {\rm [Fe/H]}&{\rm : CN}<0
\end{eqnarray*}
with reduced $\chi^2$ of 0.025 and 0.023. The inverse relations are
\begin{eqnarray*}
{\rm [Fe/H]} =& (-0.69\pm0.09)+(3.54\pm1.18)\cdot {\rm EW}&{\rm : CN}>0 \\
{\rm [Fe/H]} =& (-0.86\pm0.32)+(8.10\pm3.86)\cdot {\rm EW}&{\rm : CN}<0 
\end{eqnarray*}
with a r.m.s. of 0.115 and 0.380. The change in the slope occurs at
[Fe/H]$\sim-1.0$ dex and is significant in both parameterizations. The
metallicity sensitivity in the metal-poor part is around six times
smaller than in the metal-rich part. Only the inclusion of metal-rich
bulge globular clusters allows the sampling of the transition region
between the shallow and the steep sequence of the CN vs. [Fe/H]
relation.

\section{Galactocentric Index Variations}
\label{ln:radial}
\begin{figure*}[!ht]
 \centering 
 \includegraphics[width=18.5cm]{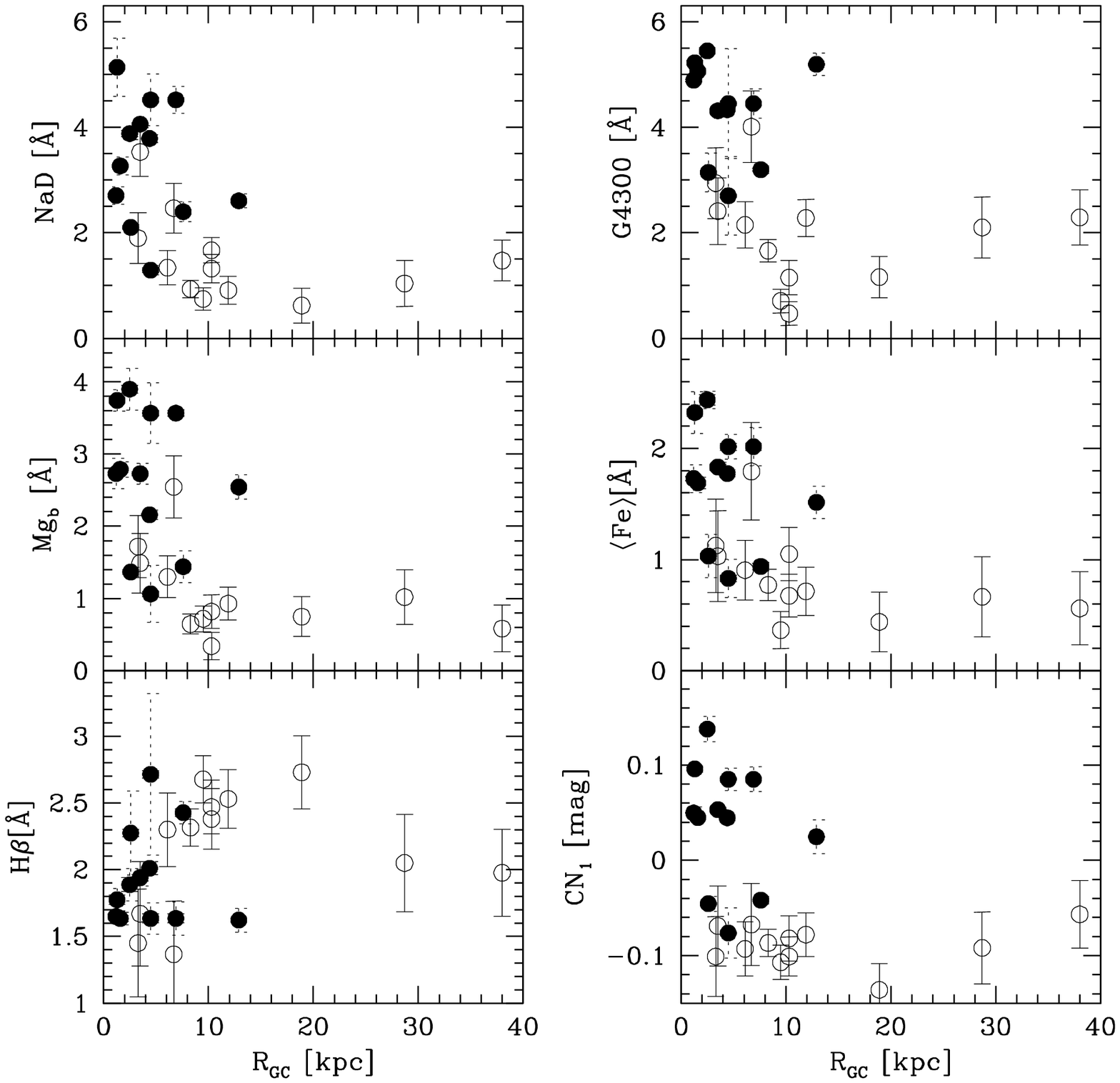}
 \caption{Various line indices as a function of galactocentric
   radius $R_{\rm GC}$. Filled dots show our sample globular clusters.
   Their error bars are split into the Poisson error (solid error bars
   which are very small) and slit-to-slit variations (dashed error
   bars) Open circles mark the globular clusters from
   \cite{trager98}.}
 \label{ps:radplot}
\end{figure*}

In Figure~\ref{ps:radplot} we plot some Lick indices as a function of
galactocentric radius $R_{\rm GC}$. To increase the range in radius,
we again merge our sample with the data for metal-poor halo globular
cluster of \cite{trager98}. The galactocentric radius was taken from
the 1999 update of the McMaster catalog of Milky Way globular clusters
\citep{harris96}. Our compilation includes now both bulge and halo
globular clusters and spans a range $\sim1-40$ kpc in galactocentric
distance. 

All metal indices show a gradually declining index strength as a
function of $R_{\rm GC}$. The inner globular clusters show a strong
decrease in each index out to $\sim10$ kpc. The sequence continues at
apparently constant low values out to large radii. Furthermore, some
indices (CN, Mgb, and $\langle$Fe$\rangle$) show a dichotomy between
the bulge and the halo globular cluster system. While the Mgb and
$\langle$Fe$\rangle$ indices clearly reflect the bimodality in the
metallicity distribution of Milky Way globular clusters, the striking
bimodality in the CN index is more difficult to understand. In the
context of Section~\ref{ln:cn} this may well be explained by
evolutionary differences between metal-rich bulge and metal-poor halo
globular clusters.

The behavior of H$\beta$ differs from that of the other indices.
There is no clear sequence of a decreasing index as a function of
$R_{\rm GC}$, as for the metal-sensitive indices. Instead we measure a
mean H$\beta$ index with $2.1\pm0.5$~\AA . The strength of the Balmer
series is a function of $T_{\rm eff}$. In old stellar populations,
relatively hot stars, which contribute significantly to the
Balmer-line strength of the integrated light, are found at the main
sequence turn-off and on the horizontal branch. The temperature of the
turn-off is a function of age and metallicity while the temperature of
the horizontal branch is primarily a function of metallicity and, with
exceptions, of the so-called ``second parameter''.

\begin{figure}[!ht]
 \centering 
 \includegraphics[width=8.0cm]{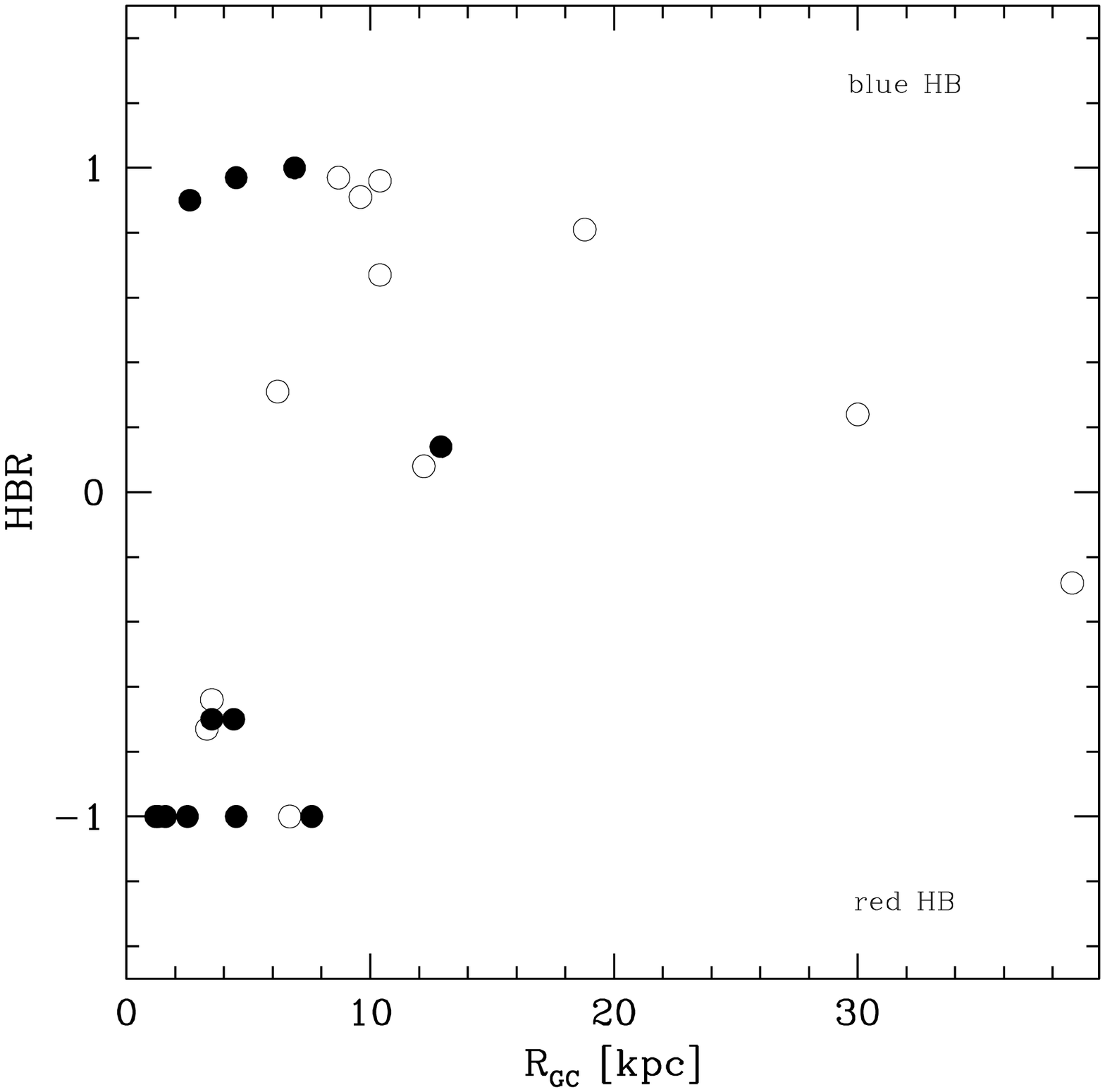}
 \includegraphics[width=8.0cm]{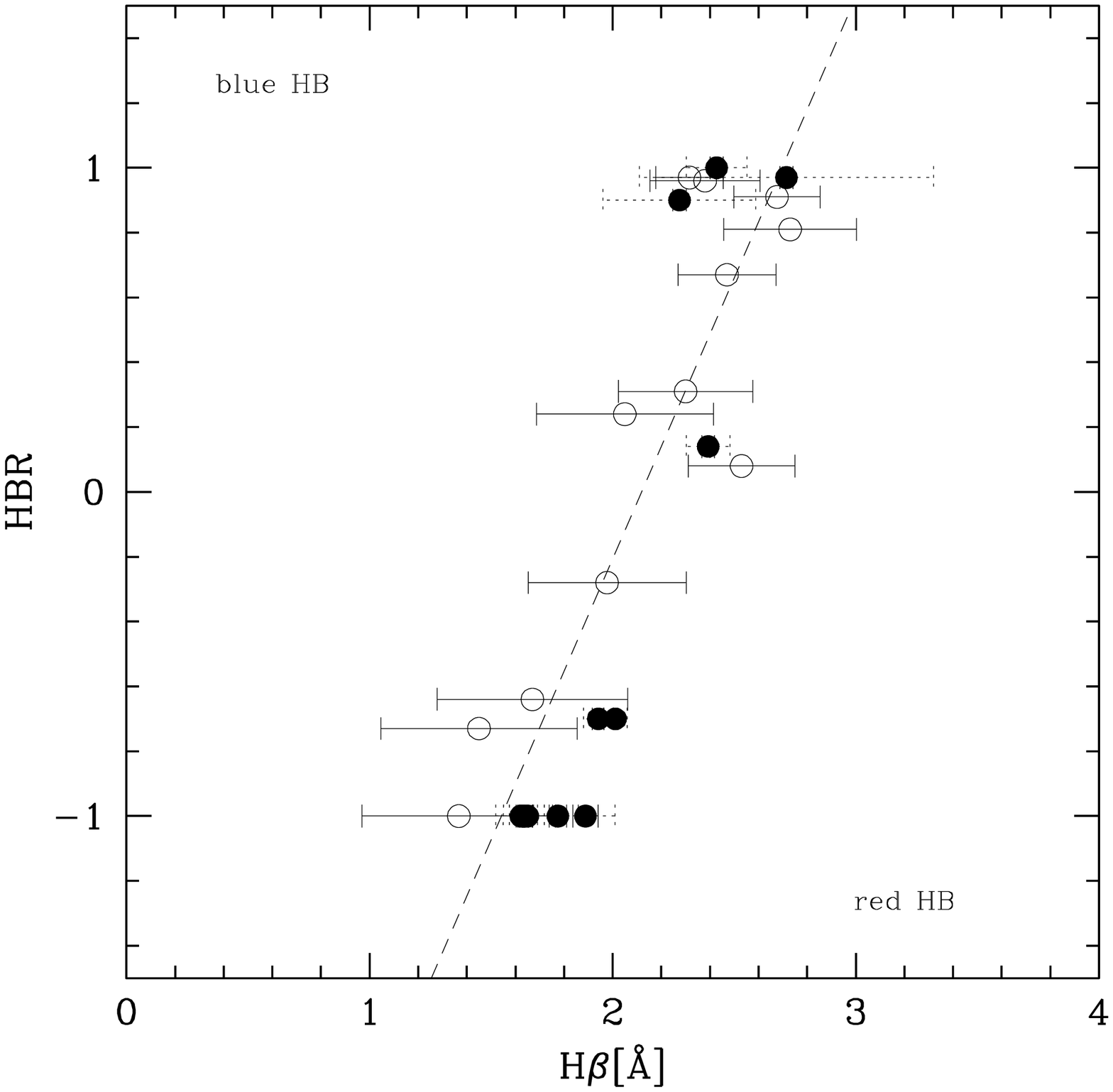}
 \caption{Horizontal branch morphology in terms of the HBR parameter
   as a function of galactocentric radius $R_{\rm GC}$ (upper panel)
   and H$\beta$ (lower panel). Filled and open circles show our
   globular cluster data and the data of \cite{trager98},
   respectively.}
 \label{ps:hb_rgc_hbeta}
\end{figure}

In the following we focus on the correlation of the horizontal branch
morphology on the H$\beta$ index. We use the horizontal branch ratio
HBR from the McMaster catalog (HBR = (B$-$R)/(B+V+R): B and R are the
number of stars bluewards and redwards of the instability strip; V is
the number of variable stars inside the instability strip) to
parameterize the horizontal branch morphology. Figure
\ref{ps:hb_rgc_hbeta} shows that the HBR parameter vs. $R_{\rm GC}$
follows a similar trend as H$\beta$ vs. $R_{\rm GC}$ in Figure
\ref{ps:radplot}. This supports the idea that the change in H$\beta$
(as a function of $R_{\rm GC}$) is mainly driven by the change of the
horizontal branch morphology as one goes to more distant halo globular
clusters with lower metallicities. Indeed, the lower panel in Figure
\ref{ps:hb_rgc_hbeta} shows that HBR is correlated with the H$\beta$
index (Spearman rank coefficient 0.77). The functional form of this
correlation is
\begin{equation}
\label{eqn:hbrhbeta}
{\rm HBR} = (-3.71\pm0.41) + (1.75\pm0.19)\cdot{\rm H}\beta
\end{equation}
with an r.m.s. of 0.39 which is marginally larger than the mean
measurement error (0.36). That is, the scatter found can be fully
explained by observational uncertainties. Note that according to this
relation the H$\beta$ index can vary by $\sim1$ \AA\ when changing the
horizontal branch morphology from an entirely red to an entirely blue
horizontal branch \citep[see also][]{defreitas95}. This behaviour is
also predicted by previous stellar population models
\citep[e.g.][]{lee00, maraston00}.

Figure~\ref{ps:hb_rgc_hbeta} implies that the change of H$\beta$ is
mainly driven by the horizontal branch morphology which itself is
influenced by the mean globular cluster metallicity. However, we know
of globular cluster pairs -- so-called ``second parameter'' pairs --,
such as the metal-poor halo globular clusters NGC~288 and NGC~362
([Fe/H]$\approx-1.2$, \citealt{catelan01}) and the metal-rich bulge
clusters NGC~6388 and NGC~6624 ([Fe/H]$\approx-0.5$, \citealt{rich97,
zoccali00}), with very similar metallicities and different horizontal
branch morphologies. In fact, NGC~6388 (and NGC~6441, another
metal-rich cluster in our sample also featuring a blue horizontal
branch) shows a stronger H$\beta$ index than other sample globular
clusters at similar metallicities (see
Section~\ref{ln:hbeta}). Clearly, metallicity cannot be the only
parameter which governs the horizontal branch morphology. In the
context of the ``second-parameter effect'' other global and non-global
cluster properties \citep{freeman81} impinging on the horizontal
branch morphology have been discussed of which the cluster age and/or
several other structural and dynamical cluster properties are
suspected to be the best candidates \citep[e.g.][]{fusipecci93,
rich97}. Our sample does not contain enough ``second parameter'' pairs
to study the systematic effects these ``second parameters'' might have
on H$\beta$, such as the correlation of the residuals of the
HBR-H$\beta$ relation as a function of globular cluster age or
internal kinematics. A larger data set would help to solve this issue.

\section{Conclusions}
\label{ln:conclusions}
For the first time the complete set of Lick indices have been measured
for a sample of metal-rich globular clusters belonging to the Galactic
bulge. In combination with data for metal-poor globular clusters this
data set has allowed us to establish an empirical calibration of the
Lick indices of old stellar populations from very low metallicities
all the way to near solar metallicity. On the one hand, these
empirical relations can be directly used to get age and chemical
composition information for the stellar populations of unresolved
galaxies. On the other hand, they can be used to submit to most
stringent tests of population synthesis models, an aspect which is the
subject of an accompanying paper \citep{maraston02}.

The comparison of the Lick indices for the Galactic bulge with those
of globular clusters shows that the bulge and the most metal-rich
globular clusters have quite similar stellar populations, with the
slightly deviating values of some of the bulge indices being the
likely result of the metallicity distribution of bulge stars, which
extends down to [Fe/H]$\simeq -1.0$ \citep{mcwilliam94,
zoccali02}. Within the uncertainties, both the metal-rich clusters and
the bulge appear to have also the same index ratios, in particular
those sensitive to [$\alpha$/Fe]. This implies similar enhancements
for individual $\alpha$-elements in clusters as in the field. Existing
spectroscopic determinations of the $\alpha$-element enhancement in
clusters and bulge field stars are still scanty, but extensive
high-resolution spectroscopy at 8--10m class telescopes will soon
provide data for a fully empirical calibration of the Lick indices at
the [$\alpha$/Fe] values of the bulge and bulge globular clusters.

Some other line index ratios, such as CN$/\langle$Fe$\rangle$, show
clear exceptions. In these cases the bulge indices are definitely
below the values for the metal-rich clusters. Several possibilities
have been discussed for the mechanism responsible for the CN index
offset between the bulge and the clusters, the environmental-pollution
being active in clusters (but not in the field) appearing as the most
likely explanation. In this scenario, globular cluster stars would
have experienced accretion of materials lost by cluster AGB stars,
early in the history of the clusters (i.e., when clusters were $\sim
10^8-10^9$ years old).

\section{Summary}
\begin{enumerate}

\item We present for the first time the full set of Lick indices for a
sample of metal-rich globular clusters in the Galactic bulge.

\item The Mg$_2$ and $\langle$Fe$\rangle$ indices of the most
metal-rich globular clusters (NGC~5927, NGC~6528, and NGC~6553) are
similar to those of the bulge, indicating that cluster and bulge
fields have a similar [Fe/H] ratio.

\item The CN index is clearly stronger in metal-rich globular
clusters, compared to the stellar population in the Galactic bulge.

\item All the metallicity-sensitive indices are tightly correlated
with the $\langle$Fe$\rangle$ index (Spearman rank coefficient
$\geq0.87$).

\item We provide empirical calibrations of several indices versus
[Fe/H] in the range $-2.29\leq$[Fe/H]$\leq-0.17$. We find that the
Mg$_2$ and the [MgFe] index are the best metallicity indicators with a
r.m.s. of 0.15 dex.

\item The H$\beta$ index shows a large scatter as a function of
galactocentric distance which can be explained by the changing
horizontal branch morphology.

\item All other indices decrease with galactocentric radius and have
constant low values beyond the radius of $\sim10$ kpc.

\end{enumerate}

\begin{acknowledgements}
  We acknowledge the use of the Lick standard star database which is
  maintained by Guy Worthey. THP gratefully acknowledges the support
  of the German \emph{Deut\-sche For\-schungs\-ge\-mein\-schaft,
  DFG\/} under the project number Be~1091/10--1. We would like to
  thank the anonymous referee for a careful and constructive report.
\end{acknowledgements}



\appendix
\section{Description and Performance Tests of the Index-Measuring Routine}
\label{ln:lickcode}
The Lick/IDS standard system is briefly described in Section
\ref{ln:licktrafo}. The passband definitions of line indices and the
index measuring prescriptions were implemented in a code\footnote{The
code (GONZO) is available upon request from T.H.~Puzia
(puzia@usm.uni-muenchen.de).}, which performs a full statistical
error treatment (see Sect.~\ref{ln:lickcoderr} for details). In
summary, a line index is defined as the missing/additional flux
between the spectrum and a pseudo-continuum which is defined by two
continuum passbands on either side of the feature passband.
\cite{trager98} defines a line index as
\begin{equation}
\label{eqn:oidxdef}
{\rm EW}_o=\int_{\lambda_{min}}^{\lambda_{max}}
\left(1-\frac{F_l(\lambda)}{F_c(\lambda)}\right)\,d\lambda,
\end{equation}
where $F_l(\lambda)$ and $F_c(\lambda)$ is the flux of the feature
passband and the pseudo-continuum, respectively. However,
\cite{gonzalez93} gives another definition of a line index
\begin{equation}
\label{eqn:tidxdef}
{\rm
EW}_t=\left(1-\frac{\int_{\lambda_{min}}^{\lambda_{max}} 
F_l(\lambda)\,d\lambda}{\int_{\lambda_{min}}^{\lambda_{max}}
  F_c(\lambda)\,d\lambda}\right)\,\cdot\Delta\lambda. 
\end{equation}
While the former is an integral of the flux ratio the latter is the
ratio of the flux integrals. We refer to the former as the observer's
definition (${\rm EW}_o$) and to the latter as the theorist's
definition (${\rm EW}_t$). For high-S/N spectra the difference between
the two definitions is negligible. EW$_t$ is a more global definition
and is, therefore, more robust for low-S/N spectra. However, since
most literature uses the observer's definition, all the measurements
which are given in this paper are EW$_o$ (Eq.~\ref{eqn:oidxdef}). To
check for systematic offsets and/or different error patterns, our code
performes index measurements with both definitions. Without exception,
we find no systematic offset between the measurements and a
value-to-value scatter of less than 0.1\%. In particular, this causes
no problems between theoretical predictions from SSP models, which
synthesise line indices using the EW$_t$ definition, and measurements.

\subsection{Accuracy Tests}
\begin{figure}[!ht]
  \centering \includegraphics[width=8.8cm]{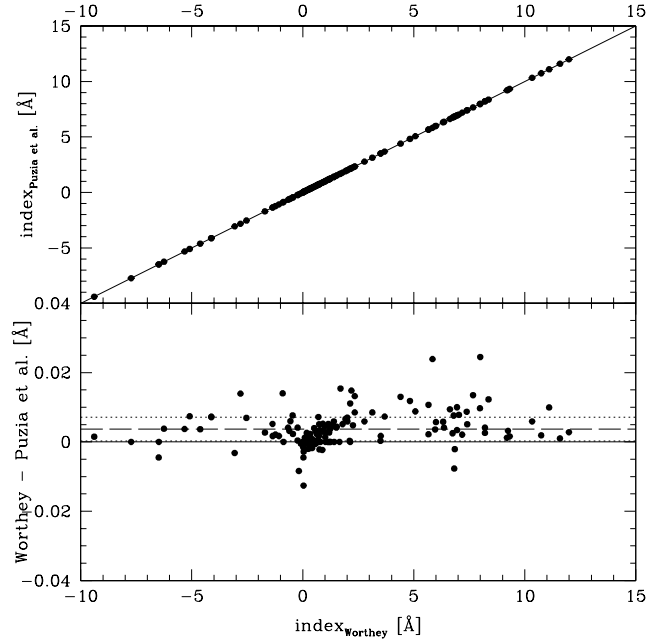}
   \caption{Comparison of index measurements of Worthey and from
     our code. The upper panel shows the direct comparison between our
     and the Worthey data. Each single point gives the measurement for
     one star in one single passband. Residuals are plotted in the
     lower panel. The measurements of passbands which are defined in
     magnitudes have been previously transformed to the \AA-scale.
     From the lower panel we determine the average statistical scatter
     with 0.0034~\AA\ and a zero offset $0.0037\pm 2\cdot10^{-5}$~\AA.}
 \label{ps:compareidx}
\end{figure}

The whole measurement procedure was tested on original Lick spectra
available from the database of Guy Worthey\footnote{The latest
passband definitions and FITS files of original Lick calibration
spectra have been obtained from
http://astro.wsu.edu/worthey/html/index.table.html}. Applying our code
to 6 original Lick spectra (HYA VB~112/sdr310007, HYA
VB~111/sdr310010, HYA VB~103 (R)/sdr310017, HYA VB~103
(R+L)/sdr310019, HYA VB~95/sdr320230, and HR~7429/sdr370421) and
comparing the results of 150 line index measurements (using the EW$_o$
definition, see above) with the original data provided by Guy Worthey,
we find excellent agreement between the Lick datasets and ours. After
transformation of molecular-band indices, which are usually given in
magnitudes, to a uniform \AA-scale, we determine an average scatter of
0.0034~\AA, which is most likely due to the different treatment of
sub-pixels at the edges of passbands. The average systematic zero
offset between the two datasets was found $0.0037\pm
2\cdot10^{-5}$~\AA. Figure \ref{ps:compareidx} shows a comparison
between measurements performed by Worthey and us based on the same
data. Since the offset between the two datasets is of the order of the
scatter of all index values, we do not consider any correction for all
measurements performed in this work. Moreover, the offset is
significantly smaller than the errors which result from Poisson noise
of the spectra themselves.

\subsection{Error Estimation}
\label{ln:lickcoderr}
The main purpose of implementing a new code for index measurements,
was the need for a robust error estimate of the indices. Since we deal
with CCD data, as opposed to the Lick spectra, which are digitized
images of a phosphor screen, we can determine the contribution of the
flux Poisson noise of each spectrum to the total error budget. Note
that due to the scanning of a spectrum off a phosphor screen the
photo-tube suffers from the correlation of photon-noise errors over a
wide wavelength range. Moreover, the photo-tube increases the noise
due to statistical fluctuations in the amplification process
\citep{robinson72}. The data used in this work is entirely free from
these effects.

Since we need to subtract background spectra from our science spectra
it is necessary to estimate the contribution of errors introduced by
the subtraction and any radial velocity uncertainties to the total
error budget. Both background and science spectra are included in the
estimation of the total index uncertainty. Radial velocity errors are
considered as systematic errors and are not included in the
statistical error budget, but listed in the paper.

The code determines the total line-index uncertainty in 100 Monte
Carlo simulations. Each simulation creates a new object and background
spectrum by adding noise according to the Poisson statistics taking
into account the detector noise. Line indices are measured on each
noise-added spectrum. Since a Monte Carlo test naturally takes into
account all possible error correlations in the line-index measurement
process (such as the correlation of errors in the background passbands
with the errors in the feature passband), the scatter in all simulated
line indices is the best estimate for their total uncertainty. We
therefore use the 1--$\sigma$ standard deviation of all Monte-Carlo
line-index measurements as the best guess for the final index
uncertainty.

The variations due to uncertain radial velocities are given
separately. They are calculated as the deviation of the initial line
index by changing the radial velocity within its error limits.

\section{Comparison with the Literature}

\begin{table*}
  \caption{Comparison of Lick indices CN1 to Fe5015 for our sample
      globular clusters with data taken from literature. The according
      errors are given in Table \ref{tab:gcindices1}. Our index
      measurements and the indices of \citeauthor{trager98} use the
      {\it new} passband definitions of
      \cite{trager98}. \citeauthor{cohen98} and \citeauthor{covino95}
      use {\it old} passband definitions of \cite{burstein84}.
}
{\tiny
         \label{tab:gcrop}
            \begin{tabular}{ll|cccccccccc}  
            \hline
            \noalign{\smallskip}
  GC  & bkg mode$^{\mathrm{a}}$ & CN1&   CN2&   Ca4227& G4300&  Fe4383& Ca4455& Fe4531& Fe4668&$H\beta$& Fe5015\\
      &              & mag&   mag&     A   &   A  &    A   &    A   &   A   &   A   &   A    &  A   \\
            \noalign{\smallskip}
            \hline
            \noalign{\smallskip}
NGC~6624 &  BE  &  0.0497& 0.0739& 0.4889& 4.8910& 2.4023& 0.5065& 2.2739& 1.4643& 1.6502& 4.1850\\
         &  w/o &  0.0635& 0.0856& 0.6070& 5.0023& 2.8872& 0.5122& 2.9936& 1.6337& 1.5347& 4.1861\\
         &  BM  &  0.0692& 0.0907& 0.6064& 5.0702& 2.7295& 0.5335& 3.1087& 1.0489& 1.4807& 4.1552\\
&\cite{trager98}& \dots  &\dots  &\dots  &\dots  &\dots  &\dots  &\dots  &\dots  &\dots  &\dots  \\
&\cite{covino95}&  0.02  & 0.02  &\dots  & 4.758 &\dots  &\dots  &\dots  &\dots  & 2.535 &\dots  \\
&\cite{cohen98} & \dots  &\dots  &\dots  &\dots  &\dots  &\dots  &\dots  &\dots  & 1.69  &\dots  \\
\noalign{\smallskip}
\hline
\noalign{\smallskip}
NGC~6218 &  BE  & -0.0763&-0.0596& 0.0586& 2.7004&-0.1175&-0.0055& 0.9504&-0.6060& 2.7147& 2.7170\\
         &  w/o & -0.0711&-0.0550& 0.0688& 2.6871&-0.1207& 0.0536& 0.8507&-0.7686& 2.5817& 2.6896\\
         &  BM  & -0.0820&-0.0661&-0.0429& 2.4039&-0.8066& 0.0178& 0.5669&-1.7789& 2.8366& 2.2599\\
&\cite{trager98}& -0.0910&-0.0490& 0.6300& 1.8700& 0.0000& 0.1000& 1.6900&-1.3200& 2.2800& 2.1100\\
&\cite{covino95}& \dots  &\dots  &\dots  &\dots  &\dots  &\dots  &\dots  &\dots  & 1.214 &\dots  \\
&\cite{cohen98} & \dots  &\dots  &\dots  &\dots  &\dots  &\dots  &\dots  &\dots  &\dots  &\dots  \\
\noalign{\smallskip}
\hline
\noalign{\smallskip}
NGC~6626 &  BE  & -0.0455&-0.0259& 0.1473& 3.1433& 0.5716& 0.0926& 1.2946&-0.0206& 2.2747& 3.1826\\
         &  w/o & -0.0425&-0.0245& 0.1859& 3.1207& 0.7364& 0.1256& 1.3939& 0.1220& 2.1582& 3.1877\\
         &  BM  & -0.0459&-0.0277& 0.1428& 3.0243& 0.4942& 0.1222& 1.3238&-0.2020& 2.2132& 3.0836\\
&\cite{trager98}& \dots  &\dots  &\dots  &\dots  &\dots  &\dots  &\dots  &\dots  &\dots  &\dots  \\
&\cite{covino95}& -0.052 &-0.052 &\dots  & 2.713 &\dots  &\dots  &\dots  &\dots  & 2.443 &\dots  \\
&\cite{cohen98} & \dots  &\dots  &\dots  &\dots  &\dots  &\dots  &\dots  &\dots  &\dots  &\dots  \\
\noalign{\smallskip}
\hline
\noalign{\smallskip}
NGC~6284 &  BE  & -0.0417&-0.0227& 0.1551& 3.1957& 0.6659& 0.1992& 1.4645&-0.1003& 2.4274& 3.1553\\
         &  w/o & -0.0347&-0.0155& 0.2107& 3.3368& 0.7969& 0.2928& 1.4811& 0.1469& 2.2370& 3.2056\\
         &  BM  & -0.0507&-0.0277& 0.0278& 2.9414&-0.5049& 0.0494& 1.0449&-2.3533& 2.8913& 2.2407\\
&\cite{trager98}& \dots  &\dots  &\dots  &\dots  &\dots  &\dots  &\dots  &\dots  &\dots  &\dots  \\
&\cite{covino95}& -0.082 &-0.082 &\dots  & 1.785 &\dots  &\dots  &\dots  &\dots  & 2.764 &\dots  \\
&\cite{cohen98} & \dots  &\dots  &\dots  &\dots  &\dots  &\dots  &\dots  &\dots  &\dots  &\dots  \\
\noalign{\smallskip}
\hline
\noalign{\smallskip}
NGC~6356 &  BE  &  0.0450& 0.0648& 0.5079& 5.0611& 2.3472& 0.5334& 2.2955& 1.3231& 1.6341& 4.0541\\
         &  w/o &  0.0432& 0.0626& 0.4911& 4.8895& 2.3269& 0.5481& 2.1968& 1.4871& 1.6190& 4.1296\\
         &  BM  &  0.0561& 0.0788& 0.4667& 5.1272& 2.0202& 0.5053& 2.1965& 0.5297& 1.7782& 3.9587\\
&\cite{trager98}&  0.0237& 0.0726& 1.3270& 4.8180& 3.9220& 1.6030& 2.6900& 2.9720& 1.4680& 4.2980\\
&\cite{covino95}& \dots  &\dots  &\dots  &\dots  &\dots  &\dots  &\dots  &\dots  & 1.646 &\dots  \\
&\cite{cohen98} & \dots  &\dots  &\dots  &\dots  &\dots  &\dots  &\dots  &\dots  & 1.62  &\dots  \\
\noalign{\smallskip}
\hline
\noalign{\smallskip}
NGC~6637 &  BE  &  0.0248& 0.0438& 0.4009& 5.1912& 2.0615& 0.4497& 2.1725& 1.3150& 1.6224& 3.9535\\
         &  w/o &  0.0223& 0.0412& 0.3859& 5.1082& 2.1333& 0.3674& 2.0921& 1.2860& 1.5773& 3.8951\\
         &  BM  &  0.0258& 0.0459& 0.3493& 5.2617& 1.9243& 0.2976& 2.0623& 0.7096& 1.6590& 3.7375\\
&\cite{trager98}& -0.0125& 0.0048& 1.0560& 5.0490& 0.2010& 1.1300& 3.4870& 1.5220& 0.8980& 4.6420\\
&\cite{covino95}& \dots  &\dots  &\dots  &\dots  &\dots  &\dots  &\dots  &\dots  & 1.15  &\dots  \\
&\cite{cohen98} & \dots  &\dots  &\dots  &\dots  &\dots  &\dots  &\dots  &\dots  &\dots  &\dots  \\
\noalign{\smallskip}
\hline
\noalign{\smallskip}
NGC~6553 &  BE  &  0.1378& 0.1619& 1.0915& 5.4464& 4.0079& 0.8316& 3.0767& 3.4849& 1.8881& 5.7254\\
         &  w/o &  0.0699& 0.0842& 0.7192& 5.0248& 4.2101& 0.9686& 2.5541& 3.5774& 1.1596& 4.9798\\
         &  BM  &  0.1107& 0.1243& 0.7044& 5.4233& 4.5439& 1.3630& 2.7317& 3.5390& 1.0138& 5.2232\\
&\cite{trager98}& \dots  &\dots  &\dots  &\dots  &\dots  &\dots  &\dots  &\dots  &\dots  &\dots  \\
&\cite{covino95}& \dots  &\dots  &\dots  &\dots  &\dots  &\dots  &\dots  &\dots  &\dots  &\dots  \\
&\cite{cohen98} & \dots  &\dots  &\dots  &\dots  &\dots  &\dots  &\dots  &\dots  & 1.63  &\dots  \\
\noalign{\smallskip}
\hline
\noalign{\smallskip}
NGC~6528 &  BE  &  0.0959& 0.1174& 0.9089& 5.2218& 4.7754& 0.8794& 2.7074& 4.2181& 1.7745& 5.1531\\
         &  w/o &  0.0696& 0.0877& 0.6629& 5.1257& 4.4139& 0.5266& 2.7904& 4.0300& 1.5097& 5.0032\\
         &  BM  &  0.1229& 0.1569& 1.3493& 6.2741& 6.1253& 1.4228& 3.7042& 4.5434& 1.0913& 6.1269\\
&\cite{trager98}& \dots  &\dots  &\dots  &\dots  &\dots  &\dots  &\dots  &\dots  &\dots  &\dots  \\
&\cite{covino95}& \dots  &\dots  &\dots  &\dots  &\dots  &\dots  &\dots  &\dots  &\dots  &\dots  \\
&\cite{cohen98} & \dots  &\dots  &\dots  &\dots  &\dots  &\dots  &\dots  &\dots  & 1.80  &\dots  \\
             \noalign{\smallskip}
            \hline
            \noalign{\smallskip}
         \end{tabular}
}
\begin{list}{}{}
\item[$^{\mathrm{a}}$] BE: background extraction; w/o: without
  background subtraction; BM: background modeling. See
  Section~\ref{ln:bkgestim} for details.
\end{list}
\end{table*}
\addtocounter{table}{-1}
\begin{table*}
  \caption{-- continued. Comparison of Lick indices Mg$_1$ --
      TiO$_2$.
}
{\tiny
 \begin{tabular}{ll|ccccccccccc}  
  \hline
  \noalign{\smallskip} 
    GC  & bkg mode$^{\mathrm{a}}$ &Mg$_1$&Mg$_2$&    Mgb&   Fe5270& Fe5335& Fe5406& Fe5709& Fe5782& Na5895&  TiO1&   TiO2\\
        &          &  mag&    mag&     A &     A   &   A   &   A   &   A   &   A   &   A   &  mag &   mag \\
  \noalign{\smallskip}
  \hline
  \noalign{\smallskip}
NGC~6624 &  BE  & 0.0707& 0.1721& 2.7280& 1.8158& 1.6403& 0.9789& 0.5009& 0.6411& 2.7063& 0.0470& 0.0628 \\
         &  w/o & 0.0696& 0.1758& 2.6730& 1.7832& 1.7237& 0.9738& 0.5322& 0.6349& 3.8927& 0.0542& 0.0658 \\
         &  BM  & 0.0642& 0.1669& 2.6254& 1.6564& 1.6780& 0.9527& 0.4837& 0.5945& 3.4454& 0.0572& 0.0719 \\
&\cite{trager98}&\dots  &\dots  &\dots  &\dots  &\dots  &\dots  &\dots  &\dots  &\dots  &\dots  &\dots   \\
&\cite{covino95}& 0.05  & 0.15  & 2.486 & 2.117 & 1.812 &\dots  &\dots  &\dots  & 2.881 &\dots  &\dots   \\
&\cite{cohen98} & 0.048 & 0.163 & 2.94  & 2.09  & 1.78  &\dots  &\dots  &\dots  & 2.20  & 0.035 &\dots   \\
\noalign{\smallskip}
\hline
\noalign{\smallskip}
NGC~6218 &  BE  & 0.0268& 0.0672& 1.0628& 0.7687& 0.8935& 0.2246&-0.1909& 0.2025& 1.2915& 0.0182& 0.0040 \\
         &  w/o & 0.0293& 0.0675& 1.4179& 0.7036& 0.7692& 0.2148&-0.1939& 0.2421& 1.2410& 0.0059&-0.0067 \\
         &  BM  & 0.0132& 0.0256& 1.0507& 0.2819& 0.4167&-0.1026&-0.5354& 0.0523&-0.2041&-0.0222&-0.0679 \\
&\cite{trager98}&-0.0060& 0.0690& 1.2800& 1.3400& 0.5000& 0.0600&-0.0600& 0.0000& 1.6300& 0.0020& 0.0000 \\
&\cite{covino95}& 0.02  & 0.07  & 1.68  & 1.125 & 0.8472&\dots  &\dots  &\dots  &\dots  &\dots  &\dots   \\
&\cite{cohen98} &\dots  &\dots  &\dots  &\dots  &\dots  &\dots  &\dots  &\dots  &\dots  &\dots  &\dots   \\
\noalign{\smallskip}
\hline
\noalign{\smallskip}
NGC~6626 &  BE  & 0.0415& 0.0919& 1.3679& 1.0900& 0.9747& 0.5413& 0.1846& 0.4735& 2.1005& 0.0288& 0.0382 \\
         &  w/o & 0.0424& 0.0956& 1.3511& 1.1049& 1.0347& 0.5392& 0.1824& 0.4674& 2.2341& 0.0299& 0.0385 \\
         &  BM  & 0.0382& 0.0841& 1.2213& 1.0144& 0.9603& 0.4811& 0.1324& 0.4479& 1.9655& 0.0245& 0.0274 \\
&\cite{trager98}&\dots  &\dots  &\dots  &\dots  &\dots  &\dots  &\dots  &\dots  &\dots  &\dots  &\dots   \\
&\cite{covino95}&-0.002 & 0.063 & 1.103 & 1.539 & 1.229 &\dots  &\dots  &\dots  & 2.565 &\dots  &\dots   \\
&\cite{cohen98} &\dots  &\dots  &\dots  &\dots  &\dots  &\dots  &\dots  &\dots  &\dots  &\dots  &\dots   \\
\noalign{\smallskip}
\hline
\noalign{\smallskip}
NGC~6284 &  BE  & 0.0427& 0.0966& 1.4403& 0.8563& 1.0216& 0.5178& 0.1110& 0.3141& 2.3978& 0.0159& 0.0039 \\
         &  w/o & 0.0463& 0.1064& 1.3460& 1.0142& 1.0683& 0.5497& 0.1973& 0.3497& 2.1540& 0.0137& 0.0023 \\
         &  BM  & 0.0063& 0.0180& 0.0513&-0.3341& 0.2294&-0.1227&-0.0984&-0.3166&-2.1392&-0.0380&-0.1080 \\
&\cite{trager98}&\dots  &\dots  &\dots  &\dots  &\dots  &\dots  &\dots  &\dots  &\dots  &\dots  &\dots   \\
&\cite{covino95}& 0.027 & 0.077 & 1.075 & 1.505 & 1.091 &\dots  &\dots  &\dots  & 2.433 &\dots  &\dots   \\
&\cite{cohen98} &\dots  &\dots  &\dots  &\dots  &\dots  &\dots  &\dots  &\dots  &\dots  &\dots  &\dots   \\
\noalign{\smallskip}
\hline
\noalign{\smallskip}
NGC~6356 &  BE  & 0.0728& 0.1773& 2.7863& 1.7187& 1.6597& 0.9557& 0.4067& 0.5493& 3.2660& 0.0333& 0.0531 \\
         &  w/o & 0.0773& 0.1851& 2.5857& 1.9996& 1.4993& 0.8985& 0.3764& 0.5451& 3.6050& 0.0390& 0.0579 \\
         &  BM  & 0.0666& 0.1656& 2.3420& 1.7049& 1.2805& 0.7273& 0.3410& 0.3598& 2.2064& 0.0327& 0.0408 \\
&\cite{trager98}& 0.0404& 0.1700& 2.9800& 1.9940& 1.4010& 1.3970& 0.6640& 0.6200& 3.3290& 0.0369& 0.0460 \\
&\cite{covino95}& 0.062 & 0.179 & 2.776 & 2.352 & 1.125 &\dots  &\dots  &\dots  &\dots  &\dots  &\dots   \\
&\cite{cohen98} & 0.070 & 0.169 & 3.09  & 2.00  & 1.69  &\dots  &\dots  &\dots  & 3.00  & 0.029 &\dots   \\
\noalign{\smallskip}
\hline
\noalign{\smallskip}
NGC~6637 &  BE  & 0.0567& 0.1542& 2.5420& 1.6335& 1.3969& 0.8222& 0.3565& 0.4906& 2.6053& 0.0381& 0.0441 \\
         &  w/o & 0.0562& 0.1541& 2.4696& 1.5448& 1.4297& 0.8195& 0.3538& 0.4789& 2.9313& 0.0418& 0.0464 \\
         &  BM  & 0.0461& 0.1369& 2.3096& 1.2681& 1.2806& 0.6971& 0.3144& 0.3349& 1.9320& 0.0359& 0.0310 \\
&\cite{trager98}& 0.0384& 0.1433& 2.3720& 1.9470& 0.9590& 0.8580& 0.3850& 0.1100& 3.2550& 0.0498& 0.0000 \\
&\cite{covino95}& 0.05  &\dots  & 2.671 & 1.642 & 1.539 &\dots  &\dots  &\dots  &\dots  &\dots  &\dots   \\
&\cite{cohen98} &\dots  &\dots  &\dots  &\dots  &\dots  &\dots  &\dots  &\dots  &\dots  &\dots  &\dots   \\
\noalign{\smallskip}
\hline
\noalign{\smallskip}
NGC~6553 &  BE  & 0.1002& 0.2552& 3.8961& 2.6091& 2.2654& 1.2371& 0.7744& 1.0970& 3.8792& 0.0689& 0.1420 \\
         &  w/o & 0.0949& 0.2513& 3.6472& 2.4476& 2.1073& 1.3023& 0.7823& 1.0423& 4.1967& 0.0640& 0.1245 \\
         &  BM  & 0.0972& 0.2606& 3.9386& 2.5740& 2.2081& 1.3565& 0.8161& 1.1643& 3.9533& 0.0640& 0.1262 \\
&\cite{trager98}&\dots  &\dots  &\dots  &\dots  &\dots  &\dots  &\dots  &\dots  &\dots  &\dots  &\dots   \\
&\cite{covino95}&\dots  &\dots  &\dots  &\dots  &\dots  &\dots  &\dots  &\dots  &\dots  &\dots  &\dots   \\
&\cite{cohen98} & 0.110 & 0.249 & 3.88  & 3.11  & 2.51  &\dots  &\dots  &\dots  & 3.40  & 0.044 &\dots   \\
\noalign{\smallskip}
\hline
\noalign{\smallskip}
NGC~6528 &  BE  & 0.1149& 0.2615& 3.7413& 2.3673& 2.2777& 1.5499& 0.8223& 0.7987& 5.1366& 0.0750& 0.1268 \\
         &  w/o & 0.1109& 0.2573& 3.4276& 2.3885& 2.0133& 1.3893& 0.7406& 0.6816& 5.5471& 0.0714& 0.1183 \\
         &  BM  & 0.1248& 0.2928& 4.1444& 2.6491& 2.4803& 1.7290& 0.9335& 0.8722& 5.4463& 0.0939& 0.1696 \\
&\cite{trager98}&\dots  &\dots  &\dots  &\dots  &\dots  &\dots  &\dots  &\dots  &\dots  &\dots  &\dots   \\
&\cite{covino95}&\dots  &\dots  &\dots  &\dots  &\dots  &\dots  &\dots  &\dots  &\dots  &\dots  &\dots   \\
&\cite{cohen98} & 0.097 & 0.247 & 3.89  & 2.96  & 2.45  &\dots  &\dots  &\dots  & 4.93  & 0.046 &\dots   \\
            \noalign{\smallskip}
            \hline
            \noalign{\smallskip}
         \end{tabular}
}
\begin{list}{}{}
\item[$^{\mathrm{a}}$] BE: background extraction; w/o: without
  background subtraction; BM: background modeling. See
  Section~\ref{ln:bkgestim} for details.
\end{list}
\end{table*}

\section{Line-Index Measurements using New Passband Definitions}
\label{ln:indexmeasurements}
Table~\ref{tab:gcindices1} summarises all our Lick index measurements
which were computed with the {\it new} set of index passband
definitions from \cite{trager98} and \cite{worthey97}.

\begin{table*}[h!]
  \caption{Lick indices CN1 -- Mg$_1$ for all sample globular clusters
    including statistical and systematic errors. Line one gives the
    index value. Line two and three document the Poisson error and the
    statistical slit-to-slit scatter. The systematic deviation of each
    index due to radial velocity uncertainties is given in line
    four. This set of indices uses the {\it new} passband definitions
    of \cite{trager98} and \cite{worthey97}.}
 \label{tab:gcindices1}
        {\tiny
        \begin{center}
        \begin{tabular}{lccccccccccccccccccccc}  
         \hline
         \noalign{\smallskip}
cluster$^{\mathrm{a}}$&  CN1  &  CN2  & Ca4227& G4300 & Fe4383& Ca4455& Fe4531& Fe4668&H$\beta$&Fe5015&  Mg$_1$   \\
                      &  mag  &  mag  &  \AA  &  \AA  &  \AA  &  \AA  &  \AA  &  \AA  &  \AA  &  \AA  &  mag   \\
\noalign{\smallskip}
\hline
\noalign{\smallskip}
 NGC~5927        & 0.0848& 0.1146& 0.7194& 4.4637& 3.0248& 0.8141& 2.3786& 2.7464& 1.6359& 4.8052& 0.0848 \\
 $\Delta\cal{B}$ & 0.0010& 0.0013& 0.0183& 0.0329& 0.0531& 0.0261& 0.0463& 0.0688& 0.0335& 0.0744& 0.0008 \\
 $\Delta\cal{S}$ & 0.0116& 0.0187& 0.1303& 1.0389& 0.7073& 0.1423& 0.3439& 0.3326& 0.1174& 0.2154& 0.0183 \\
 $\Delta v_r$    & 0.0006& 0.0033& 0.0417& 0.0943& 0.0529& 0.1206& 0.0569& 0.0367& 0.0319& 0.1213& 0.0000 \\
\noalign{\smallskip}
 NGC~6388        & 0.0446& 0.0676& 0.3498& 4.3312& 2.4211& 0.4663& 2.2645& 1.3306& 2.0111& 4.0880& 0.0568 \\
 $\Delta\cal{B}$ & 0.0003& 0.0004& 0.0057& 0.0106& 0.0171& 0.0081& 0.0130& 0.0224& 0.0098& 0.0247& 0.0002 \\
 $\Delta\cal{S}$ & 0.0032& 0.0024& 0.0496& 0.0251& 0.0930& 0.0595& 0.0463& 0.0554& 0.0489& 0.0629& 0.0026 \\
 $\Delta v_r$    & 0.0002& 0.0025& 0.0270& 0.0761& 0.0844& 0.0844& 0.0482& 0.0611& 0.0276& 0.0853& 0.0001 \\
\noalign{\smallskip}
 NGC~6528        & 0.0959& 0.1174& 0.9089& 5.2218& 4.7754& 0.8794& 2.7074& 4.2181& 1.7745& 5.1531& 0.1149 \\
 $\Delta\cal{B}$ & 0.0012& 0.0014& 0.0191& 0.0393& 0.0525& 0.0285& 0.0498& 0.0839& 0.0351& 0.0850& 0.0010 \\
 $\Delta\cal{S}$ & 0.0031& 0.0028& 0.0423& 0.1079& 0.1826& 0.0564& 0.1121& 0.7903& 0.0840& 0.0749& 0.0147 \\
 $\Delta v_r$    & 0.0004& 0.0032& 0.0402& 0.0892& 0.1177& 0.1250& 0.0868& 0.1043& 0.0314& 0.1191& 0.0002 \\
\noalign{\smallskip}
 NGC~6624        & 0.0497& 0.0739& 0.4889& 4.8910& 2.4023& 0.5065& 2.2739& 1.4643& 1.6502& 4.1850& 0.0707 \\
 $\Delta\cal{B}$ & 0.0005& 0.0006& 0.0102& 0.0178& 0.0273& 0.0154& 0.0236& 0.0417& 0.0199& 0.0429& 0.0004 \\
 $\Delta\cal{S}$ & 0.0064& 0.0062& 0.1221& 0.0299& 0.3929& 0.1254& 0.1314& 0.6138& 0.0244& 0.2230& 0.0143 \\
 $\Delta v_r$    & 0.0001& 0.0030& 0.0413& 0.0922& 0.0981& 0.1094& 0.0450& 0.0735& 0.0265& 0.1137& 0.0003 \\
\noalign{\smallskip}
 NGC~6218        &-0.0763&-0.0596& 0.0586& 2.7004&-0.1175&-0.0055& 0.9504&-0.6060& 2.7147& 2.7170& 0.0268 \\
 $\Delta\cal{B}$ & 0.0007& 0.0007& 0.0129& 0.0258& 0.0387& 0.0212& 0.0340& 0.0661& 0.0275& 0.0665& 0.0007 \\
 $\Delta\cal{S}$ & 0.0263& 0.0200& 0.1070& 0.7427& 0.5029& 0.0786& 0.3772& 0.2757& 0.6046& 0.2691& 0.0019 \\
 $\Delta v_r$    & 0.0002& 0.0094& 0.0527& 0.1904& 0.0765& 0.0615& 0.0503& 0.0537& 0.0049& 0.0455& 0.0001 \\
\noalign{\smallskip}
 NGC~6441        & 0.0532& 0.0760& 0.4629& 4.3106& 2.7542& 0.4953& 2.3203& 1.3379& 1.9406& 4.2040& 0.0721 \\
 $\Delta\cal{B}$ & 0.0008& 0.0009& 0.0169& 0.0272& 0.0360& 0.0190& 0.0348& 0.0573& 0.0240& 0.0602& 0.0007 \\
 $\Delta\cal{S}$ & 0.0025& 0.0030& 0.0111& 0.0749& 0.1047& 0.0540& 0.0446& 0.0733& 0.0612& 0.0430& 0.0011 \\
 $\Delta v_r$    & 0.0001& 0.0027& 0.0267& 0.0814& 0.0717& 0.0873& 0.0586& 0.0661& 0.0239& 0.0860& 0.0001 \\
\noalign{\smallskip}
 NGC~6553        & 0.1378& 0.1619& 1.0915& 5.4464& 4.0079& 0.8316& 3.0767& 3.4849& 1.8881& 5.7254& 0.1002 \\
 $\Delta\cal{B}$ & 0.0018& 0.0023& 0.0310& 0.0640& 0.0820& 0.0483& 0.0764& 0.1223& 0.0520& 0.1043& 0.0012 \\
 $\Delta\cal{S}$ & 0.0133& 0.0025& 0.0622& 0.0136& 1.1482& 0.1425& 0.0415& 0.7890& 0.1208& 0.2928& 0.0037 \\
 $\Delta v_r$    & 0.0007& 0.0067& 0.1065& 0.1527& 0.3085& 0.1909& 0.0892& 0.2278& 0.0633& 0.1827& 0.0007 \\
\noalign{\smallskip}
 NGC~6626        &-0.0455&-0.0259& 0.1473& 3.1433& 0.5716& 0.0926& 1.2946&-0.0206& 2.2747& 3.1826& 0.0415 \\
 $\Delta\cal{B}$ & 0.0006& 0.0009& 0.0143& 0.0272& 0.0415& 0.0192& 0.0356& 0.0688& 0.0277& 0.0656& 0.0006 \\
 $\Delta\cal{S}$ & 0.0079& 0.0067& 0.0567& 0.3647& 0.2003& 0.0768& 0.2122& 0.3109& 0.3143& 0.3829& 0.0118 \\
 $\Delta v_r$    & 0.0001& 0.0057& 0.0386& 0.1173& 0.0714& 0.0639& 0.0378& 0.0617& 0.0204& 0.0712& 0.0000 \\
\noalign{\smallskip}
 NGC~6284        &-0.0417&-0.0227& 0.1551& 3.1957& 0.6659& 0.1992& 1.4645&-0.1003& 2.4274& 3.1553& 0.0427 \\
 $\Delta\cal{B}$ & 0.0007& 0.0008& 0.0162& 0.0261& 0.0482& 0.0231& 0.0373& 0.0695& 0.0270& 0.0736& 0.0008 \\
 $\Delta\cal{S}$ & 0.0130& 0.0098& 0.0520& 0.2768& 0.3019& 0.1834& 0.0797& 0.2457& 0.1252& 0.0646& 0.0078 \\
 $\Delta v_r$    & 0.0003& 0.0053& 0.0281& 0.1449& 0.0564& 0.0794& 0.0611& 0.0369& 0.0080& 0.0952& 0.0002 \\
\noalign{\smallskip}
 NGC~6356        & 0.0450& 0.0648& 0.5079& 5.0611& 2.3472& 0.5334& 2.2955& 1.3231& 1.6341& 4.0541& 0.0728 \\
 $\Delta\cal{B}$ & 0.0006& 0.0007& 0.0111& 0.0179& 0.0295& 0.0145& 0.0275& 0.0503& 0.0189& 0.0452& 0.0005 \\
 $\Delta\cal{S}$ & 0.0040& 0.0084& 0.0586& 0.0729& 0.1553& 0.1408& 0.1940& 0.3342& 0.0839& 0.2024& 0.0118 \\
 $\Delta v_r$    & 0.0002& 0.0033& 0.0396& 0.0897& 0.1379& 0.1095& 0.0544& 0.0814& 0.0297& 0.1106& 0.0002 \\
\noalign{\smallskip}
 NGC~6637        & 0.0248& 0.0438& 0.4009& 5.1912& 2.0615& 0.4497& 2.1725& 1.3150& 1.6224& 3.9535& 0.0567 \\
 $\Delta\cal{B}$ & 0.0004& 0.0005& 0.0082& 0.0135& 0.0233& 0.0119& 0.0216& 0.0347& 0.0141& 0.0368& 0.0004 \\
 $\Delta\cal{S}$ & 0.0019& 0.0073& 0.0692& 0.0486& 0.0989& 0.0944& 0.0295& 0.1497& 0.0473& 0.0498& 0.0021 \\
 $\Delta v_r$    & 0.0002& 0.0022& 0.0643& 0.0712& 0.1074& 0.0958& 0.0622& 0.0566& 0.0267& 0.1000& 0.0002 \\
\noalign{\smallskip}
 NGC~6981        &-0.0471&-0.0381& 0.1733& 2.9404&-0.0202& 0.0870& 1.0445&-0.3391& 2.3928& 2.6586& 0.0305 \\
 $\Delta\cal{B}$ & 0.0006& 0.0007& 0.0114& 0.0216& 0.0390& 0.0195& 0.0331& 0.0580& 0.0254& 0.0594& 0.0007 \\
 $\Delta\cal{S}$ & 0.0177& 0.0101& 0.1497& 0.2136& 0.3709& 0.0865& 0.0121& 0.2190& 0.0897& 0.0631& 0.0058 \\
 $\Delta v_r$    & 0.0001& 0.0040& 0.0268& 0.0973& 0.0668& 0.0383& 0.0162& 0.0219& 0.0027& 0.0330& 0.0000 \\
\noalign{\smallskip}
Bulge $\sum$     & 0.0138& 0.0314& 0.6323& 4.4208& 3.6806& 0.4916& 2.3301& 4.0002& 1.4863& 4.4089& 0.0979 \\
 $\Delta\cal{B}$ & 0.0070& 0.0073& 0.1302& 0.2421& 0.3162& 0.1589& 0.2839& 0.4517& 0.2069& 0.4608& 0.0059 \\
 $\Delta\cal{S}$ & 0.0440& 0.0544& 0.2154& 0.7785& 0.9333& 0.7229& 0.4293& 0.8770& 0.4036& 0.4365& 0.0104 \\
 $\Delta v_r$    & 0.0001& 0.0010& 0.0168& 0.0291& 0.0208& 0.0329& 0.0209& 0.0316& 0.0172& 0.0173& 0.0000 \\
\noalign{\smallskip}
\hline
\end{tabular}
\end{center}
}
\begin{list}{}{}
\item[$^{\mathrm{a}}$] $\Delta\cal{B}$: Bootstraped 1$\sigma$ Poisson
  error, $\Delta\cal{S}$: slit-to-slit 1$\sigma$ scatter for index
  measurements in different pointings, $\Delta v_r$: systematic
  uncertainty due to radial velocity errors.
\end{list}
\end{table*}
\addtocounter{table}{-1}
\begin{table*}[h!]
      \caption{-- continued. Lick indices Mg$_2$ -- TiO$_2$.} 
        {\tiny
        \begin{center}
        \begin{tabular}{lcccccccccccccccccccc}  
         \noalign{\smallskip}
         \hline
         \noalign{\smallskip}
cluster$^{\mathrm{a}}$&Mg$_2$ &  Mgb  & Fe5270& Fe5335& Fe5406& Fe5709& Fe5782&  NaD  &TiO$_1$&TiO$_2$  \\
                      &  mag  &  \AA  &  \AA  &  \AA  &  \AA  &  \AA  &  \AA  &  \AA  &  mag  &  mag   \\
\noalign{\smallskip}
\hline
\noalign{\smallskip}
 NGC~5927        & 0.2201& 3.5692& 2.2225& 1.8085& 1.1667& 0.5023& 0.8576& 4.5095& 0.0518& 0.0950 \\
 $\Delta\cal{B}$ & 0.0010& 0.0365& 0.0417& 0.0465& 0.0360& 0.0338& 0.0268& 0.0372& 0.0009& 0.0008 \\
 $\Delta\cal{S}$ & 0.0098& 0.4197& 0.0979& 0.2000& 0.1297& 0.1372& 0.2161& 0.4903& 0.0246& 0.0501 \\
 $\Delta v_r$    & 0.0001& 0.0193& 0.0493& 0.0372& 0.0294& 0.0284& 0.0263& 0.0218& 0.0005& 0.0002 \\
\noalign{\smallskip}
 NGC~6388        & 0.1458& 2.1580& 1.8828& 1.6670& 1.0154& 0.4942& 0.6622& 3.7854& 0.0327& 0.0472 \\
 $\Delta\cal{B}$ & 0.0003& 0.0110& 0.0153& 0.0131& 0.0122& 0.0105& 0.0087& 0.0136& 0.0004& 0.0003 \\
 $\Delta\cal{S}$ & 0.0064& 0.0661& 0.0407& 0.0477& 0.0349& 0.0553& 0.0493& 0.0762& 0.0069& 0.0125 \\
 $\Delta v_r$    & 0.0001& 0.0210& 0.0256& 0.0275& 0.0225& 0.0212& 0.0470& 0.0188& 0.0003& 0.0002 \\
\noalign{\smallskip}
 NGC~6528        & 0.2615& 3.7413& 2.3673& 2.2777& 1.5499& 0.8223& 0.7987& 5.1366& 0.0750& 0.1268 \\
 $\Delta\cal{B}$ & 0.0010& 0.0402& 0.0427& 0.0417& 0.0427& 0.0313& 0.0312& 0.0374& 0.0011& 0.0009 \\
 $\Delta\cal{S}$ & 0.0288& 0.1484& 0.3639& 0.1031& 0.1437& 0.1773& 0.0230& 0.5497& 0.0274& 0.0574 \\
 $\Delta v_r$    & 0.0002& 0.0081& 0.0405& 0.0479& 0.0372& 0.0366& 0.0920& 0.0069& 0.0004& 0.0002 \\
\noalign{\smallskip}
 NGC~6624        & 0.1721& 2.7280& 1.8158& 1.6403& 0.9789& 0.5009& 0.6411& 2.7063& 0.0470& 0.0628 \\
 $\Delta\cal{B}$ & 0.0005& 0.0169& 0.0238& 0.0307& 0.0252& 0.0172& 0.0196& 0.0214& 0.0006& 0.0005 \\
 $\Delta\cal{S}$ & 0.0248& 0.2100& 0.2131& 0.1359& 0.1355& 0.0276& 0.0760& 0.1674& 0.0178& 0.0403 \\
 $\Delta v_r$    & 0.0002& 0.0222& 0.0347& 0.0295& 0.0406& 0.0256& 0.0472& 0.0210& 0.0004& 0.0001 \\
\noalign{\smallskip}
 NGC~6218        & 0.0672& 1.0628& 0.7687& 0.8935& 0.2246&-0.1909& 0.2025& 1.2915& 0.0182& 0.0040 \\
 $\Delta\cal{B}$ & 0.0009& 0.0355& 0.0357& 0.0485& 0.0390& 0.0323& 0.0382& 0.0397& 0.0011& 0.0011 \\
 $\Delta\cal{S}$ & 0.0041& 0.3940& 0.1710& 0.2960& 0.0994& 0.1382& 0.0971& 0.0895& 0.0081& 0.0078 \\
 $\Delta v_r$    & 0.0002& 0.0423& 0.0150& 0.0282& 0.0043& 0.0064& 0.0209& 0.0393& 0.0002& 0.0004 \\
\noalign{\smallskip}
 NGC~6441        & 0.1756& 2.7262& 1.9505& 1.7140& 1.0239& 0.5481& 0.7786& 4.0577& 0.0215& 0.0543 \\
 $\Delta\cal{B}$ & 0.0008& 0.0275& 0.0366& 0.0386& 0.0246& 0.0257& 0.0206& 0.0251& 0.0007& 0.0006 \\
 $\Delta\cal{S}$ & 0.0061& 0.1431& 0.0354& 0.0728& 0.0190& 0.0349& 0.0331& 0.0282& 0.0043& 0.0081 \\
 $\Delta v_r$    & 0.0001& 0.0260& 0.0268& 0.0310& 0.0297& 0.0257& 0.0426& 0.0182& 0.0004& 0.0001 \\
\noalign{\smallskip}
 NGC~6553        & 0.2552& 3.8961& 2.6091& 2.2654& 1.2371& 0.7744& 1.0970& 3.8792& 0.0689& 0.1420 \\
 $\Delta\cal{B}$ & 0.0014& 0.0524& 0.0648& 0.0656& 0.0486& 0.0427& 0.0373& 0.0535& 0.0012& 0.0010 \\
 $\Delta\cal{S}$ & 0.0114& 0.2910& 0.1071& 0.1125& 0.1042& 0.0496& 0.0138& 0.1175& 0.0130& 0.0254 \\
 $\Delta v_r$    & 0.0005& 0.0534& 0.0706& 0.0611& 0.0595& 0.0388& 0.0969& 0.0312& 0.0009& 0.0001 \\
\noalign{\smallskip}
 NGC~6626        & 0.0919& 1.3679& 1.0900& 0.9747& 0.5413& 0.1846& 0.4735& 2.1005& 0.0288& 0.0382 \\
 $\Delta\cal{B}$ & 0.0007& 0.0308& 0.0348& 0.0356& 0.0311& 0.0296& 0.0275& 0.0321& 0.0008& 0.0006 \\
 $\Delta\cal{S}$ & 0.0139& 0.0616& 0.2846& 0.2615& 0.2037& 0.1425& 0.0976& 0.0547& 0.0015& 0.0017 \\
 $\Delta v_r$    & 0.0001& 0.0241& 0.0212& 0.0200& 0.0146& 0.0198& 0.0429& 0.0210& 0.0001& 0.0002 \\
\noalign{\smallskip}
 NGC~6284        & 0.0966& 1.4403& 0.8563& 1.0216& 0.5178& 0.1110& 0.3141& 2.3978& 0.0159& 0.0039 \\
 $\Delta\cal{B}$ & 0.0010& 0.0333& 0.0411& 0.0440& 0.0378& 0.0338& 0.0271& 0.0414& 0.0010& 0.0009 \\
 $\Delta\cal{S}$ & 0.0098& 0.0494& 0.3418& 0.0506& 0.0510& 0.0486& 0.0908& 0.2560& 0.0036& 0.0051 \\
 $\Delta v_r$    & 0.0001& 0.0333& 0.0029& 0.0146& 0.0251& 0.0244& 0.0357& 0.0027& 0.0002& 0.0002 \\
\noalign{\smallskip}
 NGC~6356        & 0.1773& 2.7863& 1.7187& 1.6597& 0.9557& 0.4067& 0.5493& 3.2660& 0.0333& 0.0531 \\
 $\Delta\cal{B}$ & 0.0006& 0.0218& 0.0228& 0.0293& 0.0208& 0.0215& 0.0185& 0.0252& 0.0006& 0.0006 \\
 $\Delta\cal{S}$ & 0.0237& 0.2217& 0.0335& 0.0657& 0.1075& 0.0383& 0.0426& 0.1860& 0.0208& 0.0434 \\
 $\Delta v_r$    & 0.0001& 0.0185& 0.0318& 0.0143& 0.0185& 0.0201& 0.0416& 0.0182& 0.0004& 0.0002 \\
\noalign{\smallskip}
 NGC~6637        & 0.1542& 2.5420& 1.6335& 1.3969& 0.8222& 0.3565& 0.4906& 2.6053& 0.0381& 0.0441 \\
 $\Delta\cal{B}$ & 0.0005& 0.0167& 0.0212& 0.0218& 0.0175& 0.0141& 0.0145& 0.0221& 0.0005& 0.0005 \\
 $\Delta\cal{S}$ & 0.0059& 0.1053& 0.0650& 0.0750& 0.0131& 0.0302& 0.0585& 0.1659& 0.0030& 0.0084 \\
 $\Delta v_r$    & 0.0001& 0.0264& 0.0335& 0.0287& 0.0338& 0.0257& 0.0322& 0.0169& 0.0004& 0.0001 \\
\noalign{\smallskip}
 NGC~6981        & 0.0618& 1.0838& 0.8710& 0.5990& 0.1532&-0.1320& 0.0108& 1.4160& 0.0143&-0.0045 \\
 $\Delta\cal{B}$ & 0.0008& 0.0315& 0.0363& 0.0413& 0.0308& 0.0314& 0.0313& 0.0374& 0.0008& 0.0009 \\
 $\Delta\cal{S}$ & 0.0038& 0.1683& 0.2496& 0.1426& 0.5998& 0.0718& 0.1546& 0.1334& 0.0153& 0.0069 \\
 $\Delta v_r$    & 0.0001& 0.0121& 0.0326& 0.0165& 0.0046& 0.0110& 0.0084& 0.0074& 0.0001& 0.0000 \\
\noalign{\smallskip}
Bulge $\sum$     & 0.2281& 2.9879& 2.3942& 1.9040& 1.1906& 0.6347& 0.8276& 5.6267& 0.0466& 0.0820 \\
 $\Delta\cal{B}$ & 0.0064& 0.2415& 0.2273& 0.3218& 0.1953& 0.1871& 0.1566& 0.2401& 0.0060& 0.0063 \\
 $\Delta\cal{S}$ & 0.0305& 0.4243& 0.4991& 0.1580& 0.2601& 0.2380& 0.2212& 1.1913& 0.0319& 0.0570 \\
 $\Delta v_r$    & 0.0000& 0.0020& 0.0041& 0.0100& 0.0020& 0.0098& 0.0160& 0.0127& 0.0001& 0.0000 \\
\noalign{\smallskip}
\hline
\end{tabular}
\end{center}
}
\begin{list}{}{}
\item[$^{\mathrm{a}}$] $\Delta\cal{B}$: Bootstraped 1$\sigma$ Poisson
  error, $\Delta\cal{S}$: slit-to-slit 1$\sigma$ scatter for index
  measurements in different pointings, $\Delta v_r$: systematic
  uncertainty due to radial velocity errors.
\end{list}
\end{table*}

\addtocounter{table}{-1}
\begin{table*}[h!]
      \caption{-- continued. Lick indices H$\delta_{\rm A}$ --
        H$\gamma_{\rm F}$.} 
        {\tiny
        \begin{center}
        \begin{tabular}{lcccc}  
         \noalign{\smallskip}
         \hline
         \noalign{\smallskip}
cluster$^{\mathrm{a}}$&  H$\delta_{\rm A}$  &  H$\gamma_{\rm A}$   
                      &  H$\delta_{\rm F}$  &  H$\gamma_{\rm F}$  \\
                      &  \AA  &  \AA  &  \AA  &  \AA  \\
\noalign{\smallskip}
\hline
\noalign{\smallskip}
 NGC~5927        & -1.8817& -4.1268&  0.0876& -1.0638 \\
 $\Delta\cal{B}$ &  0.0391&  0.0429&  0.0275&  0.0272 \\
 $\Delta\cal{S}$ &  0.8170&  0.8869&  0.5176&  0.4666 \\
 $\Delta v_r$    &  0.0518&  0.0159&  0.0458&  0.0085 \\
\noalign{\smallskip}
 NGC~6388        &  0.1308& -2.6045&  1.0184&  0.0148 \\
 $\Delta\cal{B}$ &  0.0099&  0.0133&  0.0060&  0.0074 \\
 $\Delta\cal{S}$ &  0.0297&  0.0570&  0.0013&  0.0165 \\
 $\Delta v_r$    &  0.0206&  0.0022&  0.0114&  0.0065 \\
\noalign{\smallskip}
 NGC~6528        & -1.6708& -5.7216&  0.3290& -1.3866 \\
 $\Delta\cal{B}$ &  0.0441&  0.0483&  0.0296&  0.0301 \\
 $\Delta\cal{S}$ &  0.1726&  0.0633&  0.0203&  0.0815 \\
 $\Delta v_r$    &  0.0437&  0.0011&  0.0294&  0.0037 \\
\noalign{\smallskip}
 NGC~6624        & -0.5802& -3.6927&  0.6196& -0.6094 \\
 $\Delta\cal{B}$ &  0.0183&  0.0219&  0.0134&  0.0156 \\
 $\Delta\cal{S}$ &  0.0475&  0.1873&  0.0324&  0.0609 \\
 $\Delta v_r$    &  0.0324&  0.0147&  0.0313&  0.0043 \\
\noalign{\smallskip}
 NGC~6218        &  3.4626&  1.6864&  2.6734&  1.9579 \\
 $\Delta\cal{B}$ &  0.0179&  0.0254&  0.0112&  0.0149 \\
 $\Delta\cal{S}$ &  1.3602&  1.6168&  0.8272&  0.9327 \\
 $\Delta v_r$    &  0.0026&  0.0263&  0.0252&  0.0235 \\
\noalign{\smallskip}
 NGC~6441        &  0.1216& -2.7734&  1.0666& -0.0213 \\
 $\Delta\cal{B}$ &  0.0256&  0.0282&  0.0169&  0.0191 \\
 $\Delta\cal{S}$ &  0.1103&  0.0950&  0.0797&  0.0350 \\
 $\Delta v_r$    &  0.0284&  0.0056&  0.0178&  0.0037 \\
\noalign{\smallskip}
 NGC~6553        & -1.8415& -5.9315&  0.7905& -1.3923 \\
 $\Delta\cal{B}$ &  0.0709&  0.0685&  0.0422&  0.0423 \\
 $\Delta\cal{S}$ &  0.6727&  0.9318&  0.7638&  0.3674 \\
 $\Delta v_r$    &  0.0850&  0.0117&  0.0201&  0.0058 \\
\noalign{\smallskip}
 NGC~6626        &  2.6597&  0.3340&  2.3038&  1.3618 \\
 $\Delta\cal{B}$ &  0.0215&  0.0285&  0.0163&  0.0159 \\
 $\Delta\cal{S}$ &  0.4341&  0.8270&  0.2633&  0.3315 \\
 $\Delta v_r$    &  0.0045&  0.0188&  0.0070&  0.0092 \\
\noalign{\smallskip}
 NGC~6284        &  2.3486&  0.3707&  2.4889&  1.3953 \\
 $\Delta\cal{B}$ &  0.0242&  0.0295&  0.0171&  0.0193 \\
 $\Delta\cal{S}$ &  0.3979&  0.2117&  0.2231&  0.1705 \\
 $\Delta v_r$    &  0.0377&  0.0241&  0.0137&  0.0085 \\
\noalign{\smallskip}
 NGC~6356        & -0.7045& -3.8585&  0.5987& -0.7734 \\
 $\Delta\cal{B}$ &  0.0185&  0.0225&  0.0116&  0.0135 \\
 $\Delta\cal{S}$ &  0.1400&  0.1668&  0.1752&  0.0702 \\
 $\Delta v_r$    &  0.0301&  0.0147&  0.0147&  0.0030 \\
\noalign{\smallskip}
 NGC~6637        & -0.7261& -3.9037&  0.3773& -0.8734 \\
 $\Delta\cal{B}$ &  0.0154&  0.0184&  0.0107&  0.0124 \\
 $\Delta\cal{S}$ &  0.1956&  0.2869&  0.1325&  0.1962 \\
 $\Delta v_r$    &  0.0309&  0.0192&  0.0160&  0.0094 \\
\noalign{\smallskip}
 NGC~6981        &  1.8885&  0.6335&  1.7022&  1.3738 \\
 $\Delta\cal{B}$ &  0.0174&  0.0225&  0.0119&  0.0153 \\
 $\Delta\cal{S}$ &  0.6826&  0.4187&  0.4062&  0.2100 \\ 
 $\Delta v_r$    &  0.0084&  0.0023&  0.0109&  0.0091 \\
\noalign{\smallskip}
Bulge $\sum$     & -0.2080& -3.2443&  0.7604& -0.4023 \\
 $\Delta\cal{B}$ &  0.2404&  0.2509&  0.1619&  0.1558 \\
 $\Delta\cal{S}$ &  1.3701&  1.3918&  0.6590&  0.6703 \\
 $\Delta v_r$    &  0.0144&  0.0029&  0.0098&  0.0011 \\
\noalign{\smallskip}
\hline
\end{tabular}
\end{center}
}
\begin{list}{}{}
\item[$^{\mathrm{a}}$] $\Delta\cal{B}$: Bootstraped 1$\sigma$ Poisson
  error, $\Delta\cal{S}$: slit-to-slit 1$\sigma$ scatter for index
  measurements in different pointings, $\Delta v_r$: systematic
  uncertainty due to radial velocity errors.
\end{list}
\end{table*}

\section{Line-Index Measurements using Old Passband Definitions}
\label{ln:indexmeasurements_old}
Table~\ref{tab:gcindices1_old} summarises all our Lick index
measurements which were computed with the {\it old} set of index
passband definitions from \cite{worthey94etal}.

\begin{table*}[h!]
\begin{center}
  \caption{Summary of the coefficients $\alpha$ and $\beta$ for all
	1st and 2nd-order index corrections (see
	Section~\ref{ln:licktrafo}). This set of correction
	coefficients is valid for the {\it old} passband definitions
	\citep[see][]{worthey94etal}.}
\label{tab:indexlicktrafo_old}
\begin{tabular}[angle=0,width=\textwidth]{l|rrrc}
\hline
\noalign{\smallskip}
 index & z.p. -- $\alpha$ & slope -- $\beta$ & r.m.s. & units\\ 
\noalign{\smallskip}
\hline
\noalign{\smallskip}
CN$_1$   & $-0.0003$ & $ 0.0090$ & $0.0181$ &mag \\
CN$_2$   & $-0.0271$ & $ 0.1072$ & $0.0221$ &mag \\
Ca4227   & $ 0.1738$ & $-0.0696$ & $0.3295$ &\AA \\
G4300    & $-1.5890$ & $ 0.3104$ & $0.4031$ &\AA \\
Fe4384   & $ 0.4658$ & $-0.0057$ & $0.4747$ &\AA \\
Ca4455   & $ 1.1987$ & $-0.2414$ & $0.3443$ &\AA \\
Fe4531   & $ 0.2109$ & $ 0.0077$ & $0.1986$ &\AA \\
Fe4668   & $-1.1136$ & $ 0.1393$ & $0.6719$ &\AA \\
H$\beta$ & $ 0.1027$ & $-0.0648$ & $0.1054$ &\AA \\
Fe5015   & $-0.7017$ & $ 0.0811$ & $0.2113$ &\AA \\
Mg$_1$   & $ 0.0117$ & $-0.0021$ & $0.0107$ &mag \\
Mg$_2$   & $ 0.0133$ & $ 0.0300$ & $0.0117$ &mag \\
Mgb      & $ 0.2782$ & $-0.0747$ & $0.1400$ &\AA \\
Fe5270   & $-0.2250$ & $ 0.0277$ & $0.2185$ &\AA \\
Fe5335   & $-0.2182$ & $ 0.0126$ & $0.2135$ &\AA \\
Fe5406   & $ 0.0676$ & $-0.0641$ & $0.0330$ &\AA \\
Fe5709   & $-0.0406$ & $ 0.1049$ & $0.0860$ &\AA \\
Fe5782   & $ 0.1770$ & $-0.1260$ & $0.1626$ &\AA \\
NaD      & $ 0.2294$ & $-0.0432$ & $0.1994$ &\AA \\
TiO$_1$  & $ 0.0156$ & $ 0.2159$ & $0.0126$ &mag \\
TiO$_2$  & $-0.0112$ & $ 0.1047$ & $0.0258$ &mag \\
\noalign{\smallskip}
\hline
\end{tabular}
\end{center}
\begin{list}{}{}
\item[{\it Note}:] The Lick system provides a library of standard star
spectra with published Lick index measurements
\citep{worthey94etal}. We were able to reproduce the published values
of all Lick indices with the exception of Ca4227 and Ca4455. For these
two we find a significant offset and relatively large scatter compared
with the published values. Henceforth, we mark both Ca indices as
possibly unreliable Ca abundance indicators.
\end{list}
\end{table*}

\begin{table*}[h!]
  \caption{Lick indices CN1 -- Mg$_1$ for all sample globular clusters
    including statistical and systematic errors. Lines as in
    Table~\ref{tab:gcindices1}. This set of indices uses the {\it old}
    passband definitions of \cite{worthey94etal}.}
 \label{tab:gcindices1_old}
        {\tiny
        \begin{center}
        \begin{tabular}{lccccccccccccccccccccc}  
         \hline
         \noalign{\smallskip}

cluster$^{\mathrm{a}}$&  CN1  &  CN2  & Ca4227& G4300 & Fe4383& Ca4455& Fe4531& Fe4668&H$\beta$&Fe5015&  Mg$_1$   \\
                      &  mag  &  mag  &  \AA  &  \AA  &  \AA  &  \AA  &  \AA  &  \AA  &  \AA  &  \AA  &  mag   \\
\noalign{\smallskip}
\hline
\noalign{\smallskip}
 NGC~5927        & 0.0865& 0.0978& 0.9411& 3.8064& 3.5211& 1.6646& 2.7285& 2.4637& 1.6045& 4.4883& 0.0810 \\
 $\Delta\cal{B}$ & 0.0010& 0.0012& 0.0179& 0.0386& 0.0507& 0.0295& 0.0459& 0.0702& 0.0344& 0.0722& 0.0009 \\
 $\Delta\cal{S}$ & 0.0126& 0.0241& 0.1058& 1.5576& 0.9231& 0.0672& 0.3133& 0.4348& 0.1096& 0.3234& 0.0186 \\
 $\Delta v_r$    & 0.0012& 0.0047& 0.0702& 0.1200& 0.0163& 0.1565& 0.0747& 0.0503& 0.0318& 0.1213& 0.0002 \\
\noalign{\smallskip}
 NGC~6388        & 0.0469& 0.0447& 0.6527& 3.4736& 2.9890& 1.4677& 2.6294& 1.0270& 1.9557& 3.4097& 0.0515 \\
 $\Delta\cal{B}$ & 0.0003& 0.0004& 0.0057& 0.0109& 0.0139& 0.0088& 0.0156& 0.0218& 0.0112& 0.0230& 0.0003 \\
 $\Delta\cal{S}$ & 0.0033& 0.0024& 0.0500& 0.0257& 0.0776& 0.0448& 0.0532& 0.0584& 0.0456& 0.0944& 0.0026 \\
 $\Delta v_r$    & 0.0002& 0.0034& 0.0364& 0.1040& 0.1054& 0.1012& 0.0684& 0.0729& 0.0276& 0.0852& 0.0001 \\
\noalign{\smallskip}
 NGC~6528        & 0.0985& 0.1008& 1.0854& 4.8485& 5.0006& 1.6773& 2.9677& 3.9408& 1.7349& 5.0086& 0.1105 \\
 $\Delta\cal{B}$ & 0.0013& 0.0015& 0.0230& 0.0409& 0.0569& 0.0339& 0.0521& 0.0846& 0.0339& 0.0832& 0.0008 \\
 $\Delta\cal{S}$ & 0.0032& 0.0033& 0.0418& 0.1862& 0.1746& 0.0540& 0.1560& 0.7967& 0.0784& 0.1125& 0.0150 \\
 $\Delta v_r$    & 0.0006& 0.0039& 0.0809& 0.1036& 0.1806& 0.1511& 0.0951& 0.1365& 0.0314& 0.1191& 0.0001 \\
\noalign{\smallskip}
 NGC~6624        & 0.0517& 0.0521& 0.7426& 4.3356& 2.9452& 1.4807& 2.6817& 1.1594& 1.6189& 3.5552& 0.0656 \\
 $\Delta\cal{B}$ & 0.0005& 0.0006& 0.0116& 0.0181& 0.0274& 0.0127& 0.0250& 0.0380& 0.0171& 0.0414& 0.0005 \\
 $\Delta\cal{S}$ & 0.0070& 0.0082& 0.0855& 0.0576& 0.3535& 0.0754& 0.1804& 0.6350& 0.0228& 0.3347& 0.0146 \\
 $\Delta v_r$    & 0.0007& 0.0044& 0.0604& 0.1324& 0.1360& 0.1319& 0.0567& 0.0915& 0.0264& 0.1137& 0.0002 \\
\noalign{\smallskip}
 NGC~6218        &-0.0772&-0.1060& 0.3901& 1.0869& 0.7658& 1.2549& 1.4226&-0.9178& 2.6126& 1.3515& 0.0211 \\
 $\Delta\cal{B}$ & 0.0007& 0.0008& 0.0123& 0.0272& 0.0396& 0.0201& 0.0426& 0.0666& 0.0274& 0.0645& 0.0008 \\
 $\Delta\cal{S}$ & 0.0267& 0.0248& 0.0839& 1.1431& 0.4418& 0.0321& 0.3419& 0.2824& 0.5643& 0.4039& 0.0020 \\
 $\Delta v_r$    & 0.0003& 0.0108& 0.0659& 0.2294& 0.1154& 0.0688& 0.0689& 0.0654& 0.0049& 0.0454& 0.0000 \\
\noalign{\smallskip}
 NGC~6441        & 0.0553& 0.0537& 0.7283& 3.4935& 3.2861& 1.4836& 2.6659& 1.0053& 1.8900& 3.5838& 0.0671 \\
 $\Delta\cal{B}$ & 0.0007& 0.0009& 0.0129& 0.0300& 0.0404& 0.0200& 0.0335& 0.0541& 0.0270& 0.0482& 0.0006 \\
 $\Delta\cal{S}$ & 0.0024& 0.0039& 0.0065& 0.1118& 0.1000& 0.0440& 0.0453& 0.0991& 0.0571& 0.0646& 0.0011 \\
 $\Delta v_r$    & 0.0004& 0.0038& 0.0498& 0.1094& 0.1133& 0.1038& 0.0668& 0.0859& 0.0239& 0.0860& 0.0000 \\
\noalign{\smallskip}
 NGC~6553        & 0.1409& 0.1483& 1.1934& 5.1077& 4.1114& 1.6442& 3.3920& 3.0414& 1.8410& 5.8677& 0.0955 \\
 $\Delta\cal{B}$ & 0.0020& 0.0022& 0.0324& 0.0627& 0.0858& 0.0424& 0.0781& 0.1107& 0.0549& 0.1332& 0.0013 \\
 $\Delta\cal{S}$ & 0.0140& 0.0064& 0.0563& 0.0298& 0.8549& 0.1294& 0.0306& 0.7740& 0.1128& 0.4396& 0.0038 \\
 $\Delta v_r$    & 0.0014& 0.0081& 0.1601& 0.1799& 0.3711& 0.2414& 0.1395& 0.2696& 0.0632& 0.1826& 0.0007 \\
\noalign{\smallskip}
 NGC~6626        &-0.0456&-0.0667& 0.4746& 1.7188& 1.3821& 1.2905& 1.7474&-0.3440& 2.2019& 2.0504& 0.0360 \\
 $\Delta\cal{B}$ & 0.0007& 0.0008& 0.0135& 0.0262& 0.0451& 0.0226& 0.0303& 0.0645& 0.0276& 0.0568& 0.0007 \\
 $\Delta\cal{S}$ & 0.0074& 0.0070& 0.0386& 0.5166& 0.0920& 0.0277& 0.1797& 0.3180& 0.2934& 0.5749& 0.0119 \\
 $\Delta v_r$    & 0.0002& 0.0065& 0.0447& 0.1478& 0.0980& 0.0721& 0.0542& 0.0687& 0.0204& 0.0711& 0.0000 \\
\noalign{\smallskip}
 NGC~6284        &-0.0409&-0.0619& 0.4875& 1.7623& 1.4881& 1.3502& 1.8817&-0.3747& 2.3444& 2.0094& 0.0373 \\
 $\Delta\cal{B}$ & 0.0008& 0.0010& 0.0138& 0.0326& 0.0473& 0.0218& 0.0357& 0.0604& 0.0266& 0.0670& 0.0007 \\
 $\Delta\cal{S}$ & 0.0141& 0.0122& 0.0468& 0.3412& 0.2661& 0.1297& 0.0685& 0.2517& 0.1168& 0.0970& 0.0079 \\
 $\Delta v_r$    & 0.0000& 0.0069& 0.0442& 0.1951& 0.0857& 0.0932& 0.0783& 0.0438& 0.0080& 0.0952& 0.0001 \\
\noalign{\smallskip}
 NGC~6356        & 0.0465& 0.0413& 0.7848& 4.5930& 2.8455& 1.5022& 2.6618& 1.0041& 1.6038& 3.3588& 0.0678 \\
 $\Delta\cal{B}$ & 0.0006& 0.0007& 0.0088& 0.0195& 0.0315& 0.0154& 0.0252& 0.0479& 0.0213& 0.0440& 0.0005 \\
 $\Delta\cal{S}$ & 0.0040& 0.0096& 0.0370& 0.1187& 0.1224& 0.0843& 0.1665& 0.3278& 0.0783& 0.3039& 0.0119 \\
 $\Delta v_r$    & 0.0006& 0.0043& 0.0568& 0.0993& 0.1672& 0.1287& 0.0789& 0.0957& 0.0297& 0.1107& 0.0002 \\
\noalign{\smallskip}
 NGC~6637        & 0.0266& 0.0201& 0.6057& 4.8329& 2.6127& 1.4665& 2.5420& 1.0281& 1.5930& 3.2077& 0.0514 \\
 $\Delta\cal{B}$ & 0.0005& 0.0005& 0.0084& 0.0139& 0.0235& 0.0124& 0.0205& 0.0340& 0.0156& 0.0370& 0.0004 \\
 $\Delta\cal{S}$ & 0.0020& 0.0082& 0.1589& 0.0341& 0.1300& 0.0900& 0.0233& 0.1585& 0.0442& 0.0748& 0.0021 \\
 $\Delta v_r$    & 0.0004& 0.0032& 0.0886& 0.1077& 0.1419& 0.1150& 0.0743& 0.0832& 0.0267& 0.1000& 0.0001 \\
\noalign{\smallskip}
 NGC~6981        &-0.0429&-0.0781& 0.4922& 1.4464& 0.8959& 1.3249& 1.4807&-0.5578& 2.3203& 1.2395& 0.0238 \\
 $\Delta\cal{B}$ & 0.0006& 0.0007& 0.0123& 0.0218& 0.0359& 0.0201& 0.0331& 0.0572& 0.0258& 0.0652& 0.0007 \\
 $\Delta\cal{S}$ & 0.0186& 0.0127& 0.1220& 0.2878& 0.3189& 0.0487& 0.0215& 0.3210& 0.0837& 0.0947& 0.0059 \\
 $\Delta v_r$    & 0.0001& 0.0038& 0.0345& 0.0813& 0.0607& 0.0336& 0.0322& 0.0297& 0.0023& 0.0317& 0.0000 \\
\noalign{\smallskip}
Bulge $\sum$     & 0.0151& 0.0041& 0.8364& 3.6699& 4.2236& 1.4869& 2.6659& 3.7960& 1.4659& 3.8913& 0.0932 \\
 $\Delta\cal{B}$ & 0.0070& 0.0080& 0.1300& 0.2304& 0.3521& 0.1761& 0.2696& 0.4389& 0.2203& 0.4931& 0.0062 \\
 $\Delta\cal{S}$ & 0.0447& 0.0649& 0.2022& 1.1400& 1.0132& 0.3815& 0.4590& 0.8930& 0.3848& 0.6334& 0.0108 \\
 $\Delta v_r$    & 0.0001& 0.0014& 0.0298& 0.0401& 0.0366& 0.0413& 0.0311& 0.0450& 0.0161& 0.0314& 0.0000 \\
\noalign{\smallskip}
\hline
\end{tabular}
\end{center}
}
\begin{list}{}{}
\item[$^{\mathrm{a}}$] $\Delta\cal{B}$: Bootstraped 1$\sigma$ Poisson
  error, $\Delta\cal{S}$: slit-to-slit 1$\sigma$ scatter for index
  measurements in different pointings, $\Delta v_r$: systematic
  uncertainty due to radial velocity errors.
\end{list}
\end{table*}
\addtocounter{table}{-1}
\begin{table*}[h!]
      \caption{-- continued. Lick indices Mg$_2$ -- TiO$_2$.} 
        {\tiny
        \begin{center}
        \begin{tabular}{lcccccccccccccccccccc}  
         \noalign{\smallskip}
         \hline
         \noalign{\smallskip}
cluster$^{\mathrm{a}}$&Mg$_2$ &  Mgb  & Fe5270& Fe5335& Fe5406& Fe5709& Fe5782&  NaD  &TiO$_1$&TiO$_2$  \\
                      &  mag  &  \AA  &  \AA  &  \AA  &  \AA  &  \AA  &  \AA  &  \AA  &  mag  &  mag   \\
\noalign{\smallskip}
\hline
\noalign{\smallskip}
 NGC~5927        & 0.2204& 3.6656& 2.3001& 1.8048& 1.2981& 0.7175& 0.8580& 4.5614& 0.0540& 0.0977 \\
 $\Delta\cal{B}$ & 0.0008& 0.0363& 0.0437& 0.0467& 0.0420& 0.0335& 0.0292& 0.0395& 0.0010& 0.0009 \\
 $\Delta\cal{S}$ & 0.0096& 0.4042& 0.0957& 0.2183& 0.1310& 0.1160& 0.2358& 0.4793& 0.0233& 0.0470 \\
 $\Delta v_r$    & 0.0001& 0.0193& 0.0492& 0.0372& 0.0294& 0.0380& 0.0709& 0.0057& 0.0005& 0.0002 \\
\noalign{\smallskip}
 NGC~6388        & 0.1467& 2.3182& 1.9680& 1.6506& 1.1472& 0.7119& 0.6634& 3.7979& 0.0359& 0.0535 \\
 $\Delta\cal{B}$ & 0.0003& 0.0125& 0.0145& 0.0144& 0.0118& 0.0114& 0.0091& 0.0136& 0.0003& 0.0003 \\
 $\Delta\cal{S}$ & 0.0063& 0.0637& 0.0398& 0.0521& 0.0352& 0.0449& 0.0461& 0.0661& 0.0064& 0.0117 \\
 $\Delta v_r$    & 0.0000& 0.0210& 0.0256& 0.0275& 0.0225& 0.0308& 0.0720& 0.0035& 0.0004& 0.0000 \\
\noalign{\smallskip}
 NGC~6528        & 0.2608& 3.8430& 2.4415& 2.3173& 1.6869& 0.9225& 0.6725& 5.1391& 0.0758& 0.1293 \\
 $\Delta\cal{B}$ & 0.0011& 0.0368& 0.0464& 0.0475& 0.0333& 0.0287& 0.0308& 0.0361& 0.0008& 0.0008 \\
 $\Delta\cal{S}$ & 0.0284& 0.1429& 0.3556& 0.1126& 0.1450& 0.1399& 0.0070& 0.5457& 0.0258& 0.0538 \\
 $\Delta v_r$    & 0.0002& 0.0081& 0.0406& 0.0479& 0.0372& 0.0533& 0.1073& 0.0270& 0.0005& 0.0002 \\
\noalign{\smallskip}
 NGC~6624        & 0.1726& 2.8671& 1.9025& 1.6214& 1.1104& 0.7311& 0.6583& 2.6866& 0.0495& 0.0684 \\
 $\Delta\cal{B}$ & 0.0006& 0.0197& 0.0241& 0.0246& 0.0207& 0.0171& 0.0177& 0.0234& 0.0005& 0.0005 \\
 $\Delta\cal{S}$ & 0.0245& 0.2023& 0.2083& 0.1484& 0.1368& 0.0232& 0.0519& 0.1737& 0.0165& 0.0378 \\
 $\Delta v_r$    & 0.0001& 0.0222& 0.0347& 0.0295& 0.0406& 0.0353& 0.0762& 0.0099& 0.0004& 0.0001 \\
\noalign{\smallskip}
 NGC~6218        & 0.0691& 1.2634& 0.8790& 0.8061& 0.3487& 0.2373& 0.3669& 1.2326& 0.0234& 0.0137 \\
 $\Delta\cal{B}$ & 0.0008& 0.0311& 0.0392& 0.0505& 0.0345& 0.0306& 0.0293& 0.0429& 0.0010& 0.0010 \\
 $\Delta\cal{S}$ & 0.0040& 0.3795& 0.1671& 0.3232& 0.1003& 0.1060& 0.0832& 0.0898& 0.0075& 0.0075 \\
 $\Delta v_r$    & 0.0002& 0.0423& 0.0150& 0.0281& 0.0042& 0.0086& 0.0538& 0.0302& 0.0001& 0.0005 \\
\noalign{\smallskip}
 NGC~6441        & 0.1760& 2.8653& 2.0341& 1.7018& 1.1558& 0.7563& 0.7600& 4.0813& 0.0251& 0.0602 \\
 $\Delta\cal{B}$ & 0.0008& 0.0300& 0.0290& 0.0384& 0.0285& 0.0230& 0.0212& 0.0281& 0.0007& 0.0007 \\
 $\Delta\cal{S}$ & 0.0060& 0.1378& 0.0346& 0.0794& 0.0192& 0.0274& 0.0350& 0.0364& 0.0040& 0.0076 \\
 $\Delta v_r$    & 0.0000& 0.0260& 0.0268& 0.0310& 0.0298& 0.0346& 0.0742& 0.0065& 0.0004& 0.0000 \\
\noalign{\smallskip}
 NGC~6553        & 0.2545& 3.9920& 2.6779& 2.3038& 1.3711& 0.9255& 0.9838& 3.8916& 0.0695& 0.1431 \\
 $\Delta\cal{B}$ & 0.0014& 0.0533& 0.0579& 0.0665& 0.0492& 0.0446& 0.0391& 0.0488& 0.0012& 0.0011 \\
 $\Delta\cal{S}$ & 0.0112& 0.2803& 0.1047& 0.1228& 0.1052& 0.0427& 0.0216& 0.1280& 0.0122& 0.0237 \\
 $\Delta v_r$    & 0.0005& 0.0534& 0.0705& 0.0611& 0.0595& 0.0624& 0.1648& 0.0078& 0.0010& 0.0002 \\
\noalign{\smallskip}
 NGC~6626        & 0.0935& 1.5572& 1.1931& 0.8947& 0.6686& 0.4963& 0.5492& 2.0536& 0.0330& 0.0457 \\
 $\Delta\cal{B}$ & 0.0009& 0.0295& 0.0320& 0.0389& 0.0298& 0.0238& 0.0288& 0.0360& 0.0007& 0.0008 \\
 $\Delta\cal{S}$ & 0.0137& 0.0593& 0.2781& 0.2855& 0.2056& 0.0854& 0.0715& 0.0569& 0.0013& 0.0015 \\
 $\Delta v_r$    & 0.0000& 0.0241& 0.0212& 0.0200& 0.0146& 0.0282& 0.0721& 0.0089& 0.0002& 0.0003 \\
\noalign{\smallskip}
 NGC~6284        & 0.0982& 1.6270& 0.9647& 0.9459& 0.6448& 0.4366& 0.4276& 2.3230& 0.0206& 0.0128 \\
 $\Delta\cal{B}$ & 0.0009& 0.0371& 0.0451& 0.0454& 0.0349& 0.0310& 0.0313& 0.0374& 0.0010& 0.0009 \\
 $\Delta\cal{S}$ & 0.0096& 0.0476& 0.3341& 0.0552& 0.0514& 0.0558& 0.0631& 0.2339& 0.0034& 0.0049 \\
 $\Delta v_r$    & 0.0000& 0.0332& 0.0029& 0.0146& 0.0251& 0.0282& 0.0622& 0.0103& 0.0002& 0.0001 \\
\noalign{\smallskip}
 NGC~6356        & 0.1777& 2.9233& 1.8075& 1.6426& 1.0869& 0.6549& 0.6001& 3.2571& 0.0364& 0.0591 \\
 $\Delta\cal{B}$ & 0.0006& 0.0258& 0.0260& 0.0323& 0.0266& 0.0199& 0.0202& 0.0254& 0.0006& 0.0006 \\
 $\Delta\cal{S}$ & 0.0234& 0.2135& 0.0327& 0.0717& 0.1085& 0.0319& 0.0157& 0.1960& 0.0193& 0.0408 \\
 $\Delta v_r$    & 0.0001& 0.0184& 0.0318& 0.0143& 0.0185& 0.0315& 0.0670& 0.0053& 0.0004& 0.0000 \\
\noalign{\smallskip}
 NGC~6637        & 0.1549& 2.6880& 1.7243& 1.3556& 0.9521& 0.6132& 0.5669& 2.5727& 0.0409& 0.0508 \\
 $\Delta\cal{B}$ & 0.0005& 0.0171& 0.0218& 0.0227& 0.0182& 0.0172& 0.0173& 0.0195& 0.0005& 0.0005 \\
 $\Delta\cal{S}$ & 0.0058& 0.1014& 0.0636& 0.0819& 0.0132& 0.0164& 0.0664& 0.1795& 0.0029& 0.0080 \\
 $\Delta v_r$    & 0.0000& 0.0263& 0.0334& 0.0287& 0.0338& 0.0341& 0.0557& 0.0050& 0.0005& 0.0000 \\
\noalign{\smallskip}
 NGC~6981        & 0.0640& 1.2715& 1.0125& 0.4596& 0.2604& 0.2753& 0.2125& 1.3006& 0.0184& 0.0062 \\
 $\Delta\cal{B}$ & 0.0008& 0.0298& 0.0351& 0.0413& 0.0334& 0.0300& 0.0309& 0.0415& 0.0009& 0.0008 \\
 $\Delta\cal{S}$ & 0.0037& 0.1621& 0.2440& 0.1557& 0.6056& 0.0519& 0.1117& 0.1492& 0.0161& 0.0075 \\
 $\Delta v_r$    & 0.0001& 0.0114& 0.0297& 0.0162& 0.0046& 0.0081& 0.0044& 0.0052& 0.0003& 0.0001 \\
\noalign{\smallskip}
Bulge $\sum$     & 0.2279& 3.1174& 2.4678& 1.9093& 1.3241& 0.8255& 0.7953& 5.6970& 0.0489& 0.0864 \\
 $\Delta\cal{B}$ & 0.0073& 0.2229& 0.2303& 0.2526& 0.2251& 0.1816& 0.1798& 0.2335& 0.0056& 0.0045 \\
 $\Delta\cal{S}$ & 0.0315& 0.4435& 0.4931& 0.1720& 0.2542& 0.1928& 0.1954& 1.2423& 0.0299& 0.0538 \\
 $\Delta v_r$    & 0.0000& 0.0077& 0.0108& 0.0208& 0.0067& 0.0159& 0.0337& 0.0010& 0.0002& 0.0000 \\
\noalign{\smallskip}
\hline
\end{tabular}
\end{center}
}
\begin{list}{}{}
\item[$^{\mathrm{a}}$] $\Delta\cal{B}$: Bootstraped 1$\sigma$ Poisson
  error, $\Delta\cal{S}$: slit-to-slit 1$\sigma$ scatter for index
  measurements in different pointings, $\Delta v_r$: systematic
  uncertainty due to radial velocity errors.
\end{list}
\end{table*}

\end{document}